%% file: senseinsitu.tex
\documentclass[10pt,journal,compsoc]{IEEEtran}

\usepackage{ifpdf}
\ifCLASSOPTIONcompsoc
  \usepackage[nocompress]{cite}
\else
  \usepackage{cite}
\fi
\ifCLASSINFOpdf
\else
 \usepackage[dvips]{graphicx}
  \graphicspath{{figures/}{pictures/}{images/}{./}} 
\fi
\usepackage{amsmath}
\usepackage{algorithmic}
\usepackage{array}
\ifCLASSOPTIONcompsoc
  \usepackage[caption=false,font=footnotesize,labelfont=sf,textfont=sf]{subfig}
\else
  \usepackage[caption=false,font=footnotesize]{subfig}
\fi
\usepackage{stfloats}
\ifCLASSOPTIONcaptionsoff
  \usepackage[nomarkers]{endfloat}
 \let\MYoriglatexcaption\caption
 \renewcommand{\caption}[2][\relax]{\MYoriglatexcaption[#2]{#2}}
\fi
\usepackage{url}
\usepackage[dvipsnames,x11names]{xcolor}
\usepackage{adforn}
\usepackage{enumitem}
\usepackage{lscape}
\usepackage{svg}
\usepackage{comment}
\usepackage{tabularx}
\usepackage{colortbl}
\usepackage{tikz}
\usepackage{subfig}
\usepackage{multirow}
\usepackage{amssymb}
\usepackage{pifont}
\usepackage{graphics}
\usepackage{arydshln}
\usepackage{hhline}
\usepackage{booktabs}
\usepackage{wrapfig}
\usepackage{combelow}
\usepackage{soul}
\usepackage{arydshln}
\usepackage{float}
\usepackage{booktabs}
\usepackage{anyfontsize}

\definecolor{diagramPink}{RGB}{255,112,153}
\definecolor{diagramYellow}{RGB}{250,212,93}
\definecolor{myBlue}{RGB}{0,0,255}
\definecolor{tableBlue}{RGB}{109, 155, 195}

\definecolor{diagramBlue}{RGB}{144,211,245}
\definecolor{diagramYellow}{RGB}{250,212,93}
\definecolor{diagramOrange}{RGB}{244,155,127}
\definecolor{diagramGreen}{RGB}{208,230,187}

\definecolor{diagramGrey}{RGB}{100,100,100}

\definecolor{diagramRed}{RGB}{229,104,119}
\definecolor{diagramPink}{RGB}{255,112,153}

\definecolor{fulvous}{rgb}{0.86, 0.52, 0.0}

\newcounter{STc}
\newcommand{\cmark}{\cellcolor{tableBlue}}
\newcommand{\xmark}{\tiny\textcolor{white}{\ding{55}}}%

\newcommand{\smark}{\cellcolor{Thistle!60!white}}

\ADLnullwidehline

\definecolor{cIcolor}{RGB}{235,205,251} 
\definecolor{cIIcolor}{RGB}{255,198,195} 
\definecolor{cIIIcolor}{RGB}{255,242,181} 
\definecolor{cIVcolor}{RGB}{181,217,255} 
\definecolor{cVcolor}{RGB}{255,226,181} 
\definecolor{cVIcolor}{RGB}{193,242,200} 

\definecolor{OutlinerGray}{RGB}{230,230,230} 

\DeclareRobustCommand\mytikzcodebits[1]{\tikz{\draw[fill=#1]  rectangle (1em,1em);}}

\newcommand{\rev}[1]{{#1}} 

\newcommand{\AR}{AR}
\newcommand{\MR}{MR}
\newcommand{\VR}{VR}
\newcommand{\XR}{XR}
\newcommand{\SA}{SA}
\newcommand{\SV}{SV}
\newcommand{\IA}{IA}

\newlength{\Oldarrayrulewidth}
\newcommand{\Cline}[2]{%
  \noalign{\global\setlength{\Oldarrayrulewidth}{\arrayrulewidth}}%
  \noalign{\global\setlength{\arrayrulewidth}{#1}}\arrayrulecolor{black}\cline{#2}%
  \noalign{\global\setlength{\arrayrulewidth}{\Oldarrayrulewidth}}}

\begin{document}


\title{The Reality of the Situation: A Survey of Situated Analytics}

\author{Sungbok Shin, Andrea Batch, Peter W. S.\ Butcher,\\ Panagiotis D.\ Ritsos,~\IEEEmembership{Member,~IEEE} and Niklas Elmqvist,~\IEEEmembership{Senior Member,~IEEE}
\IEEEcompsocitemizethanks{
    \IEEEcompsocthanksitem Sungbok Shin, Andrea Batch, and Niklas Elmqvist are with University of Maryland, College Park, MD, United States. E-mail: \{sbshin90, ajulca, elm\}@umd.edu. 
    \IEEEcompsocthanksitem Peter W.\ S.\ Butcher and Panagiotis D.\ Ritsos are with Bangor University in the United Kingdom. E-mail: \{p.butcher, p.ritsos\}@bangor.ac.uk.}
    \thanks{Manuscript received XXX XX, 2022; revised XXX XX, 2023.}
}


\IEEEtitleabstractindextext{%
\begin{abstract}
    The advent of low-cost, accessible, and high-performance augmented reality (AR) has shed light on a situated form of analytics where in-situ visualizations embedded in the real world can facilitate sensemaking based on the user’s physical location.
    In this work, we identify prior literature in this emerging field with a focus on situated analytics.
    After collecting \rev{47} relevant situated analytics systems, we classify them using a taxonomy of three dimensions: situating triggers, view situatedness, and data depiction. 
    We then identify four archetypical patterns in our classification using an ensemble cluster analysis.
    We also assess the level which these systems support the sensemaking process. 
    Finally, we discuss insights and design guidelines that we learned from our analysis. 
\end{abstract}

}

\maketitle
\IEEEdisplaynontitleabstractindextext
\IEEEpeerreviewmaketitle

\IEEEraisesectionheading{\section{Introduction}\label{SEC:introduction}}
\input{01_intro}
\input{02_background}
\input{03_design_space}
\input{04_analysis}
\input{05_discussion}
\input{06_conclusion}

\section*{Acknowledgments}

We thank the anonymous reviewers for their feedback on this paper.
We also thank Sanghyun Hong and Hyeon Jeon for their comments on our work.
This work was partially supported by the U.S.\ National Science Foundation grant IIS-1908605, and the DSP Centre, which has been partly funded by the European Regional Development Fund through Welsh Government and also by the North Wales Growth Deal through Ambition North Wales, Welsh Government and UK Government.
Any opinions, findings, and conclusions or recommendations expressed here are those of the authors and do not necessarily reflect the views of the funding agencies.

\bibliographystyle{abbrv-doi}
\bibliography{senseinsitu}

\begin{IEEEbiography}[{\includegraphics[width=1in,height=1.25in,clip,keepaspectratio]{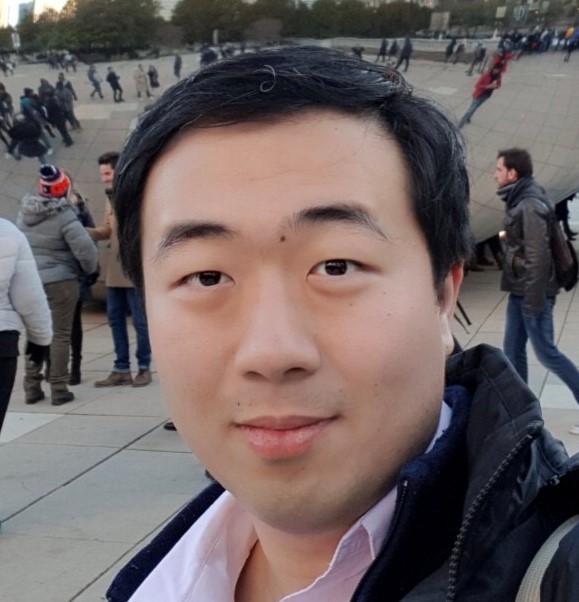}}]{Sungbok Shin} 
received the Master's degree in Computer Science and Engineering in 2019 from Korea University in Seoul, South Korea.
He is a Ph.D.\ student in the Department of Computer Science at the University of Maryland, College Park, Maryland, USA. 
He is also a student member of the Human-Computer Interaction Laboratory (HCIL) at UMD.
\end{IEEEbiography}

\begin{IEEEbiography}[{\includegraphics[width=1in,height=1.25in,clip,keepaspectratio]{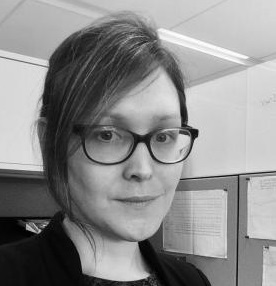}}]{Andrea Batch}
received the Ph.D.\ degree in information studies in 2022 from the University of Maryland, College Park, MD, USA.
She is an Economist at the U.S. Bureau of Economic Analysis.
She is also a member of the Human-Computer Interaction Laboratory (HCIL) at UMD.
\end{IEEEbiography}

\begin{IEEEbiography}[{\includegraphics[width=1in,height=1.25in,clip,keepaspectratio]{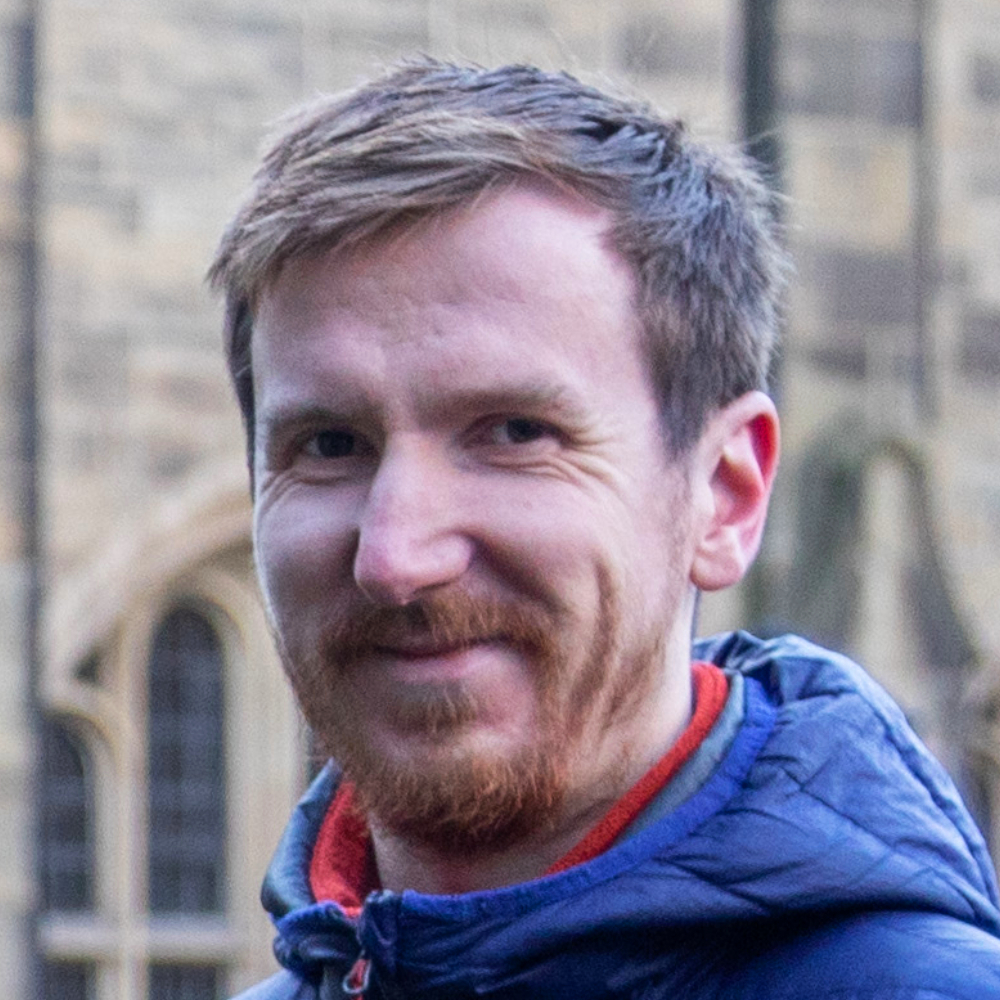}}]{Peter W.\ S.\ Butcher}
received the Ph.D. degree in 2020 from the University of Chester, UK.
He is a Lecturer in the School of Computer Science and Electronic Engineering, Bangor University, UK.
His research interests focus on immersive analytics with WebXR.
\end{IEEEbiography}

\begin{IEEEbiography}[{\includegraphics[width=1in,height=1.25in,clip,keepaspectratio]{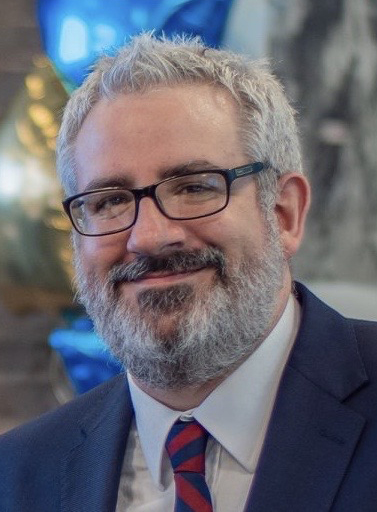}}]{Panagiotis D.\ Ritsos}
received the Ph.D. degree in 2006 from the University of Essex in Colchester, UK.
He is a Senior Lecturer in the School of Computer Science and Electronic Engineering, Bangor University, UK. 
His research interests include mixed and virtual reality, information visualization, visual analytics, and wearable computing. 
He is a member of the IEEE and the IEEE Computer Society.
\end{IEEEbiography}

\begin{IEEEbiography}[{\includegraphics[width=1in,height=1.25in,clip,keepaspectratio]{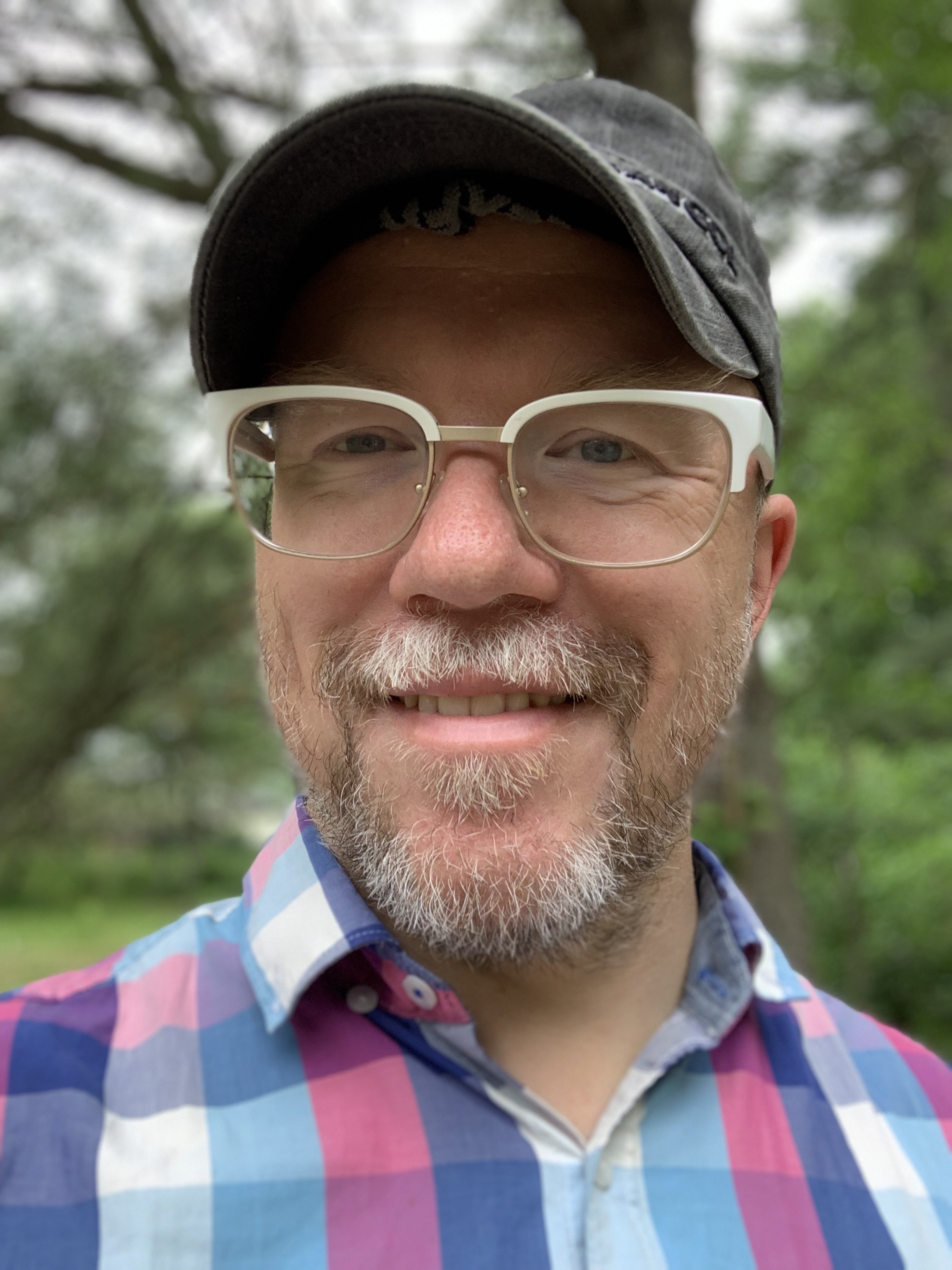}}]{Niklas Elmqvist}
received the Ph.D.\ degree in 2006 from Chalmers University of Technology in G\"{o}teborg, Sweden.
He is a professor in the College of Information Studies, University of Maryland, College Park in College Park, MD, USA. 
He is also a member of the Institute for Advanced Computer Studies (UMIACS) and formerly the director of the Human-Computer Interaction Laboratory (HCIL) at UMD.
He is a senior member of the IEEE and the IEEE Computer Society.
\end{IEEEbiography}

\vfill

\end{document}

%% file: 01_intro.tex
\IEEEPARstart{H}{umans} are fundamentally embodied beings~\cite{Shapiro2010}.
Our very own bodies, the objects we surround ourselves with, and the physical environment we live in all play a critical role in our thinking.
People tend to remember things better when they act them out bodily~\cite{Skulmowski2018}; spatial concepts abound in language (``I'm feeling \textit{up} today,'' or ``I'm on \textit{top} of this''), and thinking about the future causes people to lean slightly forwards, whereas thinking about the past causes them to lean backwards~\cite{Lynden2010}.
In fact, the theory of \textit{situated action} holds that human behavior can only be understood within its real-world context~\cite{DBLP:books/daglib/0066732}.
It follows that a \textit{situated} form of analytics~\cite{Thomas2018}, where physical space can play as much a part of sensemaking as the technology enabling it, could be particularly efficient for certain tasks.
After all, people routinely talk about being ``in the situation'' or ``on the ground'' as benefits to seeing a problem and thinking about it more clearly.
In some cases, there is information that can only be obtained by being in-situ.

\begin{figure*}[tbp]
    \centering
    \includegraphics[width=\linewidth]{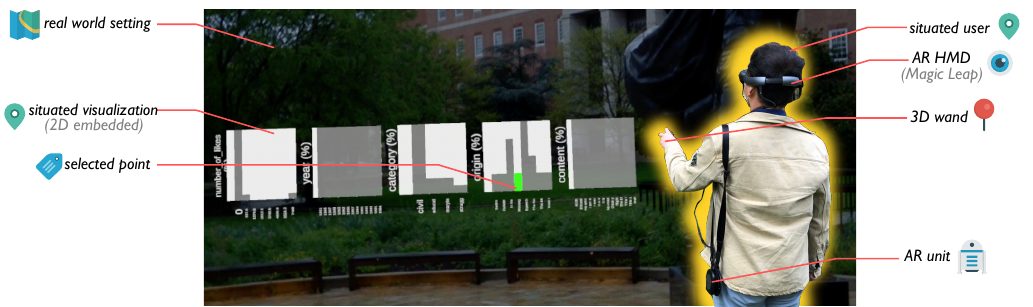}
    \caption{\textbf{Example of a situated analytics tool that uses Augmented Reality to enable sensemaking in-situ.}
    The user is exploring a set of visualizations that are contextual to (but not embedded in) their current real-world position and shown as 2D planes facing the user in the world.}
    \label{fig:teaser}
\end{figure*}

With the recent advances in Mixed/Augmented Reality (MR/AR) technology~\cite{Milgram1995}, where computer-generated imagery is overlaid and anchored on the real world, we may finally have reached a point where in-situ sensemaking is within reach for the average person.
This is the promise of \textit{situated visualization} (SV), where data relevant to a physical location is visualized directly in that location~\cite{bressa:hal-03319648, white09sitelens, Willet2017_Embedded}.
Such situated visualizations have long been a research topic in \AR{} research, dating back to the origins of the field.
The latest progress on higher-order sensemaking in mixed reality using situated visualization has now yielded the concept of \textit{situated analytics} (SA)~\cite{elsayed16situateddef, Thomas2018} (Fig.~\ref{fig:teaser}).

Recent years have seen a rise in \SA{} tools and techniques for \AR{}~\cite{elsayed16situateddef, abao18foodgo, whitlock20graphicalimmersive}, creating a mosaic of different solutions and approaches.
However, there is currently little consensus within the field of data visualization on the precise characteristics of situated analytics.
Some efforts, such as the work of Willett et al.~\cite{Willet2017_Embedded} revisit and extend terminology, whereas others, such as Ens et al.~\cite{Ens2021} focus on challenges in the broader area of \textit{immersive analytics} (IA)~\cite{Marriott2018}, as an umbrella concept of \SA{}.
And the overarching area of \textit{ubiquitous analytics}~\cite{Elmqvist2013} has been named as the future of a form of analytics that can be conducted anywhere and anytime~\cite{Elmqvist2023}.
Recently, Bressa et al.~\cite{bressa:hal-03319648} identified different perspectives of situated visualization and outlined future research directions.
However, none of these surveys and frameworks focus specifically on the \textit{situatedness} of data and visual representations and how the analytical process can be best \textit{integrated} into the real world during visual analytics.

In this paper, we address this gap in the literature by exploring \AR{} technologies and techniques developed for situated analytics.
For this, we limit our interest to data-driven visualization-based systems for sensemaking using AR techniques based on the user's physical location (i.e., situated analytics systems).
Then, we propose a design space to describe specific situated analytics techniques.
Using this design space, we study patterns of implementation, techniques, and enabling technologies.
To achieve this, we surveyed \rev{47} works on situated analytics from the human-computer interaction (HCI), visualization, and immersive technologies literature, as well as from other domains such as medicine and construction. 
We then describe four archetypes of situated analytics systems identified using multiple cluster analyses.
We close the paper by discussing observations drawn from our taxonomy.

%% file: 02_background.tex
\section{Background}
\label{sec:background}

The notion of using technology to immerse humans in an alternate world dates back to early cinematic screenings or theatrical performances.
These experiences were a precursor to the more contemporary, technology-driven, interactive flavors of Virtual Reality (VR) and \AR{}/\MR{}, which underpin both \IA{} and \SA{}.
Nonetheless, creating these alternate realities---and immersing users in them---requires a multifaceted approach, as the human perceptual and cognitive system is incredibly complex.

Inevitably, these immersive interaction paradigms, often collectively referred to as Extended Reality (XR, where X can be treated as an algebraic variable for V, A, M, etc.), have been defined in several different ways.
In an effort to avoid ambiguity and to provide a framework for our analysis, we revisit these definitions here.

\subsection{Immersive Technologies}
\label{SEC:iatech}

\textit{Virtual Reality} is defined as a computer-generated, three-dimensional (3D) environment that can be explored by a user that is predominantly isolated from their physical surrounding.
We prefer the term \textit{physical}, as opposed to ``real'' because the experience of \VR{} can be very much real.
\VR{} is typically accessed using head-mounted displays (HMDs), physical environments such as CAVEs, as well as standalone ``fishtank'' \VR{} displays.
The term is often attributed to Jaron Lanier~\cite{Lanier2001}, yet various similar definitions exist.

\textit{Augmented Reality}~\cite{azuma97AR} goes beyond \VR{} by fusing computer-generated objects with the physical world, enabling both the real and virtual worlds to be experienced and interacted with by the user.
According to Azuma~\cite{azuma97AR}, \AR{} environments have the following three characteristics: (1) combines real and virtual, (2) interactive in real time, and (3) registered in three dimensions. 

\textit{Mixed Reality}~\cite{chen17urbananalytics} is often used interchangeably with \AR{}, yet according to the definition by Milgram and Kishino~\cite{Milgram1994}, it is a superset that includes \AR{} and Augmented Virtuality (AV) where a virtual environment is enhanced with physical world information (essentially, the opposite of \AR{}).
According to this definition, \MR{} does not include \VR{} and spans between the physical (reality) and the entirely computer-generated (virtuality) environment.
This definition was elaborated further by Milgram et al.~\cite{Milgram1995} in terms of the extent of world knowledge, reproduction fidelity, and extent of presence metaphor.

These fairly graphics-centric definitions imply how dependent \XR{} flavors are on the available interface technology, and consequently how the first \VR{} systems---and \AR{} systems that followed---focused on graphics depiction.
The advent of smartphones has had a significant impact in the recent popularization and democratization of \XR{}. 
In particular, as highlighted by Barba et al.~\cite{Barba2012}, smartphones have shifted \MR{} from the lab to the real world, where technology becomes capability, space becomes place, and vision becomes perception.
At the same time, the related improvements in display technology, miniaturization, power consumption, and mobile graphic processing power make contemporary \XR{} interfaces capable of delivering upon the vast potential of \XR{} at the consumer level~\cite{Schmalstieg2016}.
Moreover, such systems often go beyond the visual modality and add virtual (or simulated) information using, auditory, somatosensory, and even olfactory modalities~\cite{Paneels2010, Patnaik2019}. 

Nevertheless, using 2D visualizations in immersive settings is challenging; several efforts explore methods for visualization placement and interactivity within such immersive information spaces~\cite{park18graphoto, kwon16immersgv}.
3D visualization, on the other hand, while initially popular in the research field~\cite{Robertson1991}, has been shunned in visualization research because of inefficiencies arising from occlusion, distortion, and navigation~\cite{Elmqvist2008, munzner14visualization}.
In addition, inherent \XR{} challenges such as interaction, registration, calibration, tracking, and visual coherence remain~\cite{Schmalstieg2016}, yet contemporary devices have somewhat ameliorated some of these issues~\cite{bach16immersiveanalytics, lee19immersive}. 

\subsection{Ubiquitous and Immersive Analytics}

\textit{Ubiquitous analytics} (UA), initially proposed in 2013~\cite{Elmqvist2013}, was one of the first concerted research efforts from the visualization domain to go beyond the regular desktop computer to encompass groups of networked mobile devices for anytime and anywhere sensemaking.
The approach is motivated by post-cognitive frameworks such as distributed and embodied cognition, which emphasize the role of the surrounding world and the user's own body for sensemaking~\cite{Elmqvist2023}.
In 2014, the approach was extended to encompass Mixed Reality as the next logical step for data visualization~\cite{Roberts2014}.

\textit{Immersive analytics} (\IA), introduced in 2015~\cite{Chandler2015}, is concerned with investigating how immersive interaction and display technologies can be used to support analytical reasoning and decision making~\cite{Marriott2018}.
\IA{} research efforts explore a plethora of challenges, such as user experiences in collaborative and multi-user applications~\cite{cordeil17immersivecollab, bonada16personimmersiv}, long-term deployment and usage~\cite{batch20econimmersive}, and evaluation of visualization techniques applied in an \IA{} context~\cite{filho20spacetimegeovis, hurter19fiberclay, simpson17take, hube18data, kraus20clusteridenttask}. 
True to its tech-centric nature, a significant focus of \IA{} is on developing toolkits and frameworks, such as DXR~\cite{sicat19dxr}, IATK~\cite{cordeil19iatk}, and VRIA~\cite{butcher20vria}.

There also exist research efforts that use \AR{}/\MR{} but which does not situate the information in the actual physical space where the data is located, collected, or leveraged.
Hence, these are not strictly \SA{} examples.
These include, for example, the study of visual cues that help spatial perception of datasets in \AR-based immersive environments~\cite{luboschik16spatialperception}, 3D spatial interaction using mobile devices~\cite{buschel173dvisualization}, effectiveness of different edge types in AR settings~\cite{buschel19argraphvis}, glyphs in AR~\cite{chen19marvist}, and studies of 3D exploration in AR~\cite{bach18hologram}.

\subsection{Situated Analytics}
\label{SEC:SA}

In \textit{situated analytics}, analytical reasoning is supported by embedding the visual representations and interaction of the resulting data in a physical environment using AR~\cite{elsayed16situateddef}.
To make the distinction between \SA{} and ``mere'' \IA{} in \AR{}, as discussed in the prior section, we typically think of SA as also requiring data and analytical tasks that are tied to the user's current physical location~\cite{Thomas2018}.

Inevitably, the real potential for \SA{} is in supporting experts in the field, as highlighted by Whitlock et al.~\cite{whitlock20fieldview}.
Although the term \SA{} is fairly contemporary, systems that potentially fulfill this definition date back to the Touring Machine~\cite{feiner97touringmachine} and similar mobile, wearable \AR{} set-ups.
A broad range of research activities in immersive technologies has addressed issues that can be challenging in \SA{} context, such as cross-device synergies~\cite{langner:2021:marvis}, perceptual challenges~\cite{SatkowskuDachselt2021}, labeling~\cite{grasset12labelAR}, highlighting~\cite{elsayed16horuseye}, etc.
One objective of this survey is to bring clarity to what constitutes an \SA{} system in both the past and the present.

\subsection{A Survey of Surveys}
\label{sec:survey_of_surveys}

There is a long history in the \AR{} field on surveying and classifying AR/MR technologies.
One of the earliest surveys from Azuma~\cite{azuma97AR}, provides a widely adopted definition of \AR{}, accompanied by examples of applications, and a discussion on the two main challenges of \AR{}, registration and tracking.
Focusing on the latter, Zhou et al.~\cite{zhou08artracking} discuss \AR{} techniques that have been studied for a 10-year period, starting from 2008, and presented at the ISMAR conference.
Van Krevelen and Poelman~\cite{van10survey} review a variety of applications of \AR{}, whereas Carmigniani et al.~\cite{carmigniani11arsurvey} classify \AR{} systems and argue that they need to be more socially acceptable to facilitate broad adoption.
Another strand of research focuses on specific areas (for example, assembly plants~\cite{wang16ARcomprehensive}) and techniques (for example, tracking technique~\cite{rabbi13survey} where \AR{} is used as a key technique.
Chatzopoulos et al.~\cite{Chatz_mobileAR17} focus on mobile \AR{}, and the influence of handheld systems in the domain, highlighting networking implications.
Finally, more inclusive surveys, such as from Billinghurst~\cite{BillinghurstSurvey_2015} and Kim et al.~\cite{kim18arsurveyismar} focus on \AR{} techniques and their challenges.

Driven by the increased interest to explore the synergies of immersive technologies and data visualization over the last five years, a number of surveys focusing specifically on the intersection of visualization and \AR{}, \IA, \SV{}, and \SA{} have emerged.
For the broader \IA{} domain, the work by Fonnet et al.~\cite{fonnet21iasurvey} explore systems that represent information related to its physical location and analyze them from five main perspectives; technologies, data mapping, interactions, collaboration, and types of user studies conducted.
They also propose three directions for the future of \IA{}, in fostering multi-sensory and embodied interactive \IA{}, converging towards best practices, and aiming at real life \IA{} systems.
Ens et al.~\cite{Ens2021} describe 17 grand challenges in the broader domain of \IA{}, deriving five specific challenges on the topic of spatially situated data visualization.

Some surveys focus specifically on visualizations embedded into the physical world.
The work of Willett et al.~\cite{Willet2017_Embedded}, although not a survey per se, offers a distinction between non-situated, situated, and embedded data representations in terms of their relationship to physical referents.
They also review enabling technologies and potential applications.
Zollmann et al.~\cite{zollmann21vistechinar} present a survey of visualization techniques used in visualizations that deploy \AR{}.
Besan\cb{c}on et al.~\cite{Bensacon_survey_2021}, in their review of spatial interfaces for 3D visualization, discuss various interaction techniques that can be applied in a \SA/\SV{} context.
Bressa et al.~\cite{bressa:hal-03319648} review prior work in terms of the situatedness, presenting five perspectives on it, including space, time, place, activity and community.
They also note that many papers do not discuss deeply their definition of \SV{}, and instead rely on the definitions from White and Feiner~\cite{white09sitelens}, and Willett et al.~\cite{Willet2017_Embedded}.
Furthermore, they also identify the dominance of \AR{} among other technologies used for \SV{}.

To that end, understanding how \AR{} techniques are used to present data in-situ is a key component for understanding \SA{} and \SV{}.
Recent work by Fr\"{o}hler et al.~\cite{FrohlerXVA:2022} discuss the concept of cross-virtuality analytics (XRA), which enable visual analysis through a combination of different devices across the reality-virtuality continuum.
Their approach identifies existing challenges and future opportunities for such cross-device visualization research.
And finally, Elmqvist~\cite{Elmqvist2023} review the history of the ubiquitous analytics research agenda and its influence on immersive and situated analytics, as well as future challenges.

\subsection{Summary and Comparison with Prior Art}

While \AR{} has been used in the context of data visualization for more than two decades, not much focus has been given on analyzing the characteristics of \AR{} techniques utilized in situated analytics, especially those of in-situ depiction. 
Our work aims to address this.
We also believe that the fact that situated analytics is based on a different platform from conventional visualizations calls for different rules for effectively presenting data in a situated context.
Our work tries to disclose these rules from a corpus of papers.
Unlike prior work, our focus is on the synergy between \AR{} technology and \SA{}: the interplay between the \textit{situatedness} of data and its \textit{integration} into the real world.
Compared to past surveys, we attempt to both be more prescriptive in qualitatively describing archetypes of \SA{} in the literature, as well as more generative of new directions for research.

The situated visualization survey by Bressa et al.~\cite{bressa:hal-03319648} is most related to the present paper.
In contrast to our work, Bressa et al.'s survey focuses on situated visualization, whereas our goal here is to survey higher-order analytics in situ based on the classic sensemaking loop by Pirolli and Card~\cite{Pirolli2005}.
We also mine our classification using dimensional reduction to extract high-level archetypes of \SA{} in the literature.
Finally, Bressa's work also mostly ignores the rich body of situated visualizations proposed by the Augmented Reality community prior to the introduction of situated analytics in 2016.

%% file: 03_design_space.tex
\section{Design Space: Situated Analytics}
\label{sec:design-space}

Here we derive the design space for \SA{} implemented using \AR{}.
We first give our definition of situated analytics.
This is followed by our survey methodology.
Then we present the design dimensions for classifying the surveyed papers.

\subsection{Situated Analytics: Definition}
\label{sec:definitions}

Thomas et al.~\cite{Thomas2018} define \textit{situated analytics} (SA) as methods \textit{``supporting analytic reasoning through the use of situated visualizations.''}
White and Feiner~\cite{white09sitelens} define \textit{situated visualization} as a visualization that \textit{``is related to and displayed in its environment.''}
Together these definitions entail that SA is (1) data-driven, (2) uses interactive visualization, (3) is based on Augmented Reality to integrate with the physical environment, (4) draws on the user's location, and (5) integrates analytical reasoning.
Table~\ref{tab:inclusion-criteria} summarizes these criteria.

In fact, we can use these criteria to distinguish between different forms of analytics and visualization. 
Table~\ref{tab:analytics-forms} gives an overview of such a classification.
Note that this classification is not intended to represent exclusive and disjoint categories; for example, it is widely accepted that \IA{} and \SA{} are subsets of visual analytics, that \SV{} is a subset of visualization, and that all forms of analytics incorporate visualizations.
In fact, this mini-taxonomy also gives rise to the concept of \textit{immersive visualization}; visualizations that form the building blocks of \IA{} applications.

It is also important to acknowledge the wealth of existing work on situated visualization even prior to situated analytics being coined in 2016~\cite{elsayed16situateddef,Schmalstieg2016}.
Most of this work was primarily done in the Augmented and Virtual Reality research areas.
While there exists some early work that fits the SA definition (and is thus included in our survey), we note that many early situated visualizations were characterized by being relatively simplistic; they tended to either use labels or simple symbols, or 3D objects with a real-world appearance.
Support for higher-level analytical reasoning is the key factor distinguishing SA from SV, similar to how visual analytics (VA) is distinguished from visualization.
We present a disambiguation of what sensemaking support means in this context in Section~\ref{SEC:sensemaking}.

\begin{table}[htb]\sffamily
    \caption{\textbf{Situated analytics criteria.}
    These criteria were used for selecting papers to include in our survey; all must be fulfilled for a paper to be selected.}
    \label{tab:inclusion-criteria} 
    \centering
    \begin{tabular}{ll}
        \toprule
        \textbf{\#}  & \textbf{Criteria} \\
        \midrule
        \rowcolor{gray!10}
        1 & Presents \textbf{data}\\ 
        2 & Uses interactive \textbf{visualization}\\ 
        \rowcolor{gray!10}
        3 & Deployed using \textbf{Augmented Reality} techniques\\
        4 & Utilizes the user's \textbf{physical location}\\
        \rowcolor{gray!10}
        5 & Integrates \textbf{analytical reasoning}\\
        \bottomrule
    \end{tabular}
\end{table}

\begin{table}[htb]\sffamily
    \caption{\textbf{Distinguishing analytics.}
    Various forms of data analytics and visualizations classified based on their use of data (Data), visualization (Vis), computing platform (Platform), user's physical location (Loc), and integration of the analysis process  (AP).}
    \label{tab:analytics-forms} 
    \centering
    \begin{tabular}{lccccc}
        \toprule
        \textbf{Topic} & \textbf{Data} & \textbf{Vis} & \textbf{Platform} & \textbf{Loc} & \textbf{AP} \\
        \midrule
        
        \rowcolor{gray!10}
        Visualization & \checkmark & \checkmark & Desktop & $\times$ & $\times$\\
                
        Visual analytics & \checkmark & \checkmark & Desktop & $\times$ & \checkmark\\
        
        \rowcolor{gray!10}
        Immersive visualization$^*$ & \checkmark & \checkmark & \VR{}/\AR{} & $\times$ & $\times$\\
        
        Immersive analytics & \checkmark & \checkmark & \VR{}/\AR{} & $\times$ & \checkmark\\
        
        \rowcolor{gray!10}
        Ubiquitous visualization & \checkmark & \checkmark & Any & \checkmark & $\times$\\

        Ubiquitous analytics & \checkmark & \checkmark & Any & \checkmark & \checkmark\\
        
        \rowcolor{gray!10}
        Situated visualization & \checkmark & \checkmark & \AR{} & \checkmark & $\times$\\
        
        Situated analytics & \checkmark & \checkmark & \AR{} & \checkmark & \checkmark\\
        
        \bottomrule
    \end{tabular}
\end{table}

\subsection{Survey Methodology}
\label{subsec:data_collection}

Our paper selection process was performed in four steps: (1) \textit{collecting} candidate papers for potential inclusion in our taxonomy, (2) \textit{selecting} the candidate papers that match our inclusion criteria, (3) \textit{classifying} the selected papers, and (4) \textit{pruning} papers that do not fit our definition of \SA{}.

\setlength{\textfloatsep}{0.3cm}

\textbf{Collection.} 
We first collect papers using the keywords listed in Table~\ref{tab:rel_kwds} from conference proceedings or journals in \AR{}/\VR{} (for example, IEEE VR, IEEE ISMAR), visualization (for example, IEEE VIS, IEEE TVCG, EuroVis, IEEE PacificVis) and human-computer interaction (HCI) (for example, ACM CHI, ACM UIST) communities.
We restrict our search from 1995 to 2021.
To retrieve papers outside of the conferences described above, such as application papers where \SA{} is deployed, we also search the ACM Digital Library, IEEE Xplore, Elsevier's ScienceDirect, and Google Scholar using the same keywords.
We also include papers that are cited by the candidate papers that have relevant titles or contain keywords from Table~\ref{tab:rel_kwds} to our list of candidates.

Using the above search process, we collected 312 candidate papers that advanced to the next step.

\textbf{Selection.} 
We select papers that match our inclusion criteria (see below) from the list of candidate papers obtained during the collection process. 
We use the definitions in Section~\ref{sec:definitions} to  formulate the inclusion criteria in Table~\ref{tab:inclusion-criteria}.

\begin{table}[htb]\sffamily
    \caption{\label{tab:rel_kwds} \textbf{Keywords relevant to situated analytics.}
    We used these keywords to search for papers during the collection process.}
    \centering
    \begin{tabular}{l}
        \toprule
        \textbf{Relevant Keywords} \\
        \midrule
        \rowcolor{gray!10}
        \textit{situated visualization (with AR), situated analytics (with AR)} \\
        \textit{visualization (with AR), augmented reality visualization}, \\
        \rowcolor{gray!10}
        \textit{immersive analytics, immersive visualization, situated information}, \\
        \textit{situated information visualization, situated infovis, situation vis,} \\
        \rowcolor{gray!10}
        \textit{point of interest visualization, situated vis, immersive vis, AR vis,} \\
        \textit{poi vis, situation visualization, augmented reality vis,} \\
        \rowcolor{gray!10}
        \textit{in situ data visualization (with AR), in situ data analysis (with AR),} \\
        \textit{in situ analytics (with AR), in situ data vis (with AR),} \\
        \rowcolor{gray!10}
        \textit{situated analysis (with AR), situated data (with AR)}. \\
        \bottomrule
    \end{tabular}
\end{table} 

The selection process was conducted by three authors.
A paper was selected only if all three authors agreed that the paper met the four criteria.
Note that we include only one mention of each system; in other words, we eliminated multiple versions of the same work by the same authors.

We selected 75 papers from the 312 candidate papers.
The earliest work in our taxonomy is the Touring Machine~\cite{feiner97touringmachine} by Feiner et al., published in 1997. 
The latest work is MARVIS~\cite{langner:2021:marvis} by Langner et al., published in 2021.

\textbf{Classification.}
We classified 75 papers using our taxonomy (Section~\ref{subsec:design_dimensions}); see Table~\ref{table:taxonomization_table}. 
We also present changes in taxonomy over 5-year periods in Table~\ref{table:stats_table}.
Note that we consider all aspects of the surveyed papers beyond than those on Table~\ref{table:taxonomization_table}.
Please refer to Sec.~\ref{sec:discussion} for an in-depth analysis of the surveyed papers, including aspects such as interaction, usage scenarios, 2D vs.\ DD, and development.

\textbf{Pruning.}
Finally, we wanted to prune candidate papers that our classification revealed to not be true situated analytics papers.
More specifically, papers that only support the very first Sensemaking task---\textsc{read}---are situated visualizations and do not support higher-level analytical reasoning.
We ended up removing 24 such papers out of the total of 75 classified papers, resulting in a remaining 51 papers.\vspace{-2mm}

\subsection{Taxonomy Dimensions}
\label{subsec:design_dimensions}

Drawing on our literature survey (Section~\ref{sec:background}) as well as our corpus of selected papers, we now describe the dimensions we use to characterize this novel research area.
We focus on what makes \SA{} unique; the \textit{situatedness} of the visualizations, the \textit{data depictions}, and the \textit{sensemaking} support.

To this end, we use a 5W1H approach---what, where, when, what, who, and how---to structure our inquiry. 
This approach yields several motivating questions that we present in Table~\ref{tab:dims_def} to describe how \SA{} tools display data in the world using \AR{}.
Note that we leave the user (\textit{who?}) and the motivation for the task (\textit{why?}) outside the scope of our design space as they are strictly application-specific; we want to be able to capture situations where two systems have vastly different uses, but identical (or at least similar) properties in terms of situatedness.

As can be seen from Table~\ref{tab:dims_def}, this yields four design dimensions: Situating Triggers, View Situatedness, Data Depiction, and Sensemaking Tasks.
The first two deal with how the data is situated in the world whereas the remaining deal with their representation and analytical support.
We describe each of these dimensions in turn below.

\begin{table}[t!]\sffamily
    \caption{\textbf{The 5W1Hs of our taxonomy.}
    Using basic problem-solving to derive our design dimensions: what, who, where, when, why, and how.}\vspace{-2mm}
    \label{tab:dims_def} 
    \centering
    \begin{tabular}{ll}
        \toprule
        \textbf{Questions}  & \textbf{Dimensions} \\
        \midrule
        \rowcolor{gray!10}
        \textbf{When} does the system react? & \textbf{Situating Triggers} \\
        \rowcolor{gray!10}
        \textbf{Where} is the view represented? & \textbf{View Situatedness} \\
        \rowcolor{gray!10}
        \textbf{What} type of data is presented? & \textbf{Data Depiction} \\
        \rowcolor{gray!10}
        \textbf{How} is sensemaking supported? & \textbf{Sensemaking Support} \\
        Why is the data represented in space? & Out of scope\\
        Who is interested in using the data? & Out of scope\\
        \bottomrule
    \end{tabular}
\end{table} 

\subsubsection{Situating Triggers}

\textbf{Situating triggers} are \textit{discrete events that lead to the instantiation or modification of an \SA{} tool}.
Can be classified as:

\smallskip
\noindent\textbf{Referents:} Spatial triggers that indicate a view should be instantiated or modified.
These can be further defined as:
\smallskip

\begin{wrapfigure}[5]{l}{2cm}
    \vspace{-1.1em}
    \includegraphics[width=2cm]{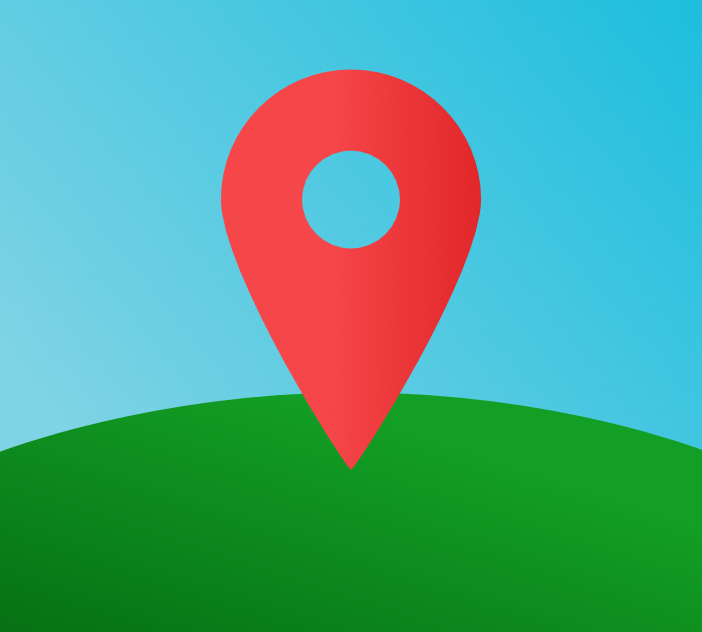}
\end{wrapfigure} 

\noindent\textit{Location:} When the instantiation of the data view or visualization is triggered by the user's location.
This is typically a physical place, such as GPS location or a distance walked from a starting point in steps as indicated by a pedometer.
For example, the Bottari app~\cite{bladuini12bottari} uses the user's location information to recommend nearby restaurants around them.
BoreholeAR~\cite{lee15boreholear} points exact locations of the boreholes in the user's view via the GPS.

\smallskip

\begin{wrapfigure}[5]{l}{2cm}
    \vspace{-1.1em}
    \includegraphics[width=2cm]{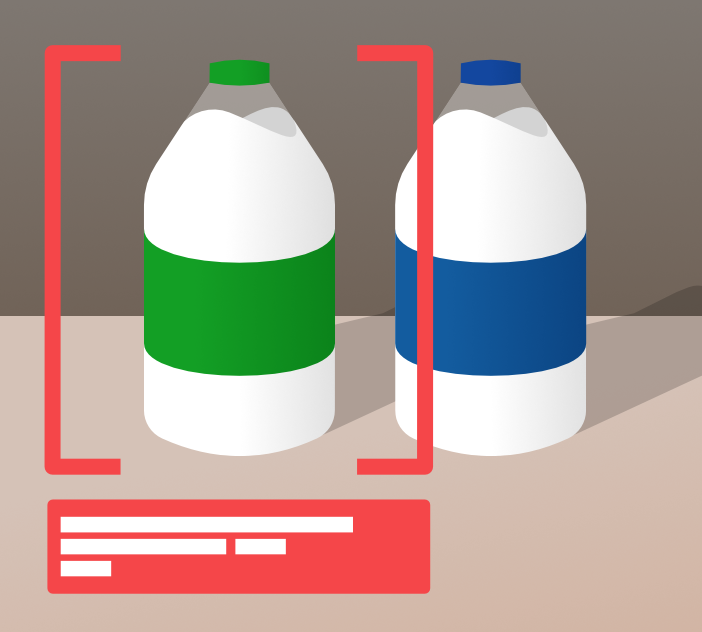}
\end{wrapfigure} 

\smallskip

\noindent\textit{Object:} Here the instantiation of the data view or visualization is triggered by an object being detected by a sensor on the user's device (for example, marker, object recognition, loud sound originating near the user).
PapARVis~\cite{chen20staticvis} detects static information displays in the real world using QR codes and extends them with virtual content in \AR{}. 
ServAR estimates the volumetric amount of food on a plate (identified by a fiducial marker)~\cite{rollo17servar}.

\smallskip

\noindent\textbf{Context:} These are situations when the instantiation of the data view or visualization is triggered by semantic or conceptual information surrounding the user at the current time and place.
The trigger may or may not be spatial.

\smallskip

\begin{wrapfigure}[5]{l}{2cm}
    \vspace{-1.1em}
    \includegraphics[width=2cm]{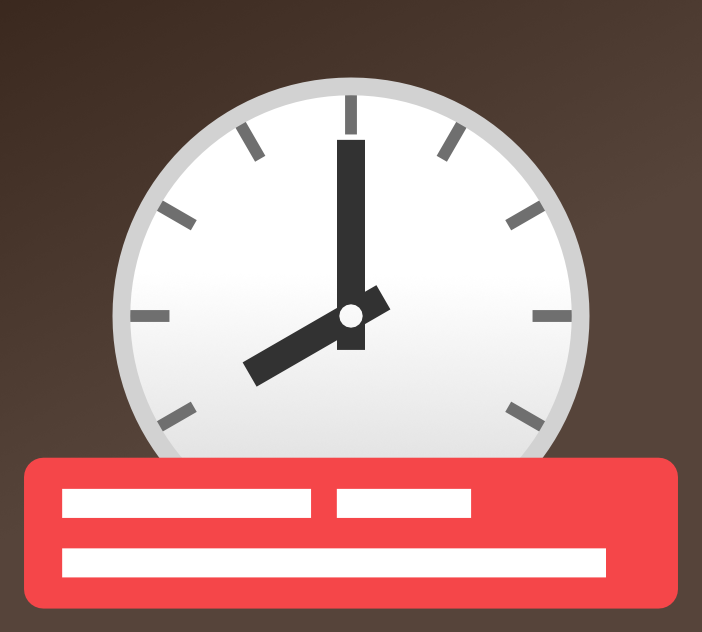}
\end{wrapfigure} 

\smallskip

\noindent\textit{Event:} 
These are situations triggered by a specific event.
They can be unplanned: this includes situational information identified by the system at a given moment in time that triggers a view instantiation or modification.
Tools of this type deliver different signals based on the situation that the user is in. 
For example, a collision detection system in a car may sound a warning when the car approaches an obstacle~\cite{kim18collisionar}, or assistive technologies for people with hearing loss may provide different sounds to assist communication during an ongoing conversation~\cite{jain15hmdvis}.
They can also be planned, such as for post-mortem analysis of a dynamic sequence; for example, Lin et al.~\cite{lin21basketballar} propose a tool for analyzing ball movements in basketball.

\smallskip

\begin{wrapfigure}[5]{l}{2cm}
    \vspace{-1.1em}
    \includegraphics[width=2cm]{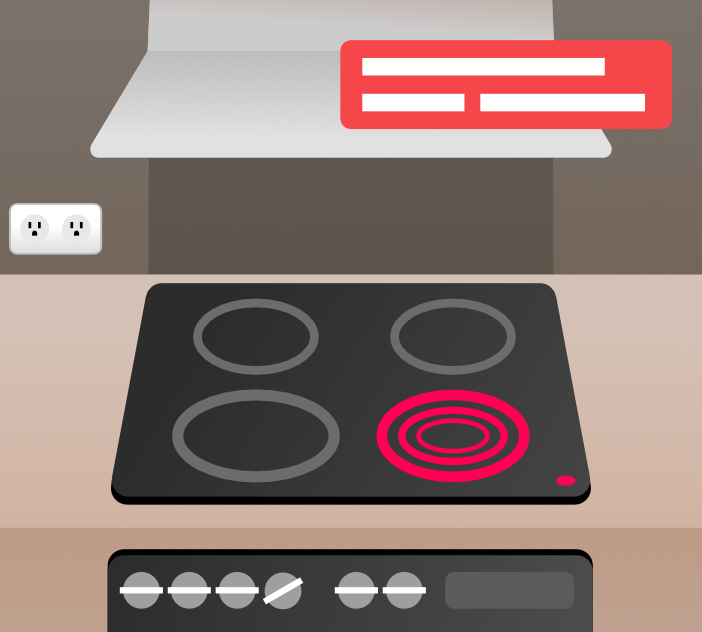}
\end{wrapfigure} 

\smallskip

\noindent\textit{Setting:} The type of space that the user is in and its state at the time they are in it---for example, a construction site causing the appropriate \SA{} software to boot up~\cite{Behzadan05constructionAR}---that is not explicitly tied to the location of the user.
Other examples include a flood simulation not tied to a specific location~\cite{haynes16floodAR} or a surgical environment augmented by \AR{} overlays~\cite{bichlmeier09keyholesurgeryAR}).

\smallskip           
            
\noindent\textbf{Non-situated:} Here the view is spawned independent of the user’s locus, context, or objects.
A system may have a non-situated view creation command, but in order for a view to be situated---that is, positioned relative to objects, locations, or contexts around the user---a situating trigger is still required.
If neither the trigger nor the view are situated, the system is not a situated analytics system.
Consequently, we do not include non-situated systems in Table~\ref{table:taxonomization_table}.

\subsubsection{View Situatedness}

\textbf{View situatedness} describes \textit{the integration of data views into the world}, potentially including the device viewport.

\smallskip

\noindent\textbf{World-Registered:} Here the data is presented so that it is positionally anchored to the real world, such as an imitation of a real object (for example, an amount of stacked sugar cubes representing the amount of sugar in food items sitting on a table near those food items).
We consider two levels:

\smallskip

\begin{wrapfigure}[5]{l}{2cm}
    \includegraphics[width=2cm]{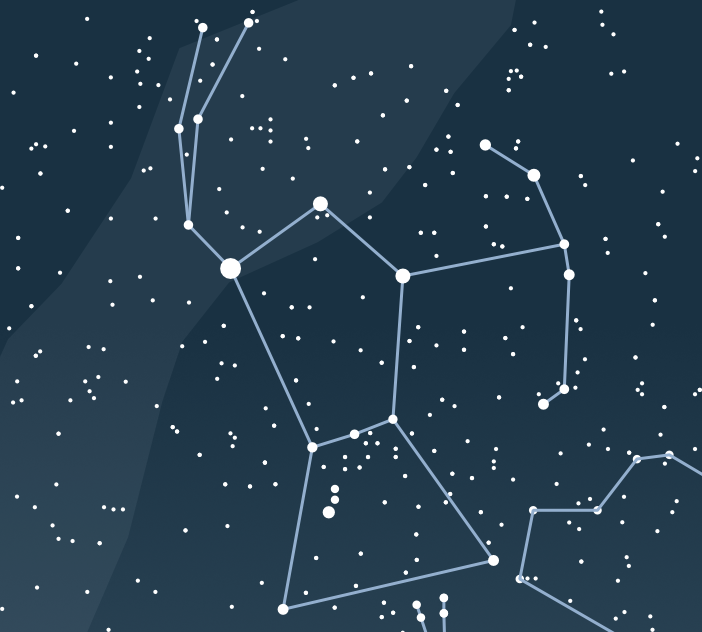}
\end{wrapfigure} 

\smallskip
        
\noindent\textit{World-absolute:} the data is situated at the exact point of interest (for example, a holographic representation of a person to reenact an event exactly on top of a marker); what Willett et al.~\cite{Willet2017_Embedded} called \textit{embedded data}.
In other words, the data is anchored to a physical location, regardless of the user's position or orientation.
This aligns well with Azuma's~\cite{azuma97AR} definition of \AR{} and in that regard it is one of the most popular flavors of \AR{}, used for navigation, tour guides, cultural heritage, etc.
For example, the Touring Machine overlays information about a university campus on the user's view~\cite{feiner97touringmachine}.

\smallskip

\begin{wrapfigure}[5]{l}{2cm}
    \vspace{-1.1em}
    \includegraphics[width=2cm]{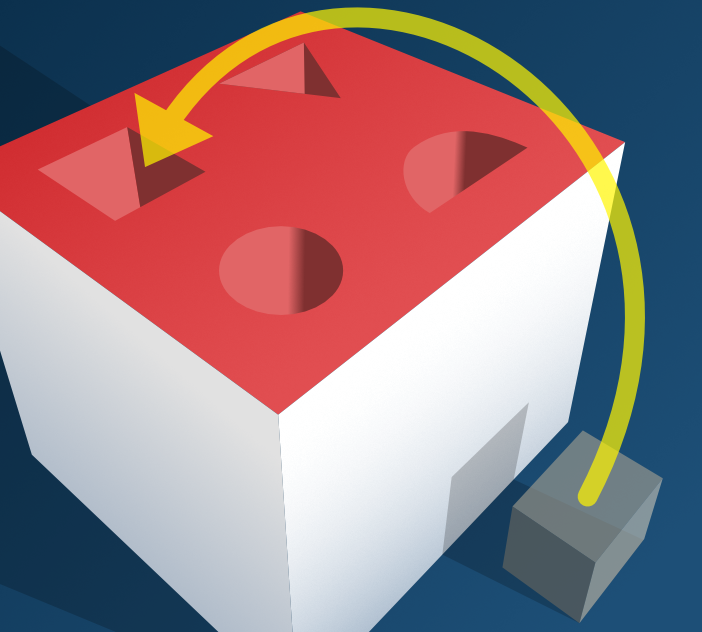}
\end{wrapfigure} 

\smallskip

\noindent\textit{World-relative:} The data is situated in a location that is relevant to a point of interest, but the actual layout of the visualizations is designed to optimize user interpretation rather than to strictly mirror real-world object positions (for example, a cluttered real-world view with labels that are neatly organized but linked to real-world objects via indicator lines).
The optimization may include user-driven interactions within the view, such as placing virtual furniture in a house~\cite{gomez08sizeestimationAR} in the IKEA Place app~\cite{ikea_app}, or estimating size to serve patients in hospital~\cite{rollo17servar}.

\smallskip

\begin{wrapfigure}[5]{l}{2cm}
    \vspace{-1.1em}
    \includegraphics[width=2cm]{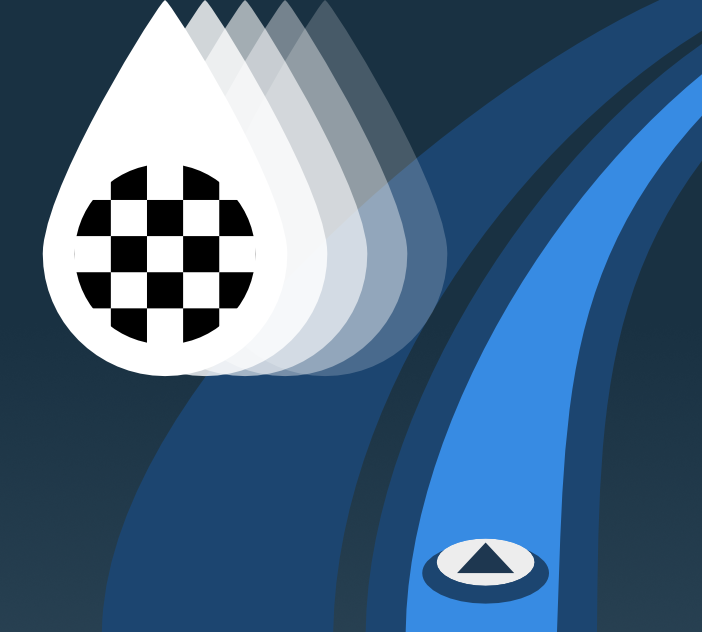}
\end{wrapfigure} 

\smallskip

\noindent\textbf{Sticky:} Sticky data views are linked to a specific location, but will persist in the periphery of the user's viewport even when this location falls outside the current field of view.
As illustrating examples, consider an arrow that guides the user to a specific destination~\cite{mulloni11indoorAR}, or labels that refer to items in the user's vicinity even when the items are outside the field of view~\cite{grasset12labelAR}.

\smallskip

\noindent\textbf{Device-Anchored:} Here the data view is linked to the user's own device, regardless of where it is in the world. 
It is situated because it was spawned by a situating trigger; for example, a notification on a smartphone when you are near a point of interest.
We consider two types of anchorings:

\smallskip

\begin{wrapfigure}[5]{l}{2cm}
    \vspace{-1em}
    \includegraphics[width=2cm]{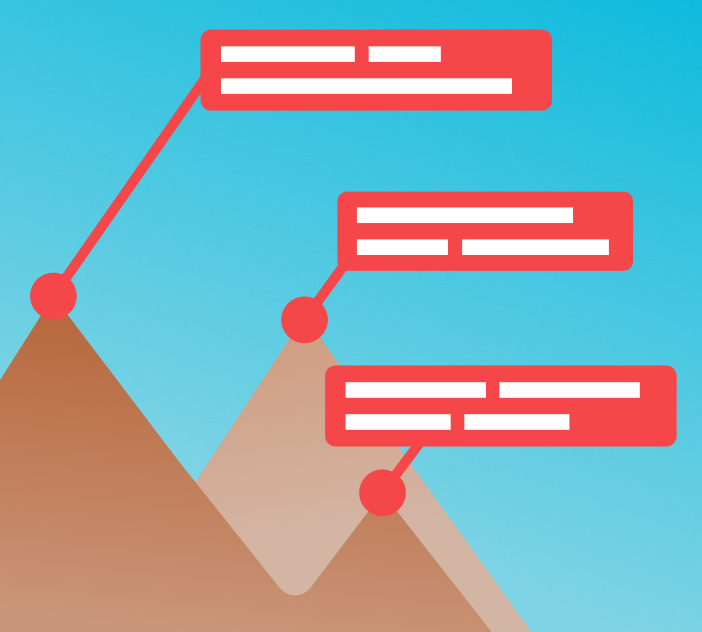}
\end{wrapfigure} 

\smallskip   

\noindent\textit{Device-relative:} The data view changes, oftentimes optimized to the user in absolute terms. 
Consider a GPS map on a smartphone that rotates and changes based on the user's position and orientation.
Various systems that utilize automated label placement algorithms~\cite{grasset12labelAR} have the data anchored to device-relative views, so that it does not disturb the user's view.

\smallskip

\begin{wrapfigure}[5]{l}{2cm}
    \vspace{-1.1em}
    \includegraphics[width=2cm]{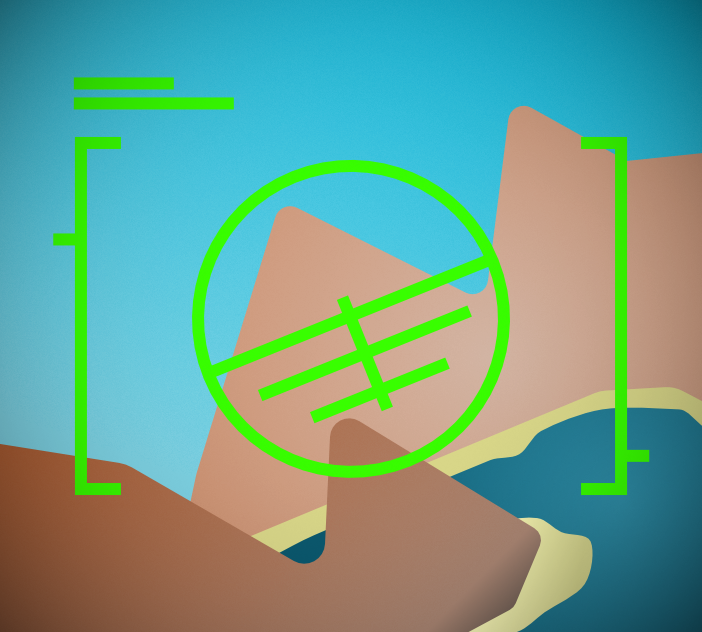}
\end{wrapfigure} 

\smallskip

\noindent\textit{Device-fixed (Overlay):} When the data is represented in front of the user’s view of the real world on their device, without being registered to the world. 
The canonical example of a device-fixed display is a heads-up display (HUD), as seen in modern (3rd generation and up) combat aircraft; the HUD is attached to the aircraft in front of the pilot, but displays situated information on top of the scene where the aircraft is pointing. 
Another example of a device-fixed view is the information display in a viewfinder of a DSLR camera, which stays fixed regardless of where the user points the camera.
An example of such views can be seen in an SA tool for shopping by Elsayed et al.~\cite{elsayed16situateddef}. 
In their system, the object in the center of the view is detected, and the analysis view is always shown on a particular area of the view.

Note that a view can be both absolute and relative. 
An object may be positioned absolutely inside an automobile, for example, but the automobile is a container that can move.
Thus, the view is relative to the world.

\subsubsection{Data Depiction} 

This dimension describes the \textit{forms that data representations take}; that is, the visual encoding and abstraction of data.
This is relevant to the taxonomy because of the way the data representations integrate with the real-world environment. 

\smallskip

\noindent\textbf{Visual Encoding:}
The visual encoding is directly relevant to how the visual representations are situated in the physical environment: as 2D, as 3D, or as nonvisual objects.

\smallskip

\begin{wrapfigure}[5]{l}{2cm}
    \vspace{-1.1em}
    \includegraphics[width=2cm]{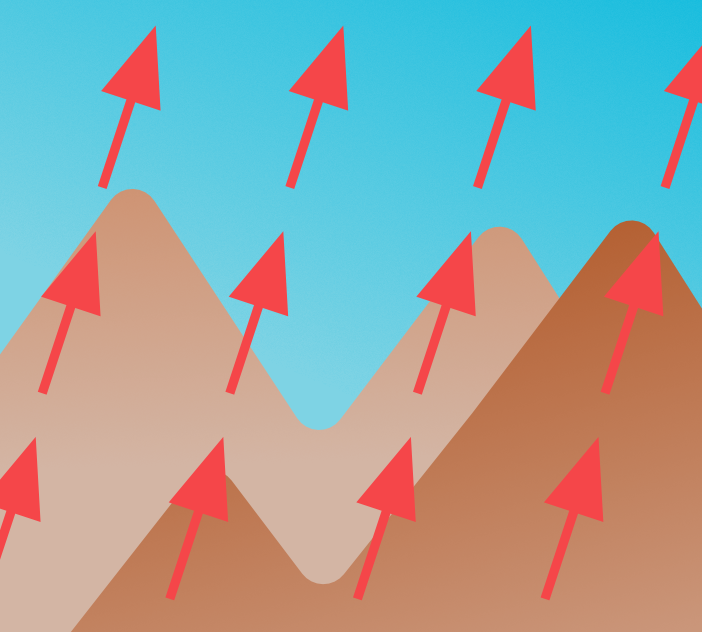}
\end{wrapfigure} 

\smallskip

\noindent\textit{2D:} The data is depicted on a two-dimensional plane inside the 3D environment.
We include 2.5D SA systems in this category, where 2D data is depicted in 3D space.
Examples include standard 2D visualizations---such as scatterplots, barcharts, and linecharts---as well as labels accompanying visualizations, in-world noticeboards, and building maps. 
ARVino~\cite{king05arvino} presents an analysis of viticulture data of a particular land using a 2D heatmap on a world-registered environment.

\smallskip

\begin{wrapfigure}[5]{l}{2cm}
    \vspace{-1.1em}
    \includegraphics[width=2cm]{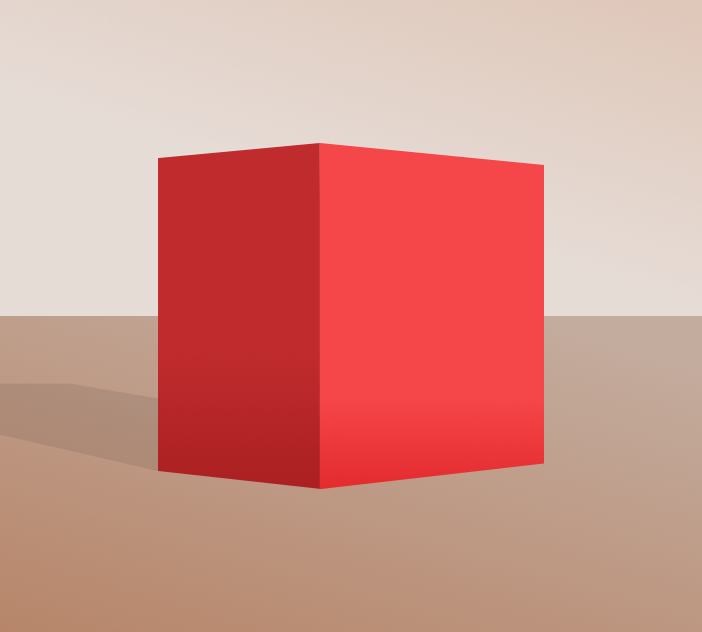}
\end{wrapfigure} 

\smallskip

\noindent\textit{3D:} The data is depicted as a three-dimensional entity; the canonical situated visualization.
These can be volumetric data, 3D visualizations (e.g, cubic bar charts, 3D scatter plots, etc.), as well as representations of physical objects. 
If the data is presented using 3D cubes, then it belongs to this category.
SiteLens~\cite{white09sitelens} and FieldView~\cite{whitlock20fieldview} embed 3D graphical cues (for example, spheres, bars) about objects within the user's view. 
Some 3D visualizations are used to simulate real-world objects within the world~\cite{Behzadan05constructionAR, schall09infravis}.

\smallskip

\begin{wrapfigure}[5]{l}{2cm}
    \vspace{-1.1em}
    \includegraphics[width=2cm]{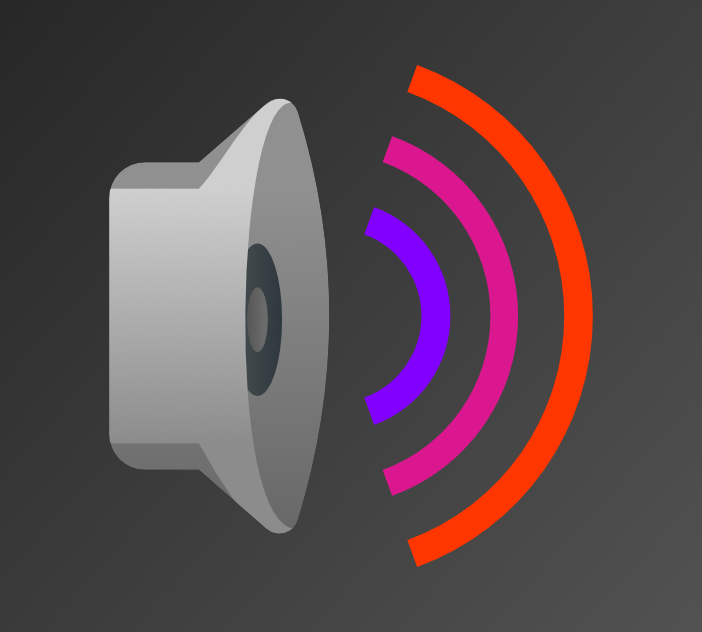}
\end{wrapfigure} 

\smallskip

\noindent\textit{Non-visual:} The data is depicted in a non-visual manner (for example, sound, smell, etc.).
In this case the non-visual data may be accompanying a visual depiction, or may be adding context to the user's situation (for example, via narration or via signals).
For example, different sound patterns are used to deliver different information to aid a microscopic precision task~\cite{roodaki17sonifeye}, or to communicate between those who are visually impaired~\cite{jiang20finedetailsar}.

\smallskip

\noindent\textbf{Data Abstraction:} 
These are the visual expressions used for data representations.
These can be classified as:
    
\smallskip

\begin{wrapfigure}[5]{l}{2cm}
    \vspace{-1.1em}
    \includegraphics[width=2cm]{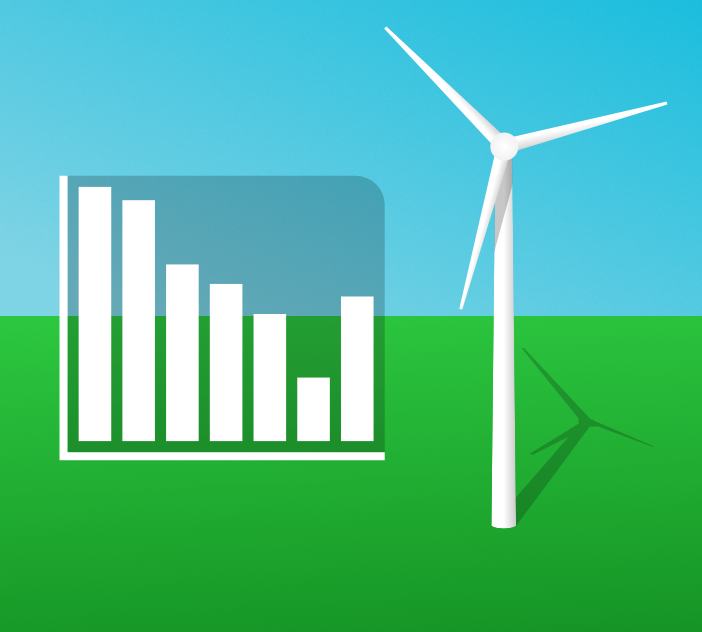}
\end{wrapfigure} 

\smallskip

\noindent\textit{Abstract:} Data is expressed using abstract graphical marks and channels that are used to encode features and values of a dataset (for example, lines, squares, spheres, etc. of varying color, position, size, etc).
This approach is most often used for data that is \textit{abstract} in nature, that is, that does not have a straightforward mapping to a visual representation.
An example of this is an analytics tool for collaboration~\cite{langner:2021:marvis}.
People share the visualization via a head-mounted device. 
Another model presents textual labels to pinpoint borehole location on top of the view~\cite{lee15boreholear}.

\smallskip

\begin{wrapfigure}[5]{l}{2cm}
    \includegraphics[width=2cm]{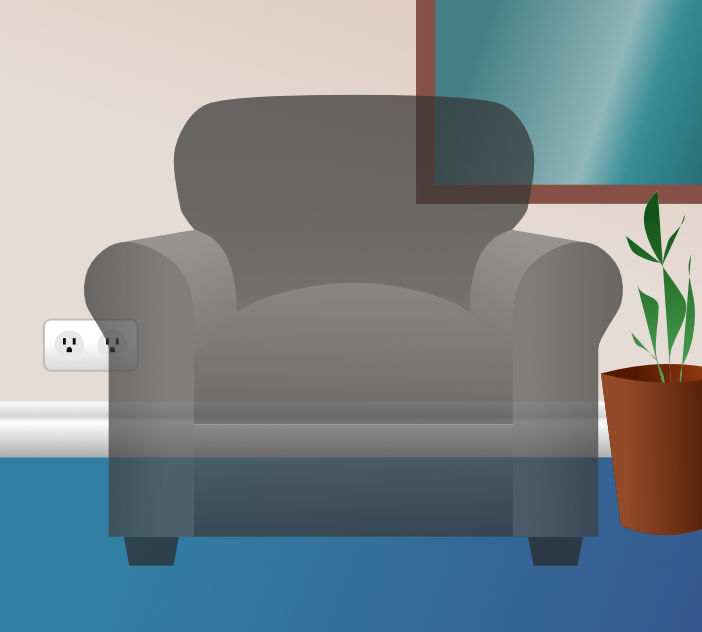}
\end{wrapfigure} 

\smallskip

\noindent\textit{Physical:} The data is expressed using the real world's spatial dimensions to represent spatial features of the data.
This could be an artificial imitation of an existing object/entity (for example, an imaginary chair in a place of interest, a heart positioned in a cutaway over a human patient).
For example, the MARVisT~\cite{chen19marvist} toolkit creates visualization glyphs of common real-world objects for data representation.

\smallskip

\begin{wrapfigure}[5]{l}{2cm}
    \vspace{-1.1em}
    \includegraphics[width=2cm]{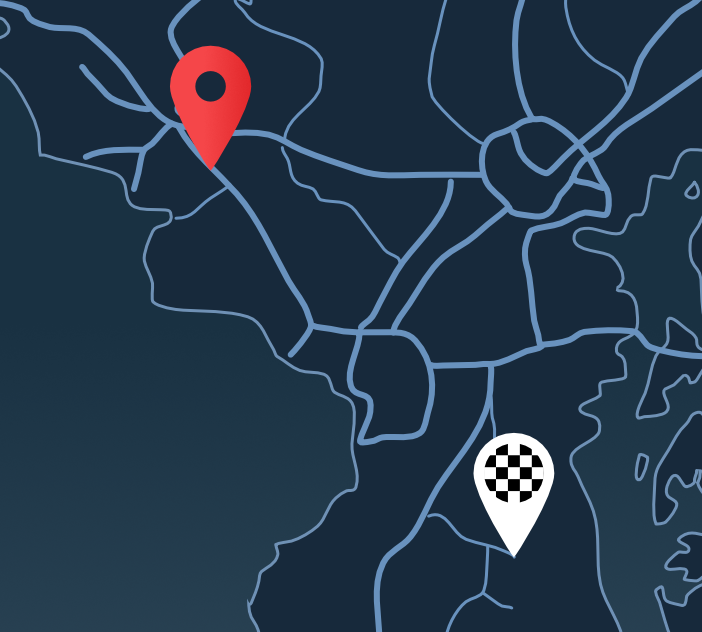}
\end{wrapfigure} 

\smallskip

\noindent\textit{Hybrid:} Hybrid abstractions include components of both abstract and physical data.
A canonical example is a map, which is an abstraction of 3D space that exists as an entity in the real world.
While 2D maps are common and useful for orientation in a 3D space, more esoteric approaches use 3D representations such as a world-in-miniature (WIM)~\cite{stoakley95virtual}, which provides the user with an easily accessible downsized version of the 3D environment.
These maps present an overview by locating the point of interest to the user during navigation~\cite{dunser12ARoutdoornavigation, williams17toarist, alqahtani17imapcampus}.

\subsubsection{Sensemaking Support}
\label{SEC:sensemaking}

Finally, according to Table~\ref{tab:inclusion-criteria}, what distinguishes situated analytics from situated visualization is the support for analytical reasoning.
While this is a vague criteria---similar to the distinction between visualization and visual analytics---we choose to base this taxonomy dimension on Pirolli and Card's \textit{sensemaking loop}~\cite{Pirolli2005}.
We model the following levels. 
\rev{Note that the distinction is based on the role the system plays in the sensemaking process}:

\smallskip

\begin{wrapfigure}[5]{l}{2cm}
    \vspace{-1.1em}
    \includegraphics[width=2cm]{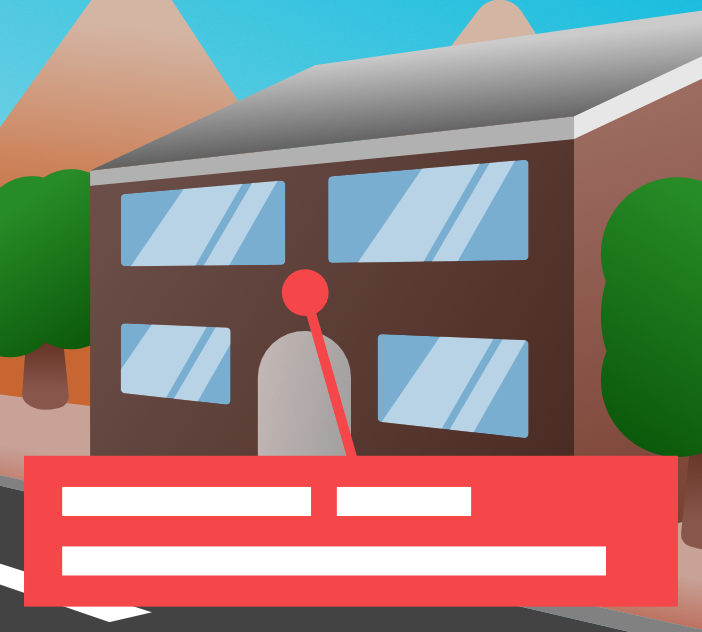}
\end{wrapfigure} 

\smallskip

\noindent\textbf{Read:} 
The technique supports extracting data from the visual representation.
\rev{The techniques that belong to this category focus on delivering information relevant to the situation.
However, reading does not entail active interaction by the user beyond merely navigating in physical space (i.e., walking and moving your head).}
Representing the lowest abstraction level in the sensemaking loop~\cite{Pirolli2005}, this property is supported by both situated visualization and analytics; in other words, it is not sufficient for a tool to qualify as \SA{}.
\rev{Examples of this task include observing the yield data overlaid on top of the vineyard as a heatmap~\cite{king05arvino}, providing guidance or information about tourist destinations (for example, attractions, restaurants and so on~\cite{williams17toarist, bladuini12bottari}), and providing information about where boreholes are within a region~\cite{lee15boreholear}.
}

\smallskip

\begin{wrapfigure}[5]{l}{2cm}
    \vspace{-1.1em}
    \includegraphics[width=2cm]{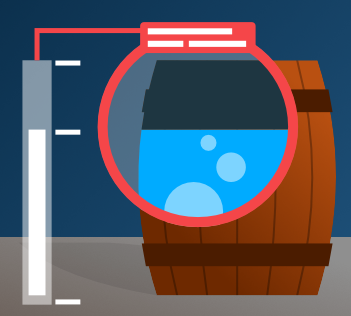}
\end{wrapfigure} 

\smallskip

\noindent\textbf{Explore:} Once a tool enables the user to interact with the data beyond mere situated viewing, we consider it to support exploration.
Exploration includes interactions such as filtering, rearranging, zooming, panning, drilling down, \rev{and other interactions that facilitate discovery}.
While the boundary between \SV{} and \SA{} is blurry, we deem a tool that only supports ``read'' as situated visualization, whereas one that also supports ``explore'' is as situated analytics.
For example, merely drawing a 3D arrow~\cite{mulloni11indoorAR} or labeling to augment the real world~\cite{grasset12labelAR} does not qualify as exploration, whereas viewing hidden features behind an \rev{occluding} wall \rev{determined by the user's interests}~\cite{kalkofen13ghostar}, or \rev{providing interactions, such as filtering and rearranging to help the user locate their items of interest during shopping experience~\cite{elsayed16situateddef}}, do. 

\begin{wrapfigure}[5]{l}{2cm}
    \vspace{-1.1em}
    \includegraphics[width=2cm]{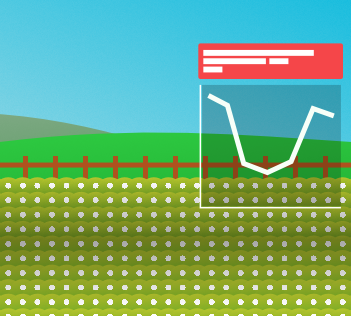}
\end{wrapfigure} 

\smallskip

\noindent\textbf{Schematize:} Pirolli and Card~\cite{Pirolli2005} define finding a schema as a marshalling or coordination action where evidence is organized into a ``small-scale story'' or some more rigid structure.
This is a step above exploration, \rev{where the system helps the analyst} to discern some underlying insight about the data being explored. 
\rev{In other words, tools that belong to this category should be capable of organizing the data inside the tool, such as creating clusters, annotating data, and tagging items---i.e., creating \textit{schemas}.}
\rev{For example, SiteLens~\cite{white09sitelens} enables the user to choose visual representation of air pollution depending on the data and context, and MARVIS~\cite{langner:2021:marvis} allows the user to arrange both physical and virtual AR displays in space depending on the structure of the underlying data.}

\smallskip

\begin{wrapfigure}[5]{l}{2cm}
    \vspace{-1.1em}
    \includegraphics[width=2cm]{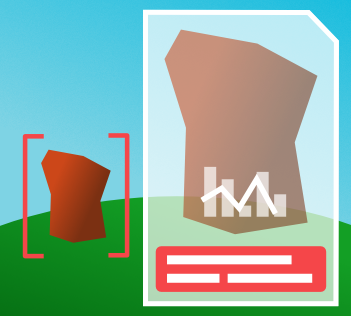}
\end{wrapfigure}

\smallskip

\noindent\textbf{Report:} Finally, at the top of the sensemaking loop, the analyst organizes the evidence into an artifact that combines all of the component findings into a coherent story that can be used to support or reject hypotheses. 
We also think of this level as a dissemination of findings to stakeholders or other analysts.
An ideal example of this would be a system that not only helps users organize evidence for a hypothesis, but also present this material in a product suitable for dissemination.

\begin{table*}[!ht]
    \def\arraystretch{1.1}
    \caption{\textbf{Classification of Augmented Reality systems for situated analytics}.
    The bottom section includes papers that were included in our selection of papers but omitted from the cluster analysis as they do not support sensemaking levels beyond \textsc{Read}.}
    \label{table:taxonomization_table}
    \centering
    \sffamily
    \tiny
    \resizebox{\textwidth}{!}{%
\setlength\arrayrulewidth{0.5pt}
\begin{tabular} {|c|l|r|c|c!{\color{diagramGrey}\vrule}c|c!{\color{diagramGrey}\vrule}c|c!{\color{diagramGrey}\vrule}c|c|c!{\color{diagramGrey}\vrule}c|c!{\color{diagramGrey}\vrule}c!{\color{diagramGrey}\vrule}c|c!{\color{diagramGrey}\vrule}c!{\color{diagramGrey}\vrule}c|c@{\hspace{-2mm}}|c|c|c|c|}
\cline{5-19}\cline{21-24}
\multicolumn{1}{c}{} &\multicolumn{1}{c}{} & \multicolumn{1}{c}{} & \multicolumn{1}{c}{} & \multicolumn{4}{|c|}{situating triggers} & \multicolumn{5}{c|}{view situatedness} & \multicolumn{6}{c|}{ data depiction} & & \multicolumn{4}{c|}{sensemaking}  \\ 
\cline{5-19}
\multicolumn{1}{c}{} & \multicolumn{1}{c}{} & \multicolumn{1}{c}{\textbf{Key}: \mytikzcodebits{diagramBlue} and \mytikzcodebits{Thistle!60!white} boxes denote dimension applies} &  \multicolumn{1}{c}{} & \multicolumn{2}{|c|}{referents} & \multicolumn{2}{c|}{context} & \multicolumn{2}{l|}{  device-} & \multicolumn{1}{c|}{} & \multicolumn{2}{l|}{world-} & \multicolumn{3}{c|}{visual encoding} & \multicolumn{3}{c|}{ data abstraction} & & \multicolumn{4}{c|}{support} \\
 \multicolumn{1}{c}{} & \multicolumn{1}{c}{} & \multicolumn{1}{c}{\hspace{0mm}Cluster colors correspond to Figure~\ref{fig:clusters_formed}} & \multicolumn{1}{c}{} & \multicolumn{2}{|c|}{} & \multicolumn{2}{c|}{} & \multicolumn{2}{l|}{anchored} & \multicolumn{1}{c|}{} & \multicolumn{2}{l|}{registered} & \multicolumn{3}{l|}{} & \multicolumn{3}{l|}{} & & \multicolumn{4}{l|}{} \\ 
 \cline{5-10} \cline{12-19} \cline{21-24}
 \multicolumn{1}{c}{\rotatebox{0}{\textbf{cluster}}} & \multicolumn{1}{c}{\textbf{n\textsuperscript{o}}} & \multicolumn{1}{c}{\textbf{name (shortened)}} & \multicolumn{1}{c}{\textbf{year}} & \multicolumn{1}{|c}{\rotatebox{90}{location}} & \multicolumn{1}{c|}{\rotatebox{90}{object}} & \multicolumn{1}{c}{\rotatebox{90}{event}} & \multicolumn{1}{c|}{\rotatebox{90}{setting}} & \multicolumn{1}{c}{\rotatebox{90}{relative}} & \multicolumn{1}{c|}{\rotatebox{90}{fixed}} & \multicolumn{1}{c|}{\rotatebox{90}{sticky}} & \multicolumn{1}{c}{\rotatebox{90}{absolute}} & \multicolumn{1}{c|}{\rotatebox{90}{relative}}  & \multicolumn{1}{c}{\rotatebox{90}{2D}} & \multicolumn{1}{c}{\rotatebox{90}{3D}} & \multicolumn{1}{c|}{\rotatebox{90}{nonvisual}} & \multicolumn{1}{c}{\rotatebox{90}{abstract}} & \multicolumn{1}{c}{\rotatebox{90}{hybrid}} & \multicolumn{1}{c|}{ \rotatebox{90}{physical } } & & \multicolumn{1}{c|}{ \rotatebox{90}{read}}  & \multicolumn{1}{c|}{ \rotatebox{90}{explore}}  & \multicolumn{1}{c|}{ \rotatebox{90}{schematize}}  & \multicolumn{1}{c|}{ \rotatebox{90}{report}}\\
\cline{1-19}\cline{21-24}

\cellcolor{cIcolor}  & 1 & Information Filtering for Mobile AR~\cite{julier02info_filtering} & 2000 &  \cmark & \cmark & \xmark & \xmark & \xmark & \xmark & \xmark & \cmark & \xmark & \xmark & \cmark & \xmark & \cmark & \xmark & \xmark & & \smark & \smark & \xmark & \xmark \\
\arrayrulecolor{diagramGrey}\cline{4-19}\cline{21-24}
\cellcolor{cIcolor}  & 2 & AR for Military in Urban Terrain~\cite{livingston02augmented} & 2002 &  \cmark & \cmark & \xmark & \xmark & \xmark & \xmark & \xmark & \cmark & \xmark & \xmark & \cmark & \xmark & \cmark & \xmark & \xmark & & \smark & \smark & \xmark & \xmark \\
\arrayrulecolor{diagramGrey}\cline{4-19}\cline{21-24}
\cellcolor{cIcolor} & 3 & The Virtual Mirror~\cite{bichlmeier09virtualmirror} & 2009 & \xmark & \cmark & \xmark & \xmark & \xmark & \xmark & \xmark & \cmark & \xmark & \cmark & \xmark & \xmark & \xmark & \xmark & \cmark & & \smark & \smark & \xmark & \xmark
\\ 
\arrayrulecolor{diagramGrey}\cline{4-19}\cline{21-24}
\cellcolor{cIcolor}  & 4 & Underground Infrastructure Vis~\cite{schall09infravis} & 2009 & \cmark & \xmark & \xmark & \xmark & \xmark & \xmark & \xmark & \cmark & \xmark & \xmark & \cmark & \xmark & \xmark & \xmark & \cmark  & &\smark & \smark & \xmark & \xmark \\
\arrayrulecolor{diagramGrey}\cline{4-19}\cline{21-24}
\cellcolor{cIcolor} & 5 & Situated Simulations~\cite{liestol09situated} & 2009 &  \cmark & \cmark & \xmark & \xmark & \xmark & \xmark & \xmark & \cmark & \xmark & \cmark & \cmark & \xmark & \cmark & \xmark & \cmark & & \smark & \smark & \xmark & \xmark \\ 
\arrayrulecolor{diagramGrey}\cline{4-19}\cline{21-24}
\cellcolor{cIcolor} & 6 & ARtifact~\cite{vanoni12artifact} & 2012 & \xmark & \cmark & \xmark & \xmark & \xmark & \cmark & \xmark & \xmark & \cmark & \cmark & \xmark & \xmark & \xmark & \xmark & \cmark & & \smark & \smark & \xmark & \xmark \\ 
\arrayrulecolor{diagramGrey}\cline{4-19}\cline{21-24}
\cellcolor{cIcolor} & 7 & Adaptive Ghost Views for AR~\cite{kalkofen13ghostar} & 2013 &  \cmark & \cmark & \xmark & \xmark & \xmark & \xmark & \xmark & \cmark & \xmark & \xmark & \cmark & \xmark & \xmark & \xmark & \cmark & & \smark & \smark & \xmark & \xmark
\\ 
\arrayrulecolor{diagramGrey}\cline{4-19}\cline{21-24}
\cellcolor{cIcolor} & 8 & In-Situ Climate Change Vis~\cite{liestol14climate} & 2014 &  \xmark & \cmark & \xmark & \xmark & \xmark & \xmark & \xmark & \cmark & \xmark & \cmark & \xmark & \xmark & \xmark & \xmark & \cmark  & & \smark & \smark & \xmark & \xmark \\
\arrayrulecolor{diagramGrey}\cline{4-19}\cline{21-24}
\cellcolor{cIcolor} & 9 & Riverwalk~\cite{cavallo16riverwalk} & 2016 &  \cmark & \cmark & \xmark & \xmark & \xmark & \xmark & \xmark & \cmark & \xmark & \cmark & \xmark & \xmark & \xmark & \xmark & \cmark & & \smark & \smark & \xmark & \xmark \\
\arrayrulecolor{diagramGrey}\cline{4-19}\cline{21-24}
\parbox[t]{1mm}{\multirow{-8}{*}{\rotatebox[origin=c]{90}{\cellcolor{cIcolor}\fontsize{6}{7}\selectfont\textbf{C1: Simulators}}}} & 10 & Corsican Twin~\cite{prouzeau20corsiantwin} & 2020 &  \xmark & \cmark & \xmark & \xmark & \xmark & \xmark & \xmark & \cmark & \xmark & \xmark & \cmark & \xmark & \xmark & \xmark & \cmark & & \smark & \smark & \smark & \xmark \\
\Cline{0.5pt}{0-18}\Cline{0.5pt}{21-24}

\cellcolor{cIIcolor}  &  11 & Situated Documentaries~\cite{hollerer99sitdocument} & 1999 & \cmark & \xmark & \xmark & \xmark & \xmark & \cmark & \xmark & \cmark & \xmark & \cmark & \cmark & \xmark & \cmark & \cmark & \cmark & & \smark & \smark & \xmark & \xmark \\
\arrayrulecolor{diagramGrey}\cline{4-19}\cline{21-24}
\cellcolor{cIIcolor}  &  12 & MARS~\cite{hollerer99mars} & 1999 & \cmark & \xmark & \xmark & \xmark & \xmark & \cmark & \xmark & \cmark & \xmark & \cmark & \cmark & \xmark & \cmark & \cmark & \cmark & & \smark & \smark & \xmark & \xmark \\
\arrayrulecolor{diagramGrey}\cline{4-19}\cline{21-24}
\cellcolor{cIIcolor}  & 13 & Virtual X-Ray Vision in Mobile AR~\cite{bane04xrayar} & 2004 &  \cmark & \xmark & \xmark & \xmark & \cmark & \cmark & \xmark & \xmark & \xmark & \cmark & \cmark & \xmark & \cmark & \xmark & \xmark & & \smark & \smark & \xmark & \xmark \\
\arrayrulecolor{diagramGrey}\cline{4-19}\cline{21-24}
\arrayrulecolor{diagramGrey}\cline{4-19}\cline{21-24}
\cellcolor{cIIcolor}  & 14 & Hypermedia Vis for AR/VR~\cite{guven06arhypermedia} & 2006 &  \cmark & \xmark & \xmark & \xmark & \xmark & \cmark & \xmark & \cmark & \xmark & \cmark & \cmark & \xmark & \cmark & \cmark & \xmark  & & \smark & \smark & \xmark & \xmark \\
\arrayrulecolor{diagramGrey}\cline{4-19}\cline{21-24}
\cellcolor{cIIcolor}  & 15 & AR on Historic Sites~\cite{guven06historicsitesAR} & 2006 &  \cmark & \xmark & \xmark & \cmark & \cmark & \cmark & \xmark & \cmark & \xmark & \cmark & \cmark & \xmark & \cmark & \cmark & \cmark & & \smark & \smark & \xmark  & \xmark \\
\arrayrulecolor{diagramGrey}\cline{4-19}\cline{21-24}
\cellcolor{cIIcolor} &  16 & SiteLens~\cite{white09sitelens} & 2009 & \cmark & \xmark & \xmark & \xmark & \xmark & \cmark & \xmark & \cmark &  \xmark & \cmark & \cmark & \xmark & \cmark & \cmark & \xmark & & \smark & \smark & \smark  & \xmark \\
\arrayrulecolor{diagramGrey}\cline{4-19}\cline{21-24}
\cellcolor{cIIcolor}  & 17 & CityViewAR~\cite{lee12cityviewAR} & 2012 &  \cmark & \xmark & \xmark & \xmark & \cmark & \cmark & \xmark & \cmark & \xmark & \cmark & \xmark & \xmark & \cmark & \cmark & \cmark & & \smark & \smark & \xmark & \xmark \\
\arrayrulecolor{diagramGrey}\cline{4-19}\cline{21-24}
\cellcolor{cIIcolor}  & 18 & Extended Overview Tech for Outdoor AR~\cite{veas12extendedAR} & 2012 & \cmark & \xmark & \xmark & \xmark & \cmark & \cmark & \xmark & \cmark & \xmark & \cmark & \cmark & \xmark & \cmark & \cmark & \xmark & & \smark & \smark & \xmark & \xmark \\
\arrayrulecolor{diagramGrey}\cline{4-19}\cline{21-24}
\cellcolor{cIIcolor}  &  19 & Wind and Uncertainty Vis~\cite{fjukstad14weatherforecast} & 2014 & \cmark & \xmark & \xmark & \xmark & \cmark & \cmark & \xmark & \cmark & \xmark & \cmark & \cmark & \xmark & \cmark &\cmark & \xmark & & \smark & \smark & \xmark & \xmark \\
\arrayrulecolor{diagramGrey}\cline{4-19}\cline{21-24}
\cellcolor{cIIcolor} & 20 & Whistland~\cite{luchetti17whistland} & 2017 &  \cmark & \xmark & \xmark & \xmark & \cmark & \cmark & \xmark & \xmark & \xmark & \cmark & \xmark & \xmark & \cmark & \cmark & \xmark & & \smark & \smark & \xmark & \xmark \\
\arrayrulecolor{diagramGrey}\cline{4-19}\cline{21-24}
\cellcolor{cIIcolor} & 21 & Pedestrian Navigation AR~\cite{joonagata17pedestrianAR} & 2017 &  \cmark & \xmark & \xmark & \xmark & \xmark & \cmark & \xmark & \xmark & \xmark & \cmark & \xmark & \xmark & \cmark & \cmark & \xmark & & \smark & \smark & \xmark & \xmark \\
\arrayrulecolor{diagramGrey}\cline{4-19}\cline{21-24}
\cellcolor{cIIcolor}  &  22 & The Urban Cobuilder~\cite{imottesjo18urbancobuilder} & 2018 & \cmark & \xmark & \xmark & \xmark & \xmark & \cmark & \xmark & \cmark & \xmark & \cmark & \cmark & \xmark & \cmark & \xmark & \xmark & & \smark & \smark & \xmark & \xmark \\
\arrayrulecolor{diagramGrey}\cline{4-19}\cline{21-24}
\cellcolor{cIIcolor} &  23 & AR for Supporting Aquaculture Farmers~\cite{xi19aquacultureAR} & 2019 & \cmark & \xmark & \xmark & \xmark & \cmark & \xmark & \xmark & \cmark & \xmark & \cmark & \cmark & \xmark & \cmark & \cmark & \xmark & & \smark & \smark & \xmark & \xmark \\ 
\arrayrulecolor{diagramGrey}\cline{4-19}\cline{21-24}
\cellcolor{cIIcolor} &  24 & Agricultural Fieldwork Vis~\cite{zheng19fieldworkar} & 2019 & \cmark & \xmark & \xmark & \xmark & \xmark & \cmark & \xmark & \xmark & \xmark & \cmark & \xmark & \xmark & \cmark & \xmark & \xmark & & \smark & \smark & \xmark & \xmark \\
\arrayrulecolor{diagramGrey}\cline{4-19}\cline{21-24}
\cellcolor{cIIcolor} & 25 & FieldView~\cite{whitlock20fieldview} & 2020 &  \cmark & \xmark & \xmark & \xmark & \cmark & \cmark & \xmark & \cmark & \xmark & \cmark & \cmark & \xmark & \cmark & \cmark & \xmark & & \smark & \smark & \smark & \xmark \\
\arrayrulecolor{diagramGrey}\cline{4-19}\cline{21-24}
 \parbox[t]{1mm}{\multirow{-18}{*}{\rotatebox[origin=c]{90}{\cellcolor{cIIcolor}\fontsize{6}{7}\selectfont\textbf{C2: Assistants}}}}  & 26 & Basketball AR~\cite{lin21basketballar} & 2021 & \cmark & \xmark & \cmark & \xmark & \xmark & \cmark & \xmark & \cmark & \xmark & \cmark & \cmark & \xmark & \cmark & \xmark & \xmark & & \smark & \smark & \xmark & \xmark \\
\Cline{0.5pt}{0-18}\Cline{0.5pt}{21-24}

\cellcolor{cIIIcolor} & 27 & AR for Manufacture Planning~\cite{doil03armanufacture} & 2003 & \xmark & \xmark & \xmark & \cmark & \xmark & \xmark & \xmark & \xmark & \cmark & \xmark & \cmark & \xmark & \xmark & \xmark & \cmark & & \smark & \smark & \xmark & \xmark \\
\arrayrulecolor{diagramGrey}\cline{4-19}\cline{21-24}
\cellcolor{cIIIcolor} & 28 & AR for Construction~\cite{Behzadan05constructionAR} & 2005 & \cmark & \xmark & \xmark & \cmark & \xmark & \cmark & \xmark & \xmark & \cmark & \xmark & \cmark & \xmark & \xmark & \xmark & \cmark & & \smark & \smark & \smark & \xmark \\
\arrayrulecolor{diagramGrey}\cline{4-19}\cline{21-24}
\cellcolor{cIIIcolor} & 29 & AR and Tangible Interface in Urban Design~\cite{seichter07ARtan} & 2007 & \xmark & \xmark & \xmark & \cmark & \xmark & \xmark & \xmark & \xmark & \cmark & \xmark & \cmark & \xmark & \xmark & \cmark & \cmark & & \smark & \smark & \smark & \xmark \\ 
\arrayrulecolor{diagramGrey}\cline{4-19}\cline{21-24}
\cellcolor{cIIIcolor} & 30 & Contextual Anatomic Mimesis~\cite{bichlmeier07mimesis} & 2007 & \xmark & \xmark & \xmark & \cmark & \xmark & \xmark & \xmark & \cmark & \xmark & \xmark & \cmark & \xmark & \cmark & \xmark & \cmark & & \smark & \smark & \xmark & \xmark \\
\arrayrulecolor{diagramGrey}\cline{4-19}\cline{21-24}
\cellcolor{cIIIcolor} & 31 & AR for Building Acceptance~\cite{schoenfelder08industacceptance} & 2008 & \xmark & \xmark & \xmark & \cmark & \xmark & \xmark & \xmark & \xmark & \cmark & \xmark & \cmark & \xmark & \xmark & \xmark & \cmark & & \smark & \smark & \xmark & \xmark \\
\arrayrulecolor{diagramGrey}\cline{4-19}\cline{21-24}
\cellcolor{cIIIcolor} & 32 & Vis for Psychomotor-Cognitive Tasks~\cite{kotranza09psychomotorAR} & 2009 & \xmark & \xmark & \xmark & \cmark & \xmark & \xmark & \xmark & \xmark & \cmark & \cmark & \xmark & \xmark & \cmark & \xmark & \cmark & & \smark & \smark & \xmark & \xmark \\
\arrayrulecolor{diagramGrey}\cline{4-19}\cline{21-24}
\cellcolor{cIIIcolor} & 33 & MR in OR~\cite{OkurMROR} & 2012 & \xmark & \xmark & \xmark & \cmark & \xmark & \xmark & \xmark & \xmark & \cmark & \xmark & \cmark & \xmark & \cmark & \xmark & \xmark & & \smark & \smark & \xmark & \xmark \\
\arrayrulecolor{diagramGrey}\cline{4-19}\cline{21-24}
\cellcolor{cIIIcolor} & 34 & In-Situ Flood Vis~\cite{haynes16floodAR} & 2016 & \xmark & \xmark & \xmark & \cmark & \xmark & \xmark & \xmark & \cmark & \xmark & \xmark & \cmark & \xmark & \xmark & \xmark & \cmark  & & \smark & \smark & \xmark & \xmark \\
\arrayrulecolor{diagramGrey}\cline{4-19}\cline{21-24}
\cellcolor{cIIIcolor} & 35 & Cues for Social Contacts~\cite{nassani17socialcontactar} & 2017 &  \xmark & \xmark & \xmark & \cmark & \xmark & \cmark & \xmark & \cmark & \xmark & \xmark & \cmark & \xmark & \cmark & \xmark & \xmark & & \smark & \smark & \xmark & \xmark \\
\arrayrulecolor{diagramGrey}\cline{4-19}\cline{21-24}
\cellcolor{cIIIcolor} & 36 & HydrogenAR~\cite{whitlock20hydrogenar} & 2020 & \xmark & \xmark & \xmark & \cmark & \xmark & \xmark & \xmark & \cmark & \xmark & \xmark & \cmark & \xmark & \cmark & \xmark & \cmark & & \smark & \smark & \xmark & \xmark \\
\arrayrulecolor{diagramGrey}\cline{4-19}\cline{21-24}
\cellcolor{cIIIcolor} & 37 & UpLift~\cite{ens21uplift} & 2021 & \xmark & \xmark & \xmark & \cmark & \xmark & \cmark & \xmark & \xmark & \cmark & \cmark & \cmark & \xmark & \cmark & \cmark & \cmark & & \smark & \smark & \smark & \xmark \\
\arrayrulecolor{diagramGrey}\cline{4-19}\cline{21-24}
\parbox[t]{1mm}{\multirow{-12}{*}{\rotatebox[origin=c]{90}{\cellcolor{cIIIcolor}\fontsize{6}{7}\selectfont\textbf{C3: Planners}}}}  & 38 & MARVIS~\cite{langner:2021:marvis} & 2021 & \xmark & \xmark & \xmark & \cmark & \xmark & \cmark & \xmark & \xmark & \cmark & \cmark & \cmark & \xmark & \cmark & \cmark & \xmark & & \smark & \smark & \smark & \xmark \\
\Cline{0.5pt}{0-18}\Cline{0.5pt}{21-24}

\cellcolor{cIVcolor} &  39 & Healthy Grocery Shopping via Mobile AR~\cite{ahn13groceryar} & 2013 & \xmark & \cmark & \xmark & \xmark & \xmark & \cmark & \xmark & \cmark & \xmark & \cmark & \xmark & \xmark & \cmark & \xmark & \xmark & & \smark & \smark & \xmark & \xmark \\
\arrayrulecolor{diagramGrey}\cline{4-19}\cline{21-24}
\cellcolor{cIVcolor} &  40 & Horus Eye~\cite{elsayed16horuseye} & 2016 & \xmark & \cmark & \xmark & \xmark & \xmark & \cmark & \xmark & \cmark & \xmark & \cmark & \cmark & \xmark & \cmark & \xmark &  \xmark  & & \smark & \smark & \xmark & \xmark \\
\arrayrulecolor{diagramGrey}\cline{4-19}\cline{21-24}
\cellcolor{cIVcolor} &  41 & SA for Shopping~\cite{elsayed16situateddef} & 2016 & \xmark & \cmark & \xmark & \xmark & \xmark & \cmark & \xmark & \cmark & \xmark & \cmark & \xmark & \xmark & \cmark & \xmark & \xmark  & & \smark & \smark & \xmark & \xmark \\
\arrayrulecolor{diagramGrey}\cline{4-19}\cline{21-24}
\cellcolor{cIVcolor} &  42 & Crime Scene Investigation AR~\cite{datcu16csiar} & 2016 & \xmark & \cmark & \xmark & \xmark & \xmark & \cmark & \xmark & \cmark & \xmark & \cmark & \xmark & \xmark & \cmark & \xmark & \xmark  & & \smark & \smark & \xmark & \xmark \\
\arrayrulecolor{diagramGrey}\cline{4-19}\cline{21-24}
\cellcolor{cIVcolor} &  43 & SA for Network Data~\cite{zhao17egocentricar} & 2017 & \xmark & \cmark & \xmark & \xmark & \xmark & \cmark & \xmark & \xmark & \xmark & \cmark & \xmark & \xmark & \cmark & \xmark & \xmark  & & \smark & \smark & \xmark & \xmark \\
\arrayrulecolor{diagramGrey}\cline{4-19}\cline{21-24}
\cellcolor{cIVcolor} & 44 & Supporting Fine Details from AR~\cite{jiang20finedetailsar} & 2020 & \xmark & \cmark & \xmark & \xmark & \cmark & \cmark & \xmark & \xmark & \xmark & \cmark & \xmark & \xmark & \cmark & \xmark & \xmark & & \smark & \smark & \xmark & \xmark \\
\arrayrulecolor{diagramGrey}\cline{4-19}\cline{21-24}
\parbox[t]{1mm}{\multirow{-7}{*}{\rotatebox[origin=c]{90}{\cellcolor{cIVcolor}\fontsize{6}{7}\selectfont\textbf{C4: Scanners}}}} & 45 & Impact of 2D Visualizations in AR~\cite{SatkowskuDachselt2021} & 2021 & \xmark & \cmark & \xmark & \xmark & \xmark & \xmark & \xmark & \cmark & \xmark & \cmark & \xmark & \xmark & \cmark & \xmark & \xmark & & \smark & \smark & \xmark & \xmark \\
\Cline{0.5pt}{0-18}\Cline{0.5pt}{21-24}

\cellcolor{OutlinerGray}\textbf{O1} & 46 & AR for Size Estimation~\cite{gomez08sizeestimationAR} & 2008 & \xmark & \cmark & \xmark & \xmark & \xmark & \xmark & \xmark & \xmark & \cmark & \xmark & \cmark & \xmark & \xmark & \xmark & \cmark & & \smark & \smark & \xmark & \xmark \\
\arrayrulecolor{diagramGrey}\cline{4-19}\cline{21-24}
\cellcolor{OutlinerGray}\textbf{O2}  & 47 & MARVisT~\cite{chen19marvist} & 2019 &  \xmark & \cmark & \xmark & \xmark & \xmark & \xmark & \xmark & \cmark & \xmark & \cmark & \cmark & \xmark & \cmark & \cmark & \cmark  & & \smark & \smark & \smark & \xmark \\
\arrayrulecolor{black}\cline{1-19}\cline{21-24} 
\multicolumn{23}{c}{}\\
\arrayrulecolor{black}\cline{1-19}\cline{21-24}
--- & 48 & A Touring Machine~\cite{feiner97touringmachine} & 1997 & \cmark & \xmark & \xmark & \xmark & \xmark & \cmark & \xmark & \cmark & \xmark & \cmark & \cmark & \xmark & \cmark & \xmark & \xmark & & \smark & \xmark & \xmark & \xmark \\
\arrayrulecolor{diagramGrey}\cline{4-19}\cline{21-24}
--- &  49 & Augmentable Reality~\cite{rekimoto98augmentable} & 1998 & \xmark & \xmark & \xmark & \cmark & \xmark & \cmark & \xmark & \cmark & \xmark & \xmark & \xmark & \cmark & \cmark & \xmark & \xmark & & \smark & \xmark & \xmark & \xmark \\
\arrayrulecolor{diagramGrey}\cline{4-19}\cline{21-24}
--- &  50 & AR on Molecular Models~\cite{gillet04moleculeAR} & 2004 & \xmark & \cmark & \xmark & \xmark & \xmark & \xmark & \xmark & \cmark & \xmark & \xmark & \cmark & \xmark & \cmark & \xmark & \xmark & & \smark & \xmark & \xmark & \xmark \\
\arrayrulecolor{diagramGrey}\cline{4-19}\cline{21-24}
--- & 51 & \rev{ARVino~\cite{king05arvino}} & 2005 &  \cmark & \xmark & \xmark & \xmark & \xmark & \cmark & \xmark & \cmark & \xmark & \cmark & \cmark & \xmark & \cmark & \xmark & \xmark & & \smark & \xmark & \xmark & \xmark \\
\arrayrulecolor{diagramGrey}\cline{4-19}\cline{21-24}
---  & 52 & AR for Opportunistic Controls~\cite{henderson10tangibleAR} & 2010 &  \xmark & \cmark & \xmark & \xmark & \xmark & \cmark & \xmark & \cmark & \xmark & \cmark & \xmark & \xmark & \cmark & \xmark & \xmark & & \smark & \xmark & \xmark & \xmark \\
\arrayrulecolor{diagramGrey}\cline{4-19}\cline{21-24}
---  & 53 & AR with Paper Maps~\cite{morrison11papermapAR} & 2011 &  \xmark & \cmark & \xmark & \xmark & \cmark & \xmark & \xmark & \xmark & \xmark & \cmark & \xmark & \xmark & \cmark & \cmark & & \xmark  & \smark & \xmark & \xmark & \xmark \\
\arrayrulecolor{diagramGrey}\cline{4-19}\cline{21-24}
---  & 54 & Handheld AR for Indoor Navigation~\cite{mulloni11indoorAR} & 2011 &  \cmark & \xmark & \xmark & \xmark & \xmark & \cmark & \cmark & \xmark & \xmark & \cmark & \xmark & \xmark & \cmark & \cmark & \xmark & & \smark & \xmark & \xmark & \xmark \\
\arrayrulecolor{diagramGrey}\cline{4-19}\cline{21-24}
---  & 55 & AR for Maintenance and Repair~\cite{henderson11maintenancerepairar} & 2011 &  \xmark & \cmark & \xmark & \xmark &  \xmark & \cmark & \xmark & \xmark & \xmark & \cmark & \cmark & \xmark & \cmark & \xmark & \cmark & & \smark & \xmark & \xmark & \xmark \\
\arrayrulecolor{diagramGrey}\cline{4-19}\cline{21-24}
---  & 56 & Bottari~\cite{bladuini12bottari} & 2012 & \cmark & \xmark & \xmark & \xmark & \cmark & \xmark & \xmark & \xmark & \xmark & \cmark & \xmark & \xmark & \cmark & \cmark & \xmark & & \smark & \xmark & \xmark & \xmark \\
\arrayrulecolor{diagramGrey}\cline{4-19}\cline{21-24}
---  & 57 & Label Placement in AR~\cite{grasset12labelAR} & 2012 &  \xmark & \cmark & \xmark & \xmark & \xmark & \cmark & \xmark & \xmark & \xmark & \cmark & \xmark & \xmark & \cmark & \xmark & \xmark & & \smark & \xmark & \xmark & \xmark \\
\arrayrulecolor{diagramGrey}\cline{4-19}\cline{21-24}
---  & 58 & AR for Outdoor Navigation~\cite{dunser12ARoutdoornavigation} & 2012 &  \cmark & \xmark & \xmark & \xmark & \cmark & \cmark & \xmark & \xmark & \xmark & \cmark & \xmark & \xmark & \cmark & \cmark & \xmark & & \smark & \xmark & \xmark & \xmark \\
\arrayrulecolor{diagramGrey}\cline{4-19}\cline{21-24}
---  & 59 & Inattentional Blindness in Surgical AR~\cite{dixon13surgeons} & 2013 & \xmark & \cmark & \xmark & \cmark & \xmark & \cmark & \xmark & \cmark & \xmark & \xmark & \xmark & \cmark & \cmark & \xmark & \xmark & & \smark & \xmark & \xmark & \xmark \\
\arrayrulecolor{diagramGrey}\cline{4-19}\cline{21-24}
---  & 60 & Orientation Measurement for Audio AR~\cite{heller14mobaudioar} & 2014 &  \xmark & \xmark & \xmark & \cmark & \xmark & \xmark & \xmark & \cmark & \xmark & \xmark & \xmark & \cmark & \cmark & \xmark & \xmark & & \smark & \xmark & \xmark & \xmark \\
\arrayrulecolor{diagramGrey}\cline{4-19}\cline{21-24}
---  & 61 & Hedgehog Labeling~\cite{tatzgern14hedgehog} & 2014 & \xmark & \cmark & \xmark & \xmark & \cmark & \xmark & \xmark & \xmark & \xmark & \cmark & \xmark & \xmark & \cmark & \xmark & \xmark & & \smark & \xmark & \xmark & \xmark \\
\arrayrulecolor{diagramGrey}\cline{4-19}\cline{21-24}
---  &  62 & HMD Vis for the Deaf~\cite{jain15hmdvis} & 2015 & \xmark & \xmark & \cmark & \cmark & \cmark & \xmark & \xmark & \xmark & \xmark & \xmark & \xmark & \cmark & \cmark & \xmark & \xmark  & & \smark  & \xmark & \xmark & \xmark \\
\arrayrulecolor{diagramGrey}\cline{4-19}\cline{21-24}
--- & 63 & Rhema~\cite{tanveer15rhema} & 2015 &  \xmark & \cmark & \xmark & \cmark & \cmark & \xmark & \xmark & \xmark & \cmark & \cmark & \cmark & \xmark & \cmark & \cmark & \xmark & & \smark & \xmark & \xmark & \xmark \\ 
\arrayrulecolor{diagramGrey}\cline{4-19}\cline{21-24}
--- & 64 & \rev{Borehole AR~\cite{lee15boreholear}} & 2015 & \cmark & \xmark & \xmark & \xmark & \cmark & \cmark & \xmark & \xmark & \xmark & \cmark & \xmark & \xmark & \cmark & \cmark & \xmark & & \smark & \xmark & \xmark & \xmark \\
\arrayrulecolor{diagramGrey}\cline{4-19}\cline{21-24}
--- & 65 & Projector-based AR for Welding~\cite{doshi17projectoar} & 2017 & \xmark & \xmark & \xmark & \cmark & \xmark & \xmark & \xmark & \xmark & \cmark & \cmark & \xmark & \xmark & \cmark & \xmark & \xmark & & \smark & \xmark & \xmark & \xmark \\ 
\arrayrulecolor{diagramGrey}\cline{4-19}\cline{21-24}
--- &  66 & ToARist~\cite{williams17toarist} & 2017 & \cmark & \xmark & \xmark & \xmark & \cmark & \cmark & \xmark & \xmark & \xmark & \cmark & \xmark & \xmark & \cmark & \cmark & \xmark & & \smark & \xmark & \xmark & \xmark \\
\arrayrulecolor{diagramGrey}\cline{4-19}\arrayrulecolor{diagramGrey}\cline{21-24}
--- & 67 & iMAP-CampUS~\cite{alqahtani17imapcampus} & 2017 &  \cmark & \xmark & \xmark & \xmark & \cmark & \cmark & \xmark & \xmark & \xmark & \cmark & \xmark & \xmark & \cmark & \cmark & & \xmark  & \smark & \xmark & \xmark & \xmark \\
\arrayrulecolor{diagramGrey}\cline{4-19}\cline{21-24}
---  & 68 & SonifEye~\cite{roodaki17sonifeye} & 2017 &  \xmark & \xmark & \cmark & \cmark & \xmark & \cmark & \xmark & \xmark & \xmark & \xmark & \xmark & \cmark & \cmark & \xmark & \xmark  & & \smark & \xmark & \xmark & \xmark \\
\arrayrulecolor{diagramGrey}\cline{4-19}\cline{21-24}
---  & 69 & ServAR~\cite{rollo17servar} & 2017 & \xmark & \cmark & \xmark & \xmark & \xmark & \xmark & \xmark & \xmark & \cmark & \xmark & \cmark & \xmark & \xmark & \xmark & \cmark & & \smark & \xmark & \xmark & \xmark \\ 
\arrayrulecolor{diagramGrey}\cline{4-19}\cline{21-24}
---  & 70 & Collision Warning AR~\cite{kim18collisionar} & 2018 & \cmark & \xmark & \cmark & \xmark & \xmark & \cmark & \xmark & \cmark & \xmark & \xmark & \cmark & \xmark & \cmark & \xmark & \xmark & & \smark & \xmark & \xmark & \xmark \\
\arrayrulecolor{diagramGrey}\cline{4-19}\cline{21-24}
--- & 71 & Information Seeking Strategy in AR~\cite{caggianese19arinformationseek} & 2019 & \cmark & \xmark & \xmark & \xmark & \xmark & \xmark & \xmark & \cmark & \xmark & \cmark & \xmark & \xmark & \cmark & \xmark & \xmark & & \smark & \xmark & \xmark & \xmark \\
\arrayrulecolor{diagramGrey}\cline{4-19}\cline{21-24}
--- & 72 & \rev{Situated Storytelling~\cite{ketchell19situatedstorytelling}} & 2019 & \cmark & \xmark & \xmark & \xmark & \xmark & \cmark & \xmark & \cmark & \xmark & \xmark & \cmark & \xmark & \cmark & \xmark & \xmark  &  & \smark & \xmark & \xmark & \xmark \\
\arrayrulecolor{diagramGrey}\cline{4-19}\cline{21-24}
--- & 73 & \rev{HypAR~\cite{engelke19hypar}} & 2019 & \xmark & \xmark & \xmark & \cmark & \xmark & \xmark & \xmark & \cmark & \cmark & \cmark & \xmark & \xmark & \cmark & \cmark & \xmark & & \smark & \xmark & \xmark & \xmark \\
\arrayrulecolor{diagramGrey}\cline{4-19}\cline{21-24}
--- & 74 & PapARVis~\cite{chen20staticvis} & 2020 & \xmark & \xmark & \xmark & \cmark & \xmark & \xmark & \xmark & \cmark & \xmark & \cmark & \xmark & \xmark & \cmark & \xmark & \xmark & & \smark & \xmark & \xmark & \xmark \\ 
\arrayrulecolor{diagramGrey}\cline{4-19}\cline{21-24}
--- & 75 & Visualizing Air Pollution (AiR)~\cite{mathews21airpollutionvis} & 2021 & \cmark & \xmark & \xmark & \xmark & \cmark & \cmark & \xmark & \xmark & \xmark & \cmark & \xmark & \xmark & \cmark & \xmark & \xmark & & \smark & \xmark & \xmark & \xmark \\
\arrayrulecolor{black}\cline{4-19}\cline{21-24}

\cline{1-19}\cline{21-23}
\end{tabular}
}
\end{table*}

\begin{table*}[!ht]
    \def\arraystretch{1.1}
    \caption{\textbf{Statistical summary of classified papers.} 
    Each value in the four dimensions has been abbreviated.}
    \label{table:stats_table}
    \centering
    \sffamily
    \resizebox{\textwidth}{!}{%
    \tiny
    \begin{tabular}{|c|c|cc|cc|cc|c|cc|ccc|ccc|cccc|}
    \cline{3-21}
    \multicolumn{1}{c}{} &\multicolumn{1}{c|}{} & \multicolumn{4}{c|}{Situating triggers} & \multicolumn{5}{c|}{View situatedness } & \multicolumn{6}{c|}{ Data Depiction }& \multicolumn{4}{c|}{} \\
    \cline{3-17}
    \multicolumn{1}{c}{} & \multicolumn{1}{c|}{} & \multicolumn{2}{c|}{Referents} & \multicolumn{2}{c|}{Context} & \multicolumn{2}{l|}{Device-} & \multicolumn{1}{c|}{} & \multicolumn{2}{l|}{World-}  & \multicolumn{3}{c|}{visual encoding} & \multicolumn{3}{c|}{data abstraction} & \multicolumn{4}{c|}{ Sensemaking Support }  \\
    \multicolumn{1}{c}{} & \multicolumn{1}{c|}{} & \multicolumn{2}{c|}{} & \multicolumn{2}{c|}{} & \multicolumn{2}{l|}{anchored} & \multicolumn{1}{c|}{} & \multicolumn{2}{l|}{registered} & \multicolumn{3}{l|}{} & \multicolumn{3}{l|}{} & \multicolumn{4}{c|}{}   \\ \cline{3-21}
    \multicolumn{1}{c}{\textbf{Year}} & \multicolumn{1}{c|}{\textbf{Total}} & \multicolumn{1}{c}{\rotatebox{0}{loc}} & \multicolumn{1}{c|}{\rotatebox{0}{obj}} & \multicolumn{1}{c}{\rotatebox{0}{time}} & \multicolumn{1}{c|}{\rotatebox{0}{set}} & \multicolumn{1}{c}{\rotatebox{0}{rel}} & \multicolumn{1}{c|}{\rotatebox{0}{fix}} & \multicolumn{1}{c|}{\rotatebox{0}{sti}} & \rotatebox{0}{abs} & \multicolumn{1}{c|}{\rotatebox{0}{rel}}  & \rotatebox{0}{2D} & \rotatebox{0}{3D} & \multicolumn{1}{c|}{\rotatebox{0}{nonv}} & \multicolumn{1}{c}{\rotatebox{0}{abs}} & \multicolumn{1}{c}{\rotatebox{0}{hybr}} & \multicolumn{1}{c|}{ \rotatebox{0}{phys}} & 
    \multicolumn{1}{c}{\rotatebox{0}{read}} & \multicolumn{1}{c}{\rotatebox{0}{expl}} & \multicolumn{1}{c}{\rotatebox{0}{sche}} & \multicolumn{1}{c|}{ \rotatebox{0}{repo}}   \\
    
    \hline
    
   \rowcolor{gray!10}  bef. 2000     & \textbf{2}  & 2 & 0 & 0 & 0 & 0 & 2 & 0 & 2 & 0 & 2 & 2 & 0 & 2 & 2 & 2 & 2 & 2 & 0 & 0  \\
    2000 - 2004   & \textbf{4}  & 3 & 2 & 0 & 1 & 1 & 1 & 0 & 2 & 1 & 1 & 4 & 0 & 3 & 0 & 1 & 4 & 4 & 0 & 0   \\
   \rowcolor{gray!10}  2005 - 2009  & \rev{\textbf{12}}  & \rev{6} & 3 & 0 & 6 & 1 & \rev{4} & 0 & \rev{7} & 5 & \rev{6} & \rev{10} & 0 & \rev{6} & 4 & 10 & \rev{12} & \rev{12} & \rev{3} & 0   \\
    2010 - 2014  & \textbf{7}  & 4 & 3 & 0 & 1 & 3 & 4 & 0 & 5 & 2 & 5 & 4 & 0 & 4 & 3 & 4 & 7 & 7 & 0 & 0  \\
    \rowcolor{gray!10} 2015 - 2019  & \rev{\textbf{14}} & \rev{6} & 7 & 0 & \rev{2} & \rev{2} & \rev{10} & 0 & \rev{10} & \rev{0} & \rev{12} & \rev{6} & 0 & \rev{12} & \rev{4} & 3 & \rev{14} & \rev{14} & \rev{1} & 0   \\
    2020 - onw.  & \textbf{8}  & 2 & 3 & 1 & 3 & 2 & 5 & 0 & 5 & 3 & 6 & 6 & 0 & 7 & 3 & 3 & 8 & 8 & \rev{4} & 0  \\
    \hline
    
    \rowcolor{AntiqueWhite1} \textbf{Total} & \rev{\textbf{47}} & \rev{\textbf{23}} & \textbf{18} & \textbf{1} & \rev{\textbf{13}} & \rev{\textbf{9}} & \rev{\textbf{26}} & \textbf{0} & \rev{\textbf{31}} & \rev{\textbf{11}} & \rev{\textbf{32}} & \rev{\textbf{32}} & \rev{\textbf{0}} & \rev{\textbf{34}} & \rev{\textbf{16}} & \textbf{23} & \rev{\textbf{47}} & \rev{\textbf{47}} & \rev{\textbf{8}} & \textbf{0}\\
    
    \hline
    \end{tabular}
    }
\end{table*}

%% file: 04_analysis.tex
\section{Meta-Analysis of Taxonomy}
\label{sec:taxonomy_analysis}

Here we explore our classification of different \SA{} systems through an ensemble combination of multiple clustering algorithms. 
This allows us to identify four archetypical patterns as well as a few outliers. 
We describe each of the four clusters as well as the outliers, focusing on their characteristics and elaborating on them using examples.

\begin{figure}[tbp]
    \centering
    \includegraphics[width=0.98\linewidth]{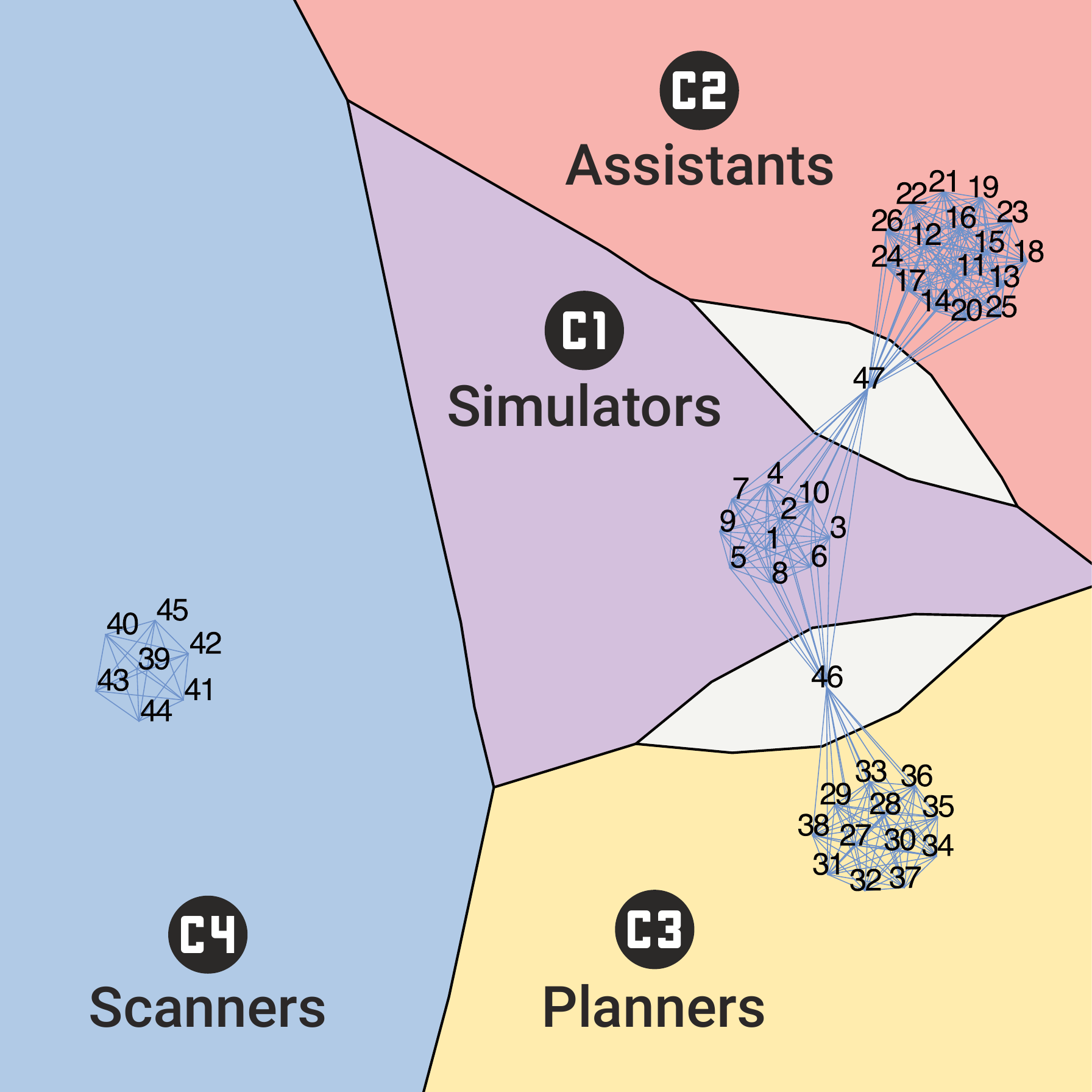}
    \caption{\textbf{Ensemble combination of multiple cluster analyses performed on our classification.}
    The resulting cluster network has been visualized through a variant of a self-organizing map~\cite{DBLP:journals/pieee/Kohonen90}: it has been laid out using a force-directed graph algorithm and the regions have been tesselated using a Voronoi diagram.
    This exposes the structure of the design space and current research in this area.
    Note that while the graph layout algorithm is non-deterministic, this has no effect on the resulting clusters, only the visual appearance of this self-organizing map.
    Thus, the specific layout does not impact the validity of our work.}
    \label{fig:clusters_formed}
\end{figure}

\subsection{Analysis Method}

Unlike traditional survey papers, where the authors manually cluster the surveyed papers, we take a data-driven approach using an ensemble of multiple cluster algorithms.
We do this to reduce the impact of author bias and increase the validity of our resulting clusters.
Furthermore, by using an ensemble combination of multiple cluster analyses, we can eliminate idiosyncracies from individual cluster algorithms.

After the automatic cluster analysis, all authors validated the resulting clusters and outliers.
Any discrepancies raised by one or more authors were discussed as a group and the classification redone.
This manual validation step allowed us to confirm the findings using our own expertise.

To perform cluster analysis, we first transform each surveyed paper into a 15-dimensional binary vector. 
Note that we do not use the dimension ``sensemaking support'' in the analysis; it is used only to distinguish SA from SVs. 
A value of each dimension in a paper is 1 if the marked value is ``yes'' and 0 if the value is ``no'' in the taxonomy table.
We have three goals in selecting a clustering algorithm:

\begin{enumerate}
    \item \textbf{Eliminate algorithm idiosyncracies.} We need a method that combines multiple cluster analyses.  
    \item \textbf{Yield high clustering quality.} The quality of created clusters should be optimal.
    \item \textbf{Represents both large and small clusters.} The algorithm should not directly dismiss small clusters. 
\end{enumerate}

Accordingly, our strategy is to create an ensemble model of five clustering algorithms, where each algorithm is optimized to yield high quality clusters. 
The five different clustering methods are k-means, Hierarchical Clustering, Spectral Clustering, Birch, and UMAP initialized with PCA.

\textbf{Optimizing models.}
We optimize all hyperparameters except the number of clusters (the $k$-value) via random grid search.
To choose the optimal number of clusters, we run each model for different number of clusters, from $k=2$ to 11, to avoid dismissing small but significant clusters. 
In choosing the optimal number of clusters, we use a combination of three metrics: Silhouette~\cite{rousseeuw87silhouette}, Calinski-Harabasz~\cite{calinski74harabaszscore}, and Davies-Bouldin~\cite{davies79daviesbouldin} scores.
These metrics are widely used for evaluating the quality of clusters that do not have predetermined labels.
Since they have different units, for each $k$ we rank them ordinally, giving 10 to the highest and 1 to the lowest ones. 
Then, we sum these three scores and select the one with the lowest score.
As a result, the optimal $k$ for k-means is 4, \rev{11} for Hierarchical Clustering, \rev{7} for Spectral Clustering, and \rev{4} for Birch.

UMAP is a model optimized for obtaining local structure. 
We find the optimal hyperparameters via random grid search.
We initialize UMAP using PCA~\cite{wold87PCA}, as recent work~\cite{kobak21inittsneumap} suggests that such initialization helps preserve global structure. 
We categorize an element to a group if the Euclidean distance between the element and any element of the group is less than the threshold of 0.1. 
That is, if an element has a neighbor within 0.1 distance, then it is classified as the same group as that neighbor. 
This results in a total of seven clusters. 

\textbf{Ensemble Model.} 
We use an ensemble model to formulate clusters using the five selected clustering algorithms.
More specifically, we use a graph-based consensus clustering method.
First, we create five \rev{$47 \times 47$} similarity matrices $S_i~(i=\text{1 to 5})$, where each $i$ represent different clustering algorithms. 
These similarity matrices assume that papers belonging to the same cluster in a clustering algorithm are related. 
Accordingly, given $p_x$ as the $x^{th}$ paper of the taxonomy, and $C_i$ as the clustering results of $i^{th}$ cluster, each entry in the similarity matrix is defined as follows:\\[-2em]

\begin{equation} \label{eq:1}
S_{i}(p_x,p_y) = 
    \begin{cases}
      1 & \text{if $C_i(p_x)$ = $C_i(p_y)$, $x \geq y$.}\\
      0 & \text{otherwise.}\\
    \end{cases}
\end{equation}

By adding all $S_i$s, we formulate a consensus matrix~\cite{fred05consensusmat} $CSS(p_x,p_y)$ that represents the relationship between papers based on clusters from five clustering algorithms. 
We do not define the lower triangular part of the matrix, as it is redundant to the upper triangular one.
Then, we create a graph $G = (V,E)$ where the vertices $v_x$, $v_y$ are the set of \rev{47} papers and the edge $v_{ij}$ that connect between papers $p_x$ and $p_y$ are defined as $CSS(p_x,p_y)$. 
In identifying the clusters, we utilize the normalized cut algorithm~\cite{shi00normalizedcut} to the graph.
This yields four clusters representing common patterns.

\textbf{Validation Process.} Three of our authors validated the results from the final cluster.
To understand the topological structure of our graph, we created a force-directed node-link diagram, as shown in Fig.~\ref{fig:clusters_formed}. 
The diagram is drawn using the same set of vertices and weighted edges of the graph $G$.
Using the diagram and the information from Table~\ref{table:taxonomization_table}, we checked papers within the clusters and identified the similarities between them. 
We also distinguish outliers and ambiguous works that do not clearly belong to a cluster.

\textbf{Implementation.}
We conducted the clustering in a Jupyter Notebook with Python 3.9 environment, on a Macbook Pro 2021 version that has a M1 Chip, 1TB SSD, and 16GB RAM.
We provide details of our analysis---classification, cluster analysis, and ensemble combination---in an OSF repository: \url{https://osf.io/3wxv2/}

\subsection{Resulting Clusters}
\label{sec:umap_results}

The graph structure of surveyed \SA{} systems resulted in four clusters and four outliers; see Fig.~\ref{fig:clusters_formed}.
Information about each paper is shown in Table~\ref{table:taxonomization_table}.
We also present the composition of these clusters in 5-year periods in Table~\ref{table:stats_table}. 
Below we describe the characteristics of these clusters (C1-C4) in detail.

\begin{figure*}[htb]
    \centering
    \subfloat[]{\includegraphics[height=0.17\textwidth]{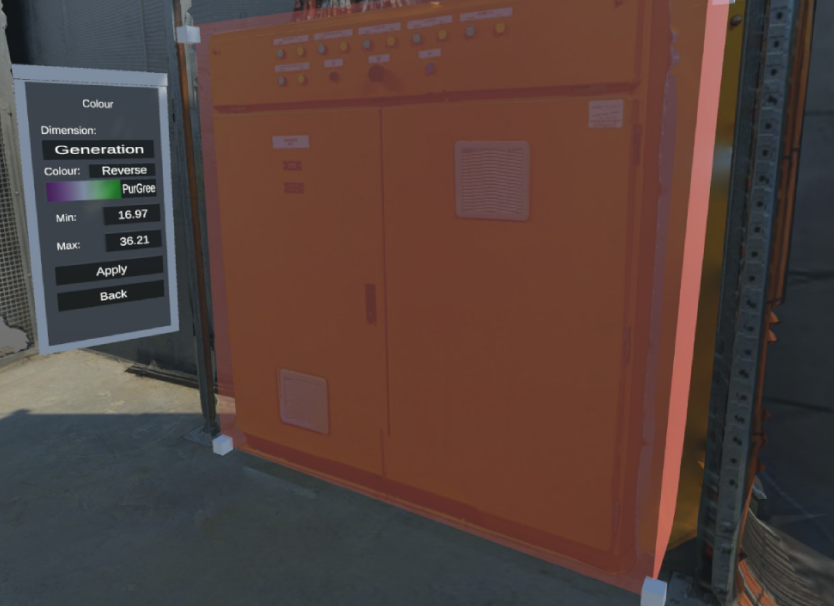}\label{subfig:corsicantwin}}\hspace{0.2mm} 
    \subfloat[]{\includegraphics[height=0.17\textwidth]{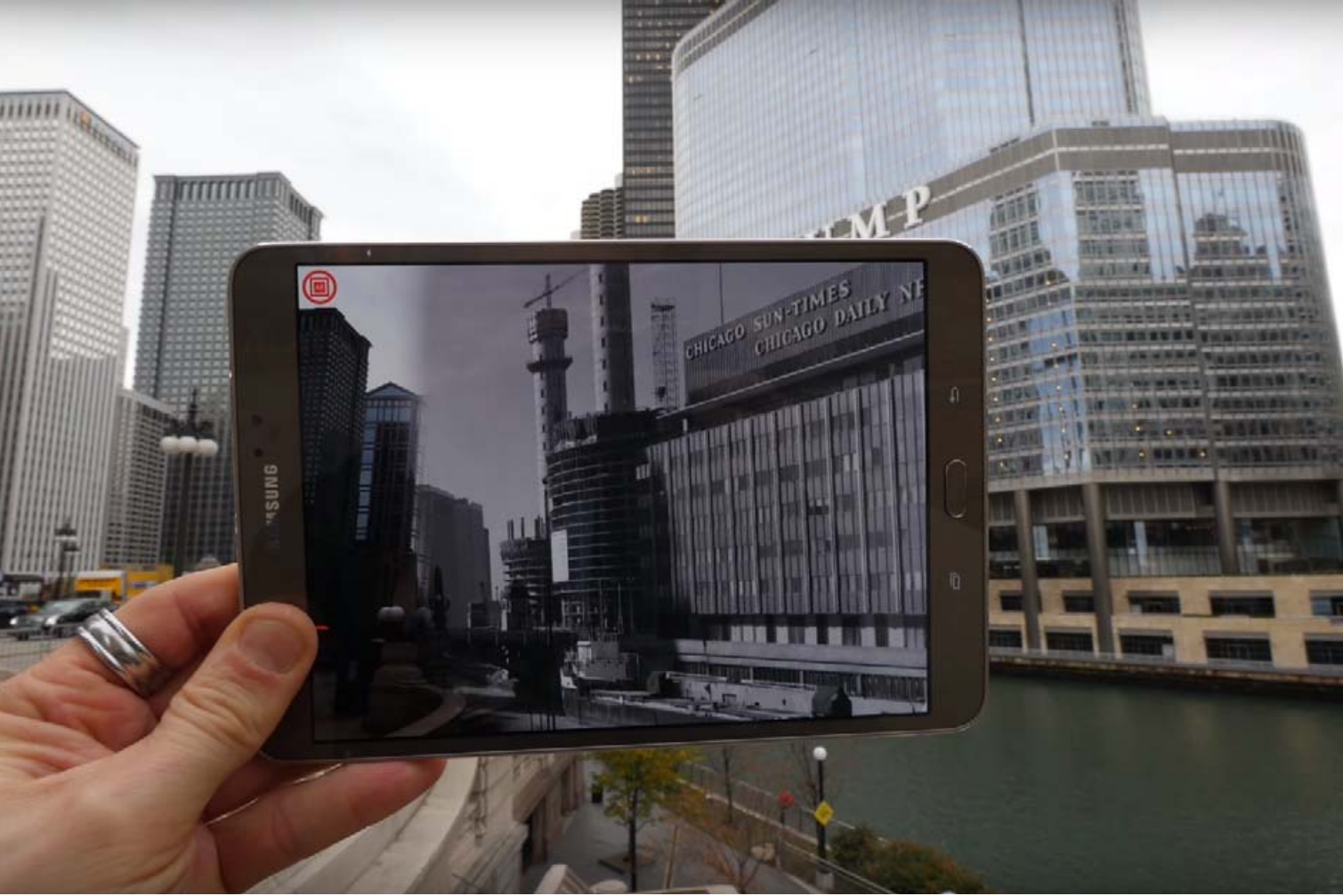}\label{subfig:riverwalk}}\hspace{0.1mm}
    \subfloat[]{\includegraphics[height=0.17\textwidth]{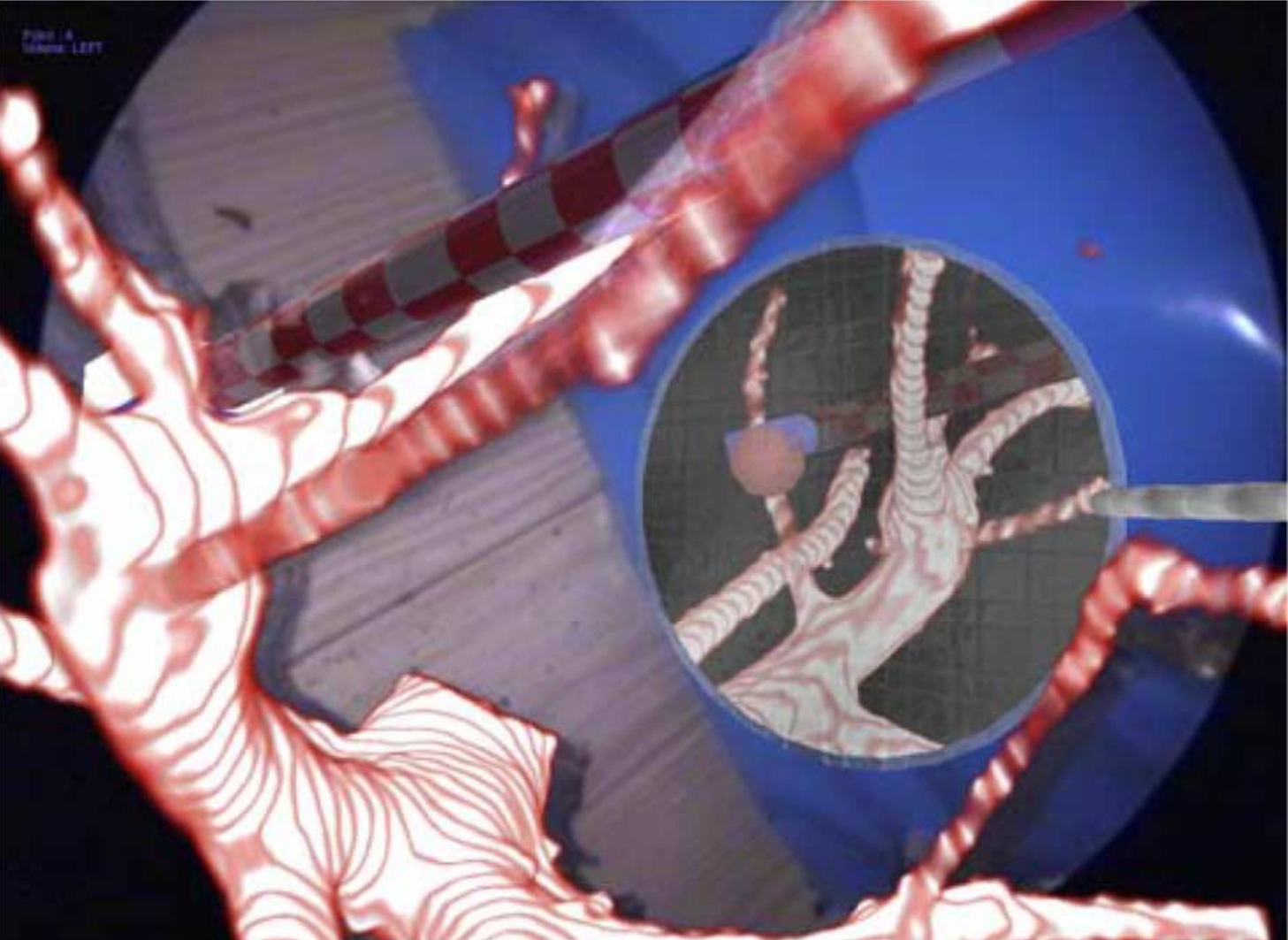}\label{subfig:virtual_mirror}}\hspace{0.1mm}
    \subfloat[]{\includegraphics[height=0.17\textwidth]{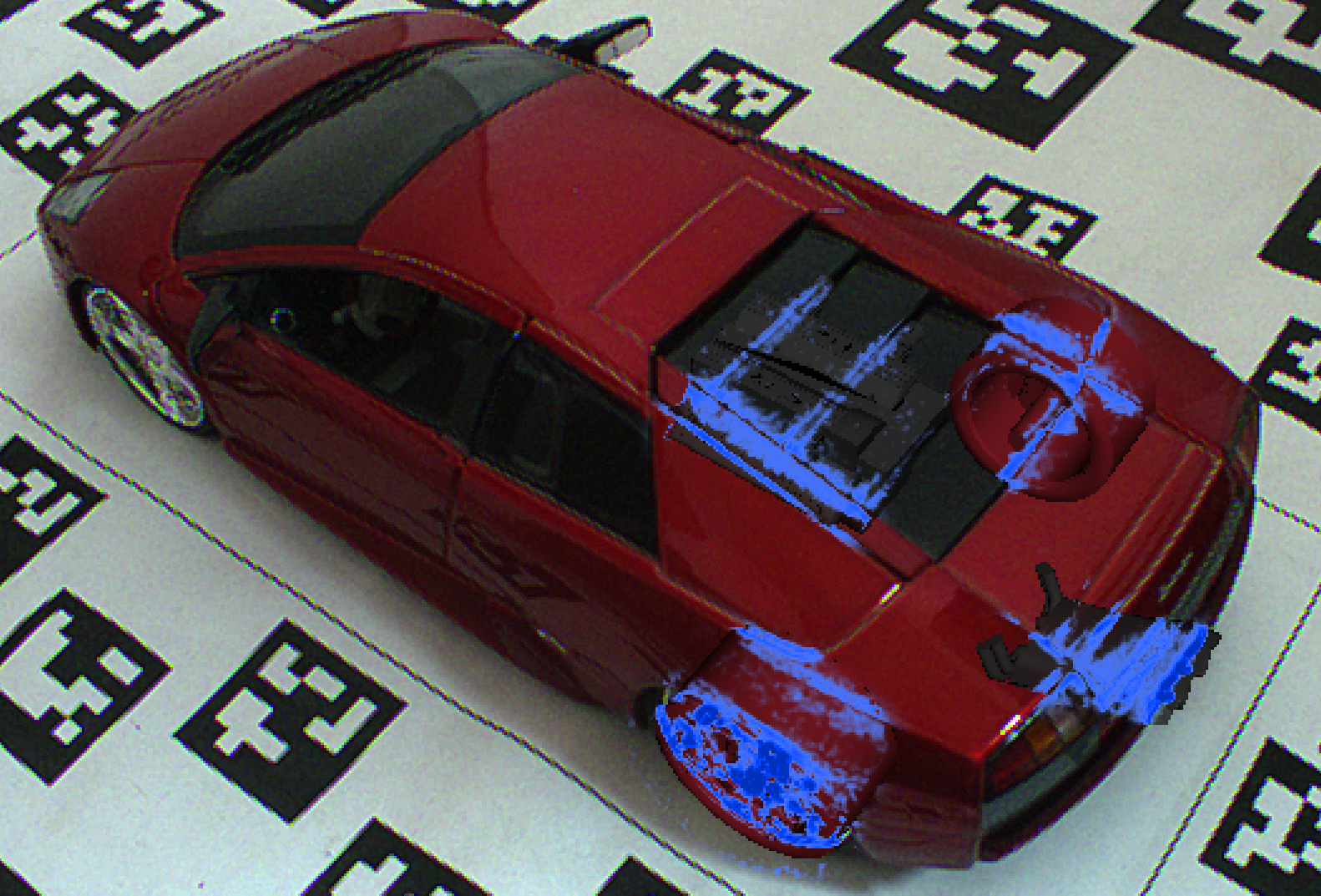}\label{subfig:ghost}}
    \caption{\textbf{Prototypical examples of \textit{Simulators} (C1).}
    \protect\subref{subfig:corsicantwin}~Corsican Twin: a tool for designing situated data visualizations~\cite{prouzeau20corsiantwin}.
    \protect\subref{subfig:riverwalk}~RiverWalk: a situated analytics tool that superimposes historical images onto matching views~\cite{cavallo16riverwalk}.
    \protect\subref{subfig:virtual_mirror}~The Virtual Mirror~\cite{bichlmeier09virtualmirror}: an AR-based mirror that lets users view part of a virtual object that is occluded in the user's view. 
    \protect\subref{subfig:ghost}~Adaptive ghosted view~\cite{kalkofen13ghostar}: a technique that helps users look at the interior of an object from outside, in a see-through manner}.
    \label{fig:c1_images}
\end{figure*}

\begin{figure*}[htb]
    \centering
    \subfloat[]{\includegraphics[height=0.171\textwidth]{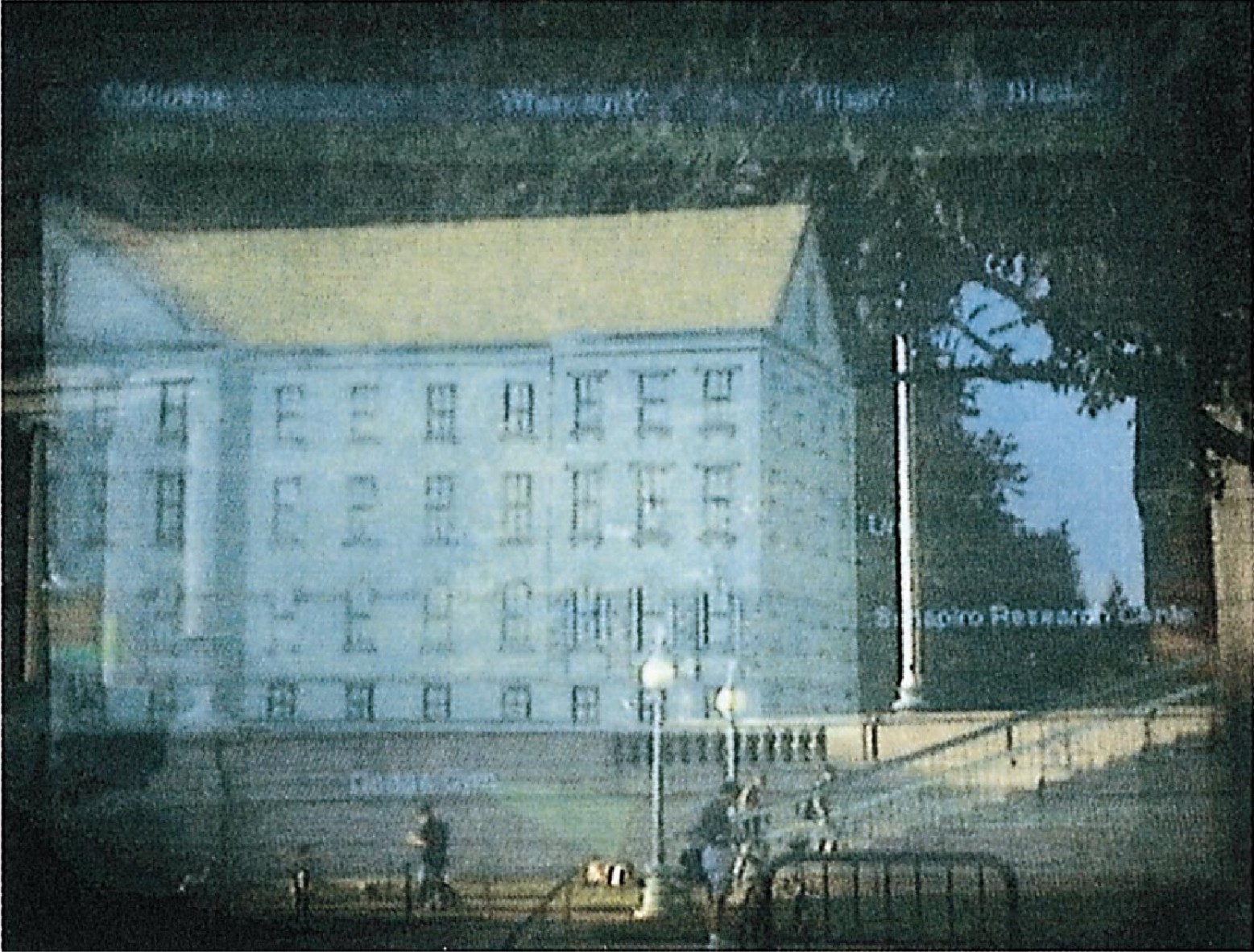}
    \label{subfig:mars}}
    \subfloat[]{\includegraphics[height=0.171\textwidth]{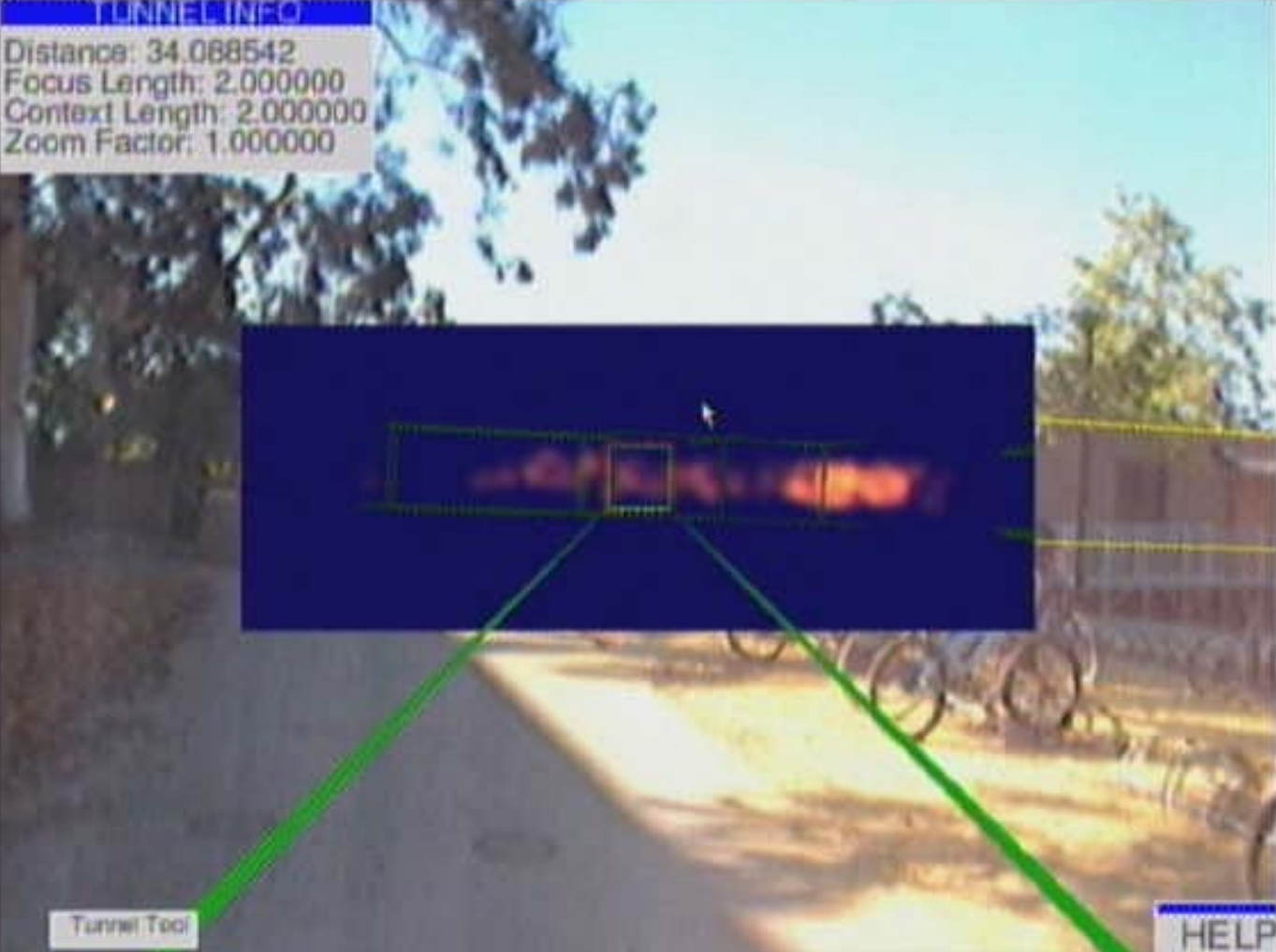}
    \label{subfig:xrayvision}}
    \subfloat[]{\includegraphics[height=0.171\textwidth]{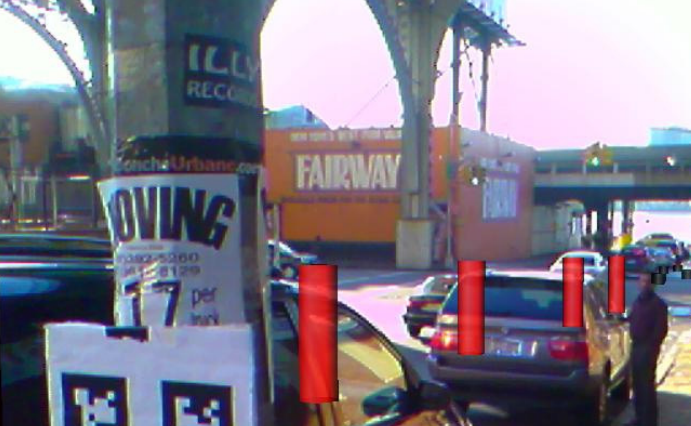}
    \label{subfig:sitelens}}
    \subfloat[]{\includegraphics[height=0.171\textwidth]{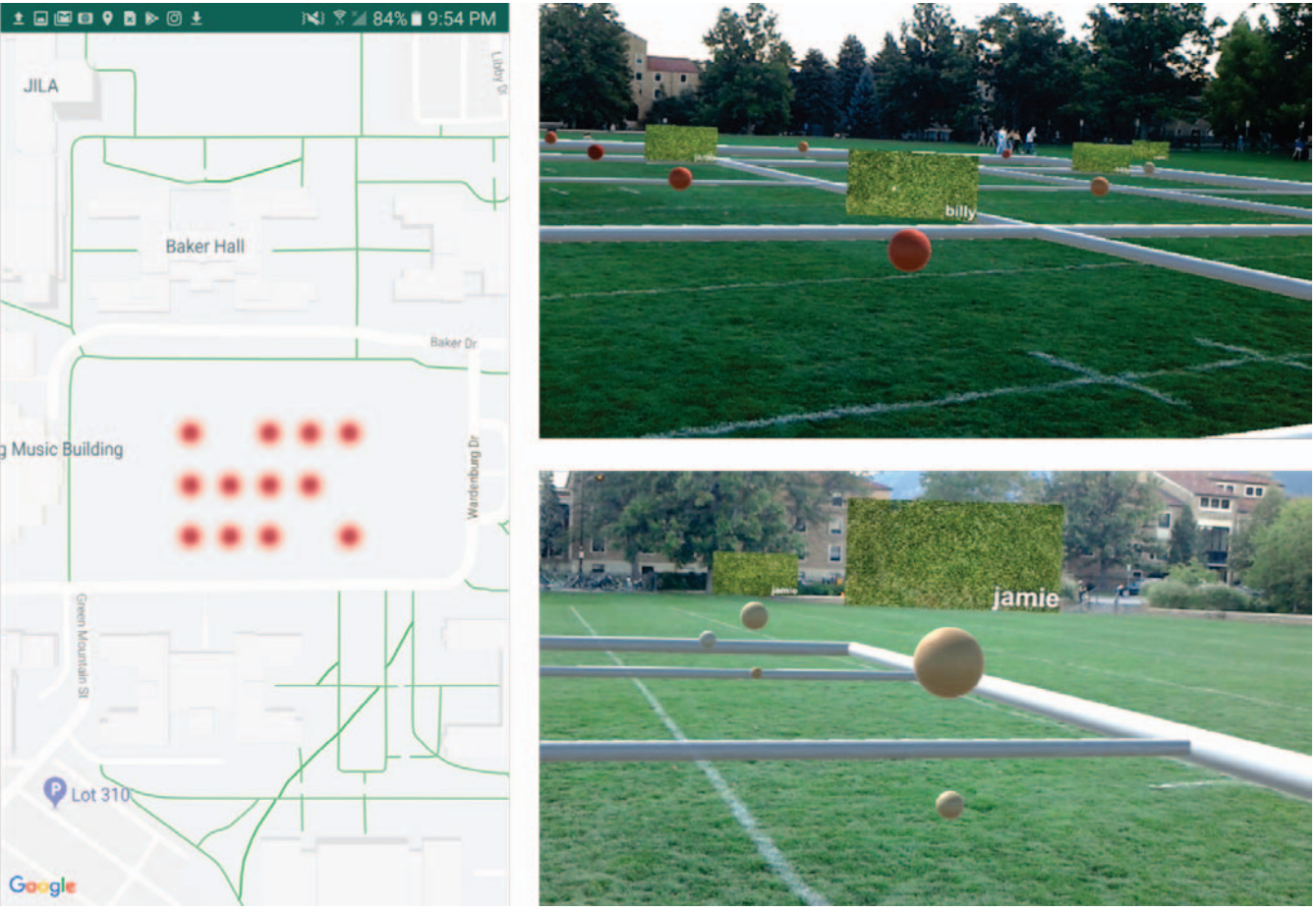}
    \label{subfig:fieldview}}
    \caption{\textbf{Prototypical examples of \textit{Assistants} (C2).}
    \protect\subref{subfig:mars}~MARS: a situated analytics tool that helps the user to explore about different locations of Columbia University campus~\cite{hollerer99mars}.
    \protect\subref{subfig:xrayvision}~A situated analytics tool showing a virtual X-Ray vision of a location~\cite{bane04xrayar}.
    \protect\subref{subfig:sitelens}~SiteLens: a 3D visualization that directly visualizes information related to the user's physical location~\cite{white09sitelens}.
    \protect\subref{subfig:fieldview}~FieldView: a situated analytics tool that supports professional activities conducted in the field~\cite{whitlock20fieldview}.}
    \label{fig:c2_images}
\end{figure*}

\subsubsection*{C1 -- Simulators}

\begin{table}[!ht]
    \sffamily
    \footnotesize
    \centering
    \begin{tabular}{p{1.7cm}p{2.5cm}p{1.3cm}p{1.5cm}} 
    \toprule
    \textbf{Situating trigger} & \textbf{View\newline  situatedness} & \textbf{Visual\newline encoding} & \textbf{Data\newline abstraction} \\ 
    \midrule
    referents/ location, object & world-registered/ absolute & 3D/2D & physical  \\ 
    \bottomrule
    \end{tabular}
\end{table}

The \textit{Simulators} cluster presents simulated data in situ (Figure~\ref{fig:c1_images}).
Such data representations add authenticity and immersion for the user.
To achieve this effect, many systems avoid using device-anchored views and instead register data to the world. 
These tools are absolute in their view situatedness. 
Works that belong to this cluster are triggered by mainly referents, either by objects or locations.
The data in this cluster are presented as a form of 2D or 3D glyph that emulates a real object. 
They are registered to the World in an absolute manner.
This may mean that current \SA{} systems are more focused in locating the data than providing changes, or interactions within the view. 

One example of this is Corsican Twin~\cite{prouzeau20corsiantwin}, an \SA{} tool for embedding data visualizations for building maintenance purposes~(Fig.~\ref{fig:c1_images}(a)). 
There are also simulators that utilize 2D visualization by superimposing another processed view on top of the scene the user is currently seeing. 
One example of this is RiverWalk, a visualization that superimposes historical images onto matching views of a city~\cite{cavallo16riverwalk}~(Fig.~\ref{fig:c1_images}(b)).

\subsubsection*{C2 -- Assistants} 

\begin{table}[ht!]
    \sffamily
    \footnotesize
    \centering
    \begin{tabular}{p{1.5cm}p{2.5cm}p{1.5cm}p{1.5cm}} 
    \toprule
    \textbf{Situating trigger} & \textbf{View\newline situatedness} & \textbf{Visual\newline encoding} & \textbf{Data\newline abstraction} \\ 
    \midrule
    referents/ location & device-anchored/ fixed & 2D/3D & abstract/ hybrid \\ 
    & world-registered/ absolute &  &  \\
    \bottomrule
    \end{tabular}
\end{table}

\textit{Assistants} aid the user with location-specific information (Fig.~\ref{fig:c2_images}).
The 19 works belonging to this cluster are all triggered by location.
Assistants attempt to support user's decision making processes about by presenting data relevant to the user's location. 
These systems often use both 2D and 3D components to represent data.
The task is observational, either being fixed to the device or being positioned in an absolute location in the world. 

One example of this type is FieldView~(Fig.~\ref{fig:c2_images}(d))~\cite{whitlock20fieldview}, an \SA{} tool for field analysts that uses HMDs and mobile tablet PCs to communicate and investigate field data.
The system presents various types of data that would be helpful for field analysts to conduct analysis of the land.
The data is presented not only as a chart in tablets, but also as a 3D object to the points of interest.
\rev{Another example is MARS~\cite{hollerer99mars}~(Fig.~\ref{fig:c2_images}(a)), which helps users explore locations of interest at Columbia University. 
The tool provides advanced interactions such as labeling places, superimposing past buildings at a location, linking relevant websites, and so on.}

\subsubsection*{C3 -- Planners}

\begin{table}[ht]
    \sffamily
    \footnotesize
    \centering
    \begin{tabular}{p{1.5cm}p{2.5cm}p{1.5cm}p{1.5cm}} 
    \toprule
    \textbf{Situating trigger} & \textbf{View\newline situatedness} & \textbf{Visual\newline encoding} & \textbf{Data\newline abstraction} \\ 
    \midrule
    context/ setting & world-registered/ relative & 3D & physical \\ 
    \bottomrule
    \end{tabular}
\end{table}

The third cluster, denoted as \textit{Planners}, shares some similarities with Simulators, yet its main difference is that the visualizations are not registered in specific locations, but are relative to a setting, such as a factory floor, a classroom or an operating theater. 
In that regard, they do not merely simulate the physical space in its absolute location, but enhance the sensemaking processes involved with that space.

\begin{figure*}[ht]
    \centering
    \subfloat[]{\includegraphics[height=0.153\textwidth]{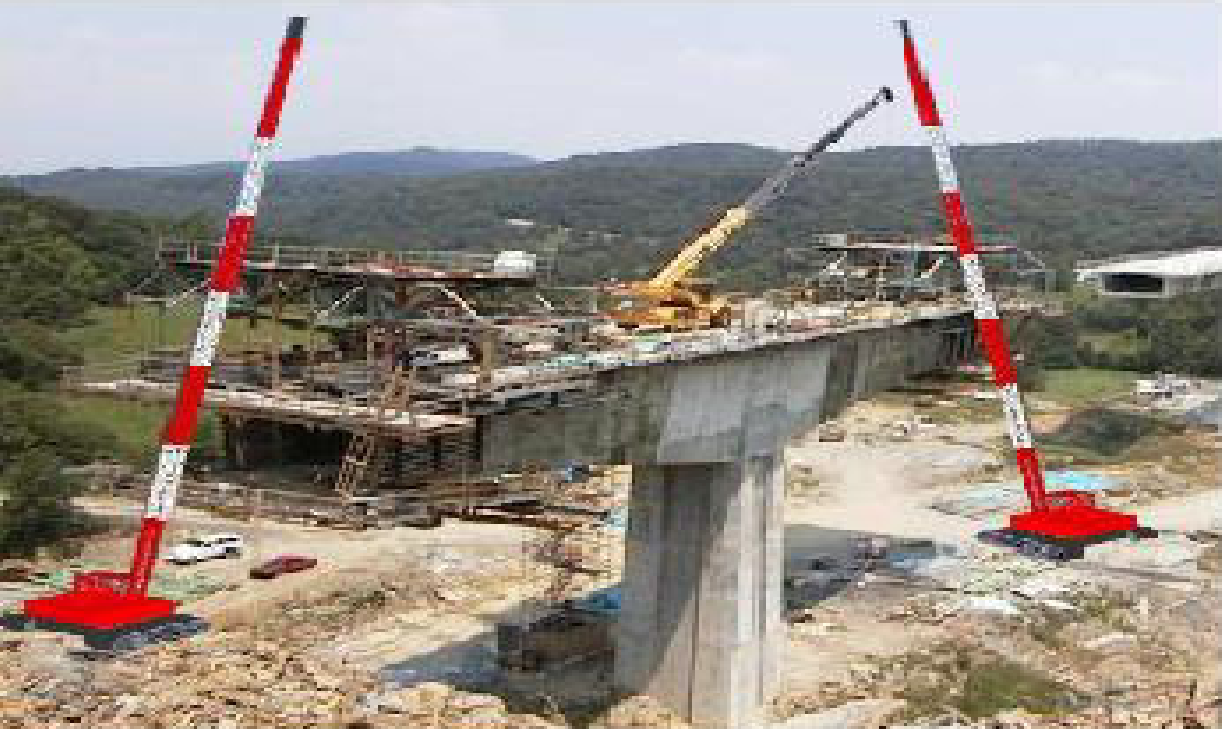}\label{subfig:construction}}\hspace{0.2mm}
    \subfloat[]{\includegraphics[height=0.153\textwidth]{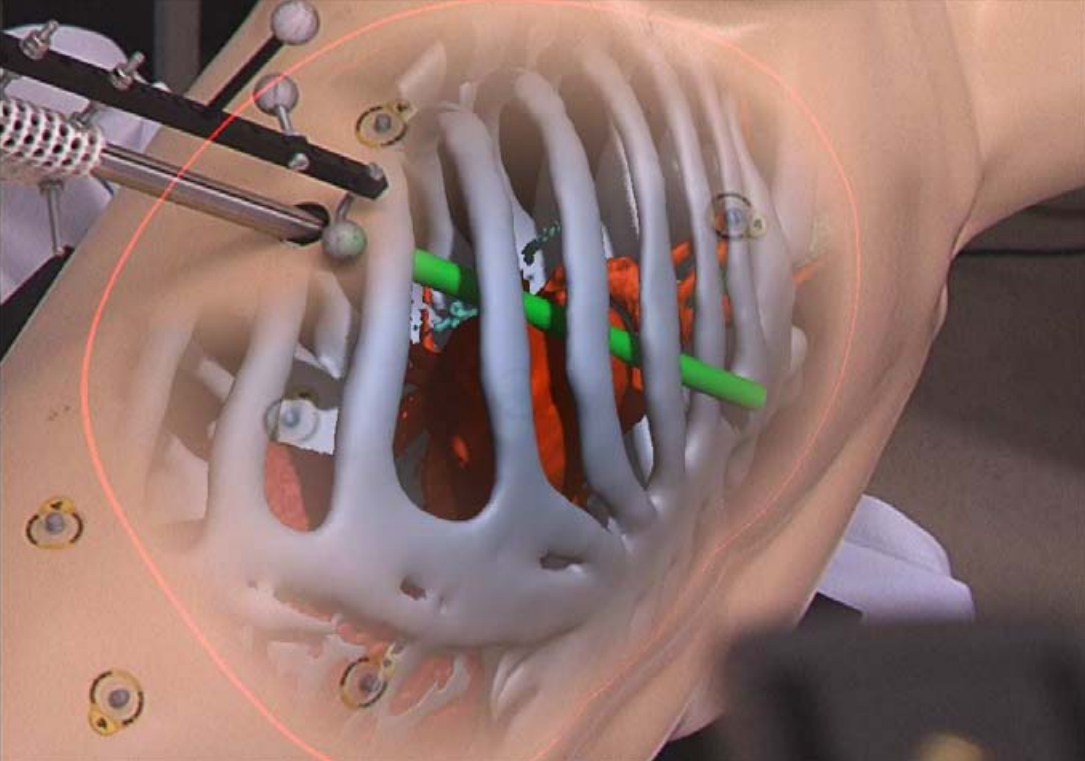}\label{subfig:mimesis}}\hspace{0.2mm}
    \subfloat[]{\includegraphics[height=0.153\textwidth]{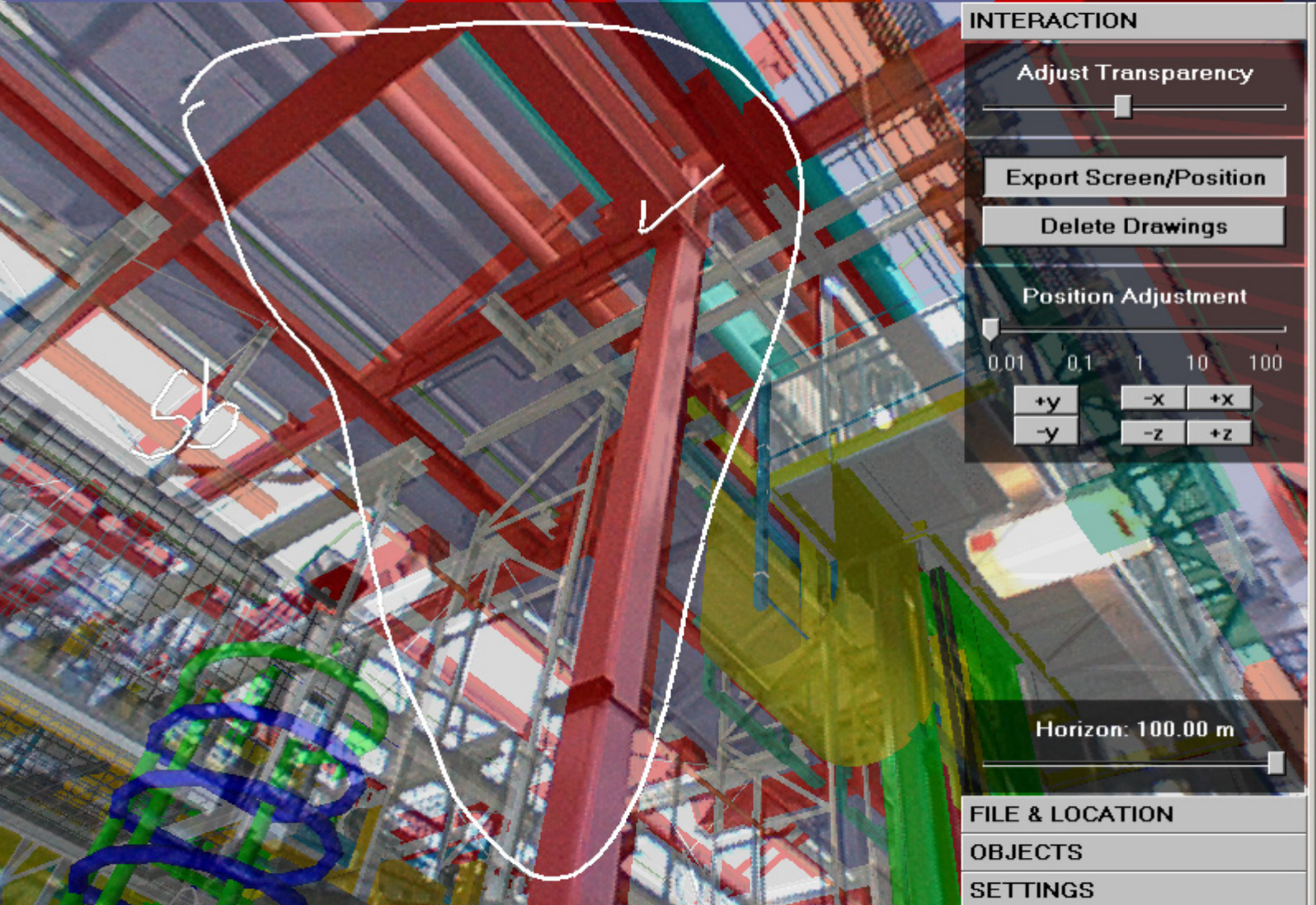}\label{subfig:buildacceptance}}\hspace{0.2mm}
    \subfloat[]{\includegraphics[height=0.153\textwidth]{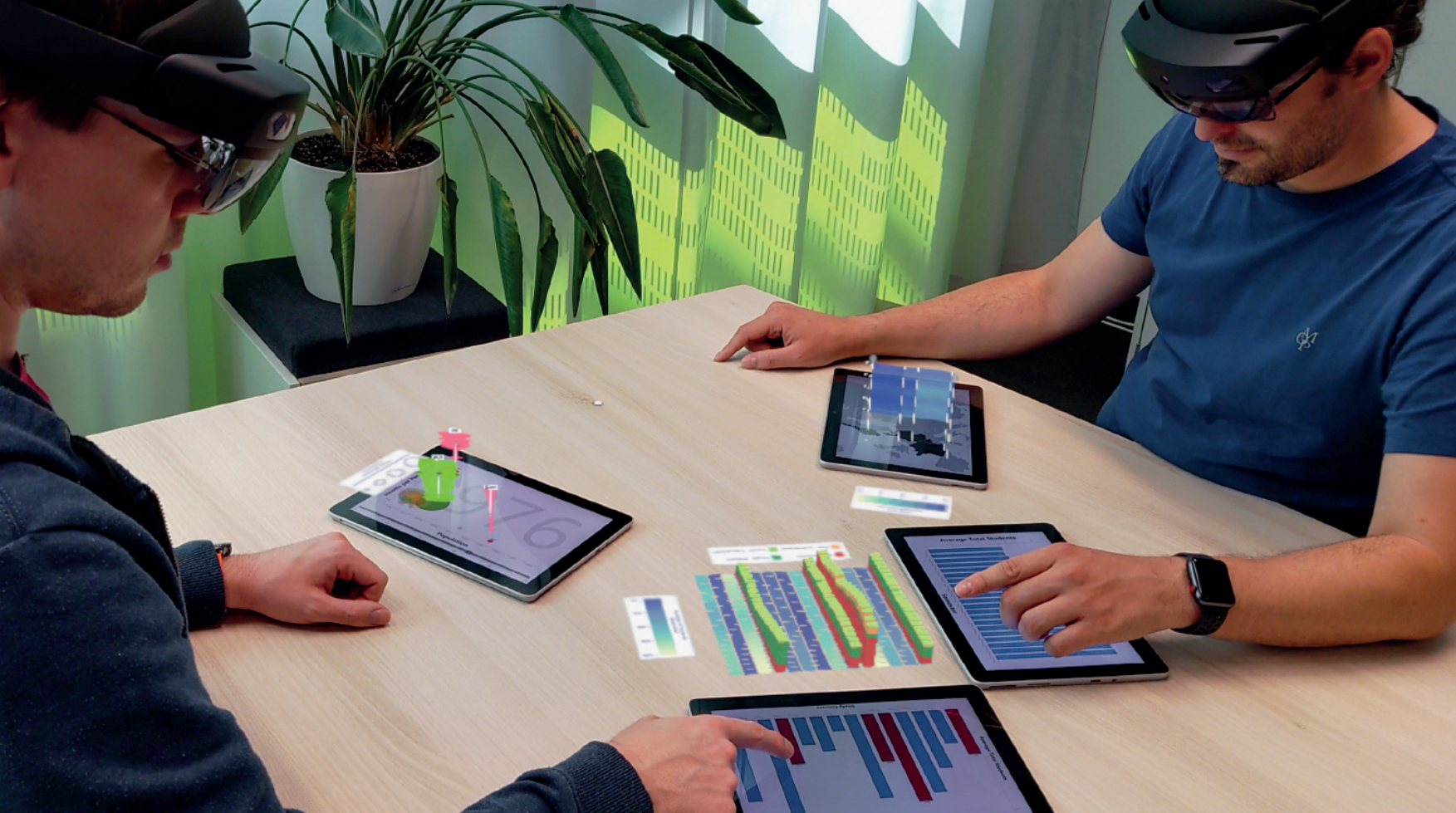}\label{subfig:marvis}}
    \caption{\textbf{Prototypical examples for \textit{Planners} (C3).}
    \protect\subref{subfig:construction}~\SA{} for construction site: a \SA{} tool where the user can find the optimal location for construction facilities prior to proceeding the operation~\cite{Behzadan05constructionAR}.
    \protect\subref{subfig:mimesis}~Contextual Anatomic Mimesis: an in-situ visualization to improve depth perception of doctors during medical surgery.~\cite{bichlmeier07mimesis}.
    \protect\subref{subfig:buildacceptance}~\SA{} for building acceptance~\cite{schoenfelder08industacceptance} A user can compare the condition of the building with a prototype of the plan.  
    \protect\subref{subfig:marvis}~MARVIS combines mobile devices and Augmented Reality for data analysis to enable collaboration between users~\cite{langner:2021:marvis}.
    }
    \label{fig:c3_images}
\end{figure*}

One example is a visualization of construction sites by Behzadan et al.~\cite{Behzadan05constructionAR} (Fig.~\ref{fig:c3_images}(a)). 
The user can move around physical 3D cues around the space to identify the optimal location for construction devices, such as cranes or excavators. Other examples include medical operation planning~\cite{OkurMROR} and real-time feedback prompts for joint psychomotor-cognitive training~\cite{kotranza09psychomotorAR}.

\subsubsection*{C4 -- Scanners}

\begin{table}[ht]
    \sffamily
    \footnotesize
    \centering
    \begin{tabular}{p{1.5cm}p{2.5cm}p{1.5cm}p{1.5cm}} 
    \toprule
    \textbf{Situating trigger} & \textbf{View\newline situatedness} & \textbf{Visual\newline encoding} & \textbf{Data\newline abstraction} \\ 
    \midrule
    referents/ object & device-anchored/ fixed & 2D & abstract \\ 
    \bottomrule
    \end{tabular}
\end{table}

\begin{figure*}[ht]
    \centering
    \subfloat[]{\includegraphics[height=0.172\textwidth]{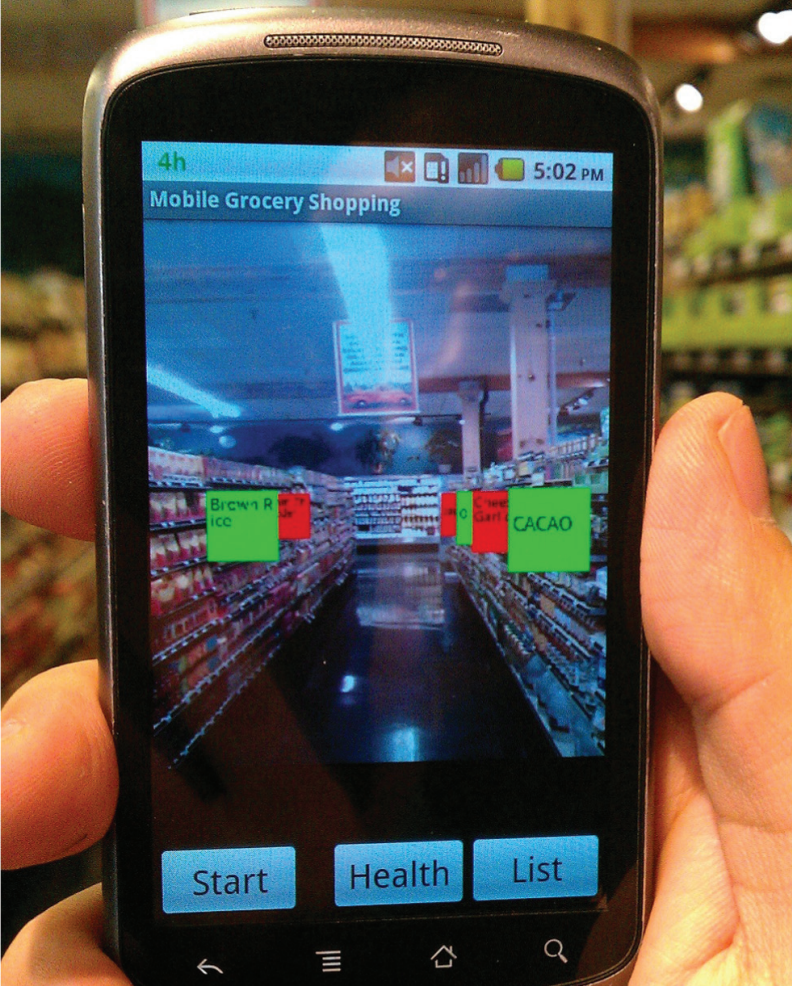}\label{subfig:groceryshopping}}\hspace{0.2mm}
    \subfloat[]{\includegraphics[height=0.172\textwidth]{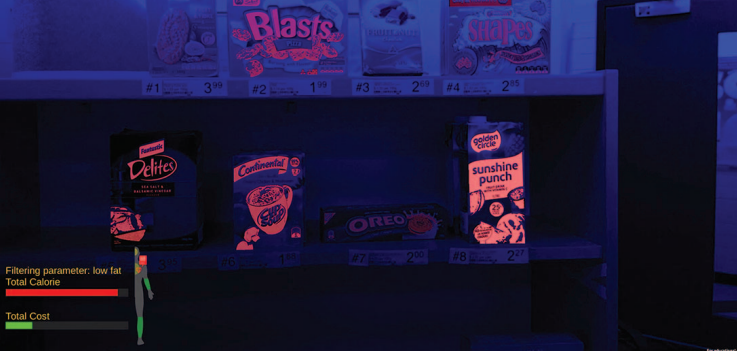}\label{subfig:horuseye}}\hspace{0.2mm}
    \subfloat[]{\includegraphics[height=0.172\textwidth]{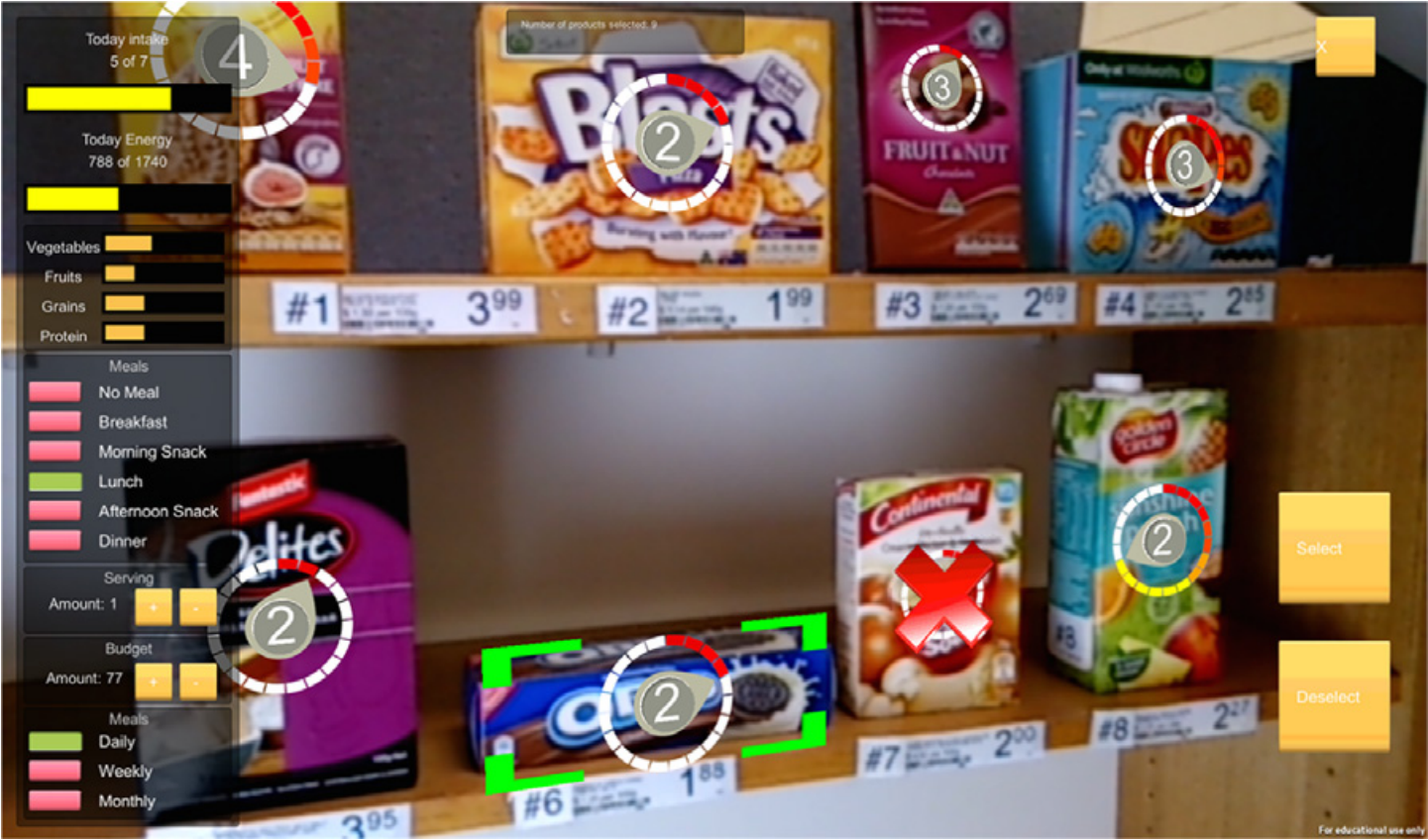}\label{subfig:shopping}}\hspace{0.2mm}
    \subfloat[]{\includegraphics[height=0.172\textwidth]{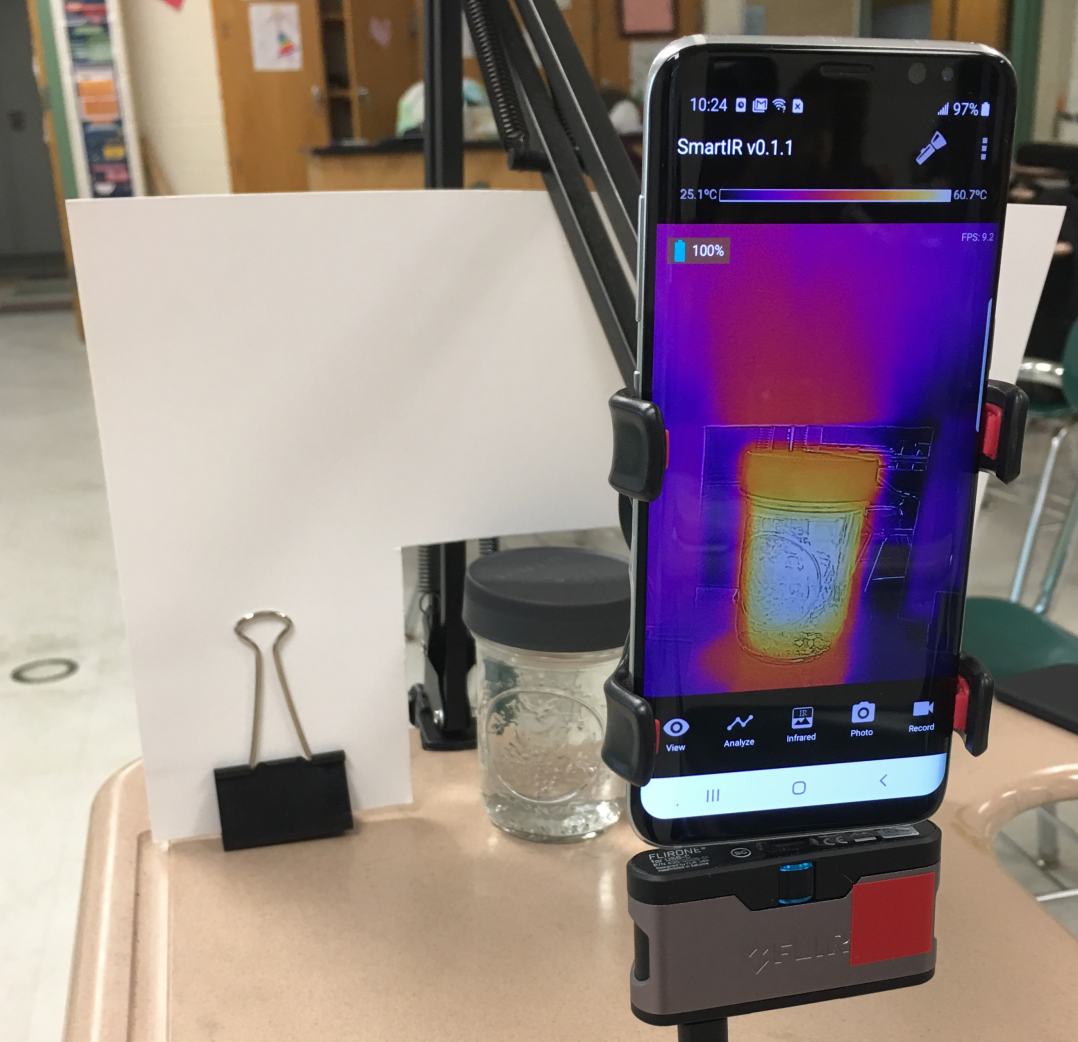}\label{subfig:ar_temp}}
    \caption{\textbf{Prototypical examples of \textit{Scanners} (C4).}
    \protect\subref{subfig:groceryshopping}~A tool that helps grocery shopping by presenting recommendations on customized healthy products~\cite{ahn13groceryar}. 
    \protect\subref{subfig:horuseye}~HorusEye: A technique that imitates bird and snake vision to highlight data of interest shown in objects~\cite{elsayed16horuseye}. 
    \protect\subref{subfig:shopping}~An \SA{} tool that lets users choose products by identifying and comparing products~\cite{elsayed16situateddef}.
    \protect\subref{subfig:ar_temp}~A system that shows changes in thermal images over time~\cite{jiang20finedetailsar}. 
    }
    \label{fig:c4_images}
\end{figure*}

The \textit{Scanners} cluster augments real-world objects with additional information, akin to a science fiction ``scanner'' sensor (Figure~\ref{fig:c4_images}).
The cluster is composed of 7 systems.
Triggered by objects, systems in this group present abstract information mainly in the form of 2D representations. 
The view is mainly fixed to the device.

One example of a Scanner is a \SA{} tool for shopping by Elsayed et al.~(Fig.~\ref{fig:c4_images}(c))~\cite{elsayed16situateddef}. 
When the user points the device camera at a product, the system displays the product's nutritional values, tracks the total amount of calories bought, and keeps a record of the amount of money spent based on a given budget.
Other Scanner examples include systems for crime scene investigation~\cite{datcu16csiar} as well as for visualizing the invisible fine scientific details~\cite{jiang20finedetailsar}~(Fig.~\ref{fig:c4_images}(d)).

\begin{figure}[ht]
    \centering
    \subfloat[]{\includegraphics[height=0.162\textwidth]{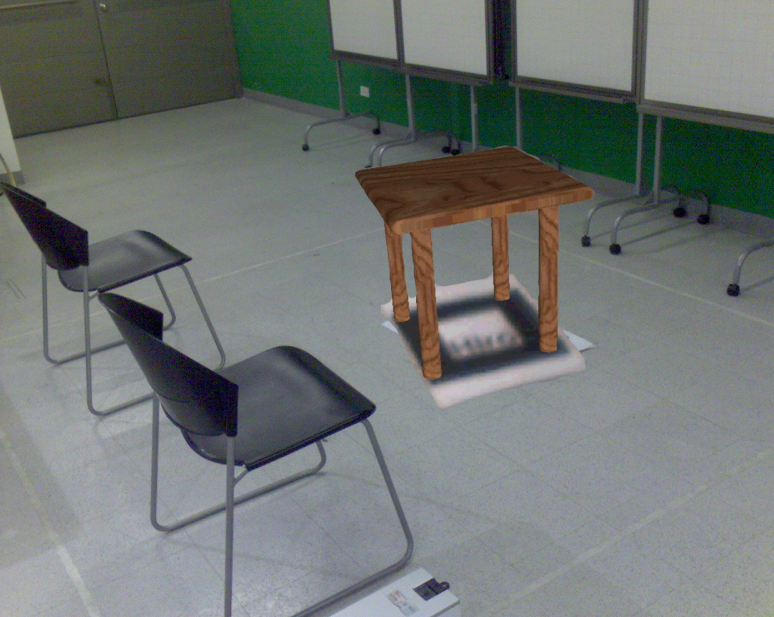}\label{subfig:sizeestimation}}\hspace{0.2mm}
    \subfloat[]{\includegraphics[height=0.162\textwidth]{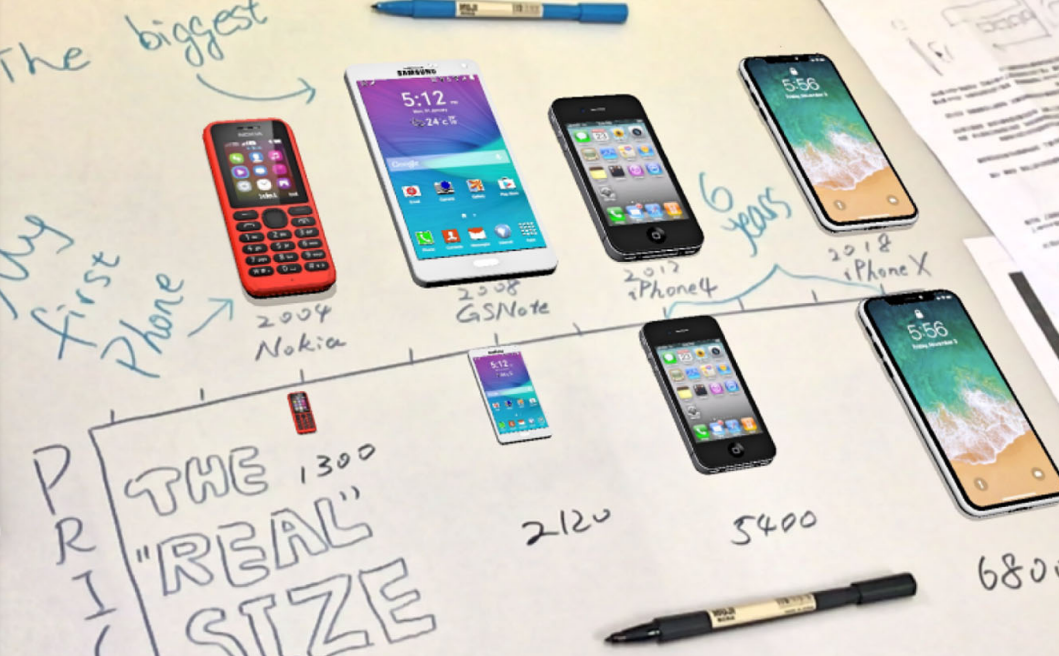}\label{subfig:marvist}}\hspace{0.2mm}
    \caption{\textbf{Examples of Outliers \rev{(O1-O2)}.}
    \protect\subref{subfig:sizeestimation}~A \SA{} tool that superimposes an augmented object at the user's point of interest~\cite{gomez08sizeestimationAR}.
    \protect\subref{subfig:marvist}~MARVisT: A technique that highlights the data of interest by using virtual objects as glyphs and bing data on them~\cite{chen19marvist}.
    \label{fig:A1_images}
    }
\end{figure}


\subsection{Outliers}

Beyond the four clusters discussed above, we also find \rev{two} outliers that do not conform to any specific cluster (see Table~\ref{table:taxonomization_table}).
\rev{Below we describe these two outliers.}

\textbf{Size estimation (O1).} 
The first type of outliers is a size estimation tool. 
Gomez and Figueroa~\cite{gomez08sizeestimationAR} enable estimating the exact size of furniture by visualizing it in the intended space using \AR{}~(Fig.~\ref{fig:A1_images}(a)).
This is also a type of a simulation, but most simulations in Simulators (C1) are triggered by referents anchored to an absolute position, without the ability to alter its size or location. 
Such systems allow a form of spatial exploration.

\textbf{Highlighters (O2).} 
\rev{This outlier is} capable of highlighting both abstract and physical information about a context. 
MARVisT~\cite{chen19marvist} effectively highlights the data of interest by using virtual objects as glyphs and by binding data on them~\rev{(Fig.~\ref{fig:A1_images}(b))}.



%% file: 05_discussion.tex
\section{Discussion}
\label{sec:discussion}

Here we discuss several notable observations drawn from our taxonomy as well as the surveyed papers that would help future situated analytics research and practice.

\subsection{General Observations}

Despite our efforts to collect historical situated analytics tools, \rev{almost half of the \SA{} systems (22/47) in this review} are from 2015 or later (Table~\ref{table:stats_table}).
There was less research on interactive data exploration in \AR{} before this time, possibly because the \AR{} equipment needed was highly specialized and expensive, making it mostly inaccessible to the data visualization field.
As a case in point, the commercial release of the Microsoft HoloLens was in 2016 and the situated analytics concept grew out of the immersive analytics movement in data visualization in that same year~\cite{elsayed16situateddef, Schmalstieg2016}.
\rev{Such improvement of AR technology has also contributed to the proliferation of \SA{} tools, as most systems with schematizing support were only developed after 2015.}

Overall, most of our surveyed \SA{} systems are triggered by referents, whether they are location or object; very few are triggered by context (see Table~\ref{table:stats_table}).
We speculate that this is because referents are generally easier to detect than context.
For view situatedness, systems have a slight preference for device-anchored rather than world-registered views.
\textit{Registration}, or the alignment of virtual objects with the real world, is a key challenge in \AR{}, which may explain the focus on device-anchored views that do not require accurate registration.
Note that whether device-anchored or world-registered, systems tend to place their data in a fixed, absolute rather than a relative location. 
Finally, abstract forms of data is the dominant type in data abstraction. 

\subsection{When to Use Situated Analytics?}

We may be approaching a future where wearable \AR{} devices become ubiquitous. 
Imagining such a future, it is clear that essentially \textit{all} computing tasks, from email to gaming, can be seen as being situated simply by virtue of being conducted in the field and on the go (in fact, with today's mobile computing technologies, this is very much the truth). 
However, situated analytics comes into its own when representing situated data.
The question is, what are the real ``killer apps'' of such situated computing capabilities?

Based on our survey, \textit{simulation}---where a situated representation of data in the world can contribute to immersion and understanding---is one such killer app.
Simulations are used to present data as a three-dimensional object that imitates the form of its original physical shape.
For example, MARVisT~\cite{chen19marvist} renders information using a glyph that looks like a real object.
Such imitation of real-world objects both increases the user's immersion as well as provides easy real-world references; for example, by showing the flood level of a historical hurricane in an afflicted area~\cite{haynes16floodAR}.
To that end, simulation techniques can be beneficial for training purposes especially for sophisticated tasks involving manual dexterity~\cite{henderson10tangibleAR}, such as medical surgery~\cite{bichlmeier09keyholesurgeryAR, bichlmeier07mimesis}, and construction~\cite{behzadan15reviewarconstruction}, or in hazardous conditions~\cite{liestol14climate}, such as when training first responders for an emergency or military personnel~\cite{livingston02augmented} for an upcoming operation.

Furthermore, \AR{} can also provide a see-through view not visible to our naked eyes. 
This is in line with the ``Enhanced Vision'' concept proposed by Willett et al.~\cite{willett22superpower}.
Examples include displaying underground infrastructure~\cite{schall09infravis} on top of the real world, or imagery of a patient's internal organs superimposed on their body to a surgeon.

At the same time, simulations are a canonical use-case of \AR{} in general, and so their ``killer app'' status is not surprising.
The real question is what other such critical and well-suited applications are in store for the future of \SA{}?
Based on our survey of the literature, we think that future \SA{} tools will excel at truly situated tasks where place itself can be used as an index into a dataset; for example, in space planning such as organizing the layout of a factory floor based on usage data, or in real-time command and control such as controlling emergency responders during a developing natural disaster.

\subsection{Interaction in Situated Analytics}

Interaction is a key component in visualization~\cite{Yi2007}.
However, current \SA{} systems are more focused on displaying the data rather than providing powerful ways for the user to interact with it.
This is a microcosm of the history of visualization, where interaction initially received short shrift largely because visual challenges were dominant.
Similarly, for situated analytics, registration and depiction are still significant challenges in \AR{}/\MR{} research, thus consuming much research attention.
Furthermore, efficient 3D interaction is generally an intractable problem, with an entire scientific community devoted to such techniques.

Nevertheless, analyzing our surveyed corpus of situated analytics systems, we find that many use innovative forms of interaction that leapfrog traditional visualization interaction techniques.
Some of this is due to the mobile setting, which necessitates using interaction methods beyond the traditional mouse and keyboard common with personal computers~\cite{Roberts2014}.
For example, many of the surveyed systems use so-called ``natural'' user interfaces~\cite{Wigdor2011} such as touch~\cite{elsayed16situateddef}, gestures~\cite{langner:2021:marvis}, and specialized controllers~\cite{livingston02augmented} to enable selecting, filtering, and drilling into data.

At the same time, interaction is also a fundamental element when implementing human input for sensemaking process.
However, while many tools are focused on providing interactions to read and explore the situation and their relevant data, they lack the ability for generating hypotheses and validating them (such as tools for authoring visualizations of the user's interest or understanding patterns from data in situ).
We speculate that this could inspire interesting future work. 
We envision that the development of automated chart authoring tools~\cite{moritz19draco}, natural language interfaces~\cite{srinivasan21nlutterance,kim21chartsandcaptions} for data analysis, and recent large language models (for example, GPT-3~\cite{brown20gpt3}) can help develop \SA{} tools that facilitate generating insights from the situation.

\subsection{Anchoring to the Device or the World}

\AR{} technology is evolving to provide better immersion and visual quality while increasing user mobility and decreasing unit weight. 
In that regard, device-anchored views, whether fixed or relative, can be seen as an antiquated form that estranges the user from being completely immersed into the environment.
Hence in the long term, we expect that the developers of situated analytics will be inclined to use world-registered interfaces in places that previously used device-anchored views. 
That said, we anticipate that device-anchored views will still have a role over world-registered views, such as when (1) reading text~\cite{ketchell19situatedstorytelling}, and when (2) interacting with controls~\cite{elsayed16situateddef, bladuini12bottari, whitlock20fieldview}.

Reading text yields the best performance when the text is both (a) facing the reader and (b) is a convenient size: not too large so that head movements are needed for reading, and not too small so that letters are not legible.
It follows that a device-anchored view is optimal for this task because fulfilling the above two conditions is trivial when the text is attached to the device.
In contrast, 3D text anchored to the world can run afoul of both conditions.
Using a billboard and scaling the text to mitigate them is not realistic and will thus inevitably impact the user's presence in the world.
This is doubly true for interface elements; a button placed in 3D not only must be facing the user and be at an appropriate distance to be seen, it must also not be obscured by other objects and be within reach for easy manipulation.
This is again trivial for a 2D button attached to the user's own device.
In all of these cases, placing elements on the device will break immersion, but can serve to remind the user that the view is ``augmented'' and not fully ``real.''

\subsection{Utility of 3D in Situated Analytics}
 
2D visualizations are the most common way of depicting data in our survey corpus, but 3D is slowly gaining popularity. 
Standard visualizations tend to be 2D in nature, and designing 3D representations that integrate well with the surrounding world is a challenge.
Furthermore, 3D is still seen with some skepticism in the visualization community (and rightly so); consider, for example, Munzner's rule of thumb of ``no unmotivated 3D''~\cite{munzner14visualization}.

Many of our observations from the survey are consistent with the benefits of presenting data in 3D by Willett et al.~\cite{marriott18immersiveanalytics}.
Intuitively, 3D representations that mimic real-world objects provide better immersiveness for the user.
This allows users to identify the impact of the data onto the world, such as estimating the impact of flood by different precipitation levels~\cite{haynes16floodAR}.
However, 3D visualization is considered a risky choice for representing data because of its propensity for occlusion~\cite{Elmqvist2008}, legibility of text, and utility on a 2D monitor screen, etc.
Recently, Satkowski and Dachselt~\cite{SatkowskuDachselt2021} found that reading 2D charts in an \AR{} environment is still effective, although slightly less so than when using 2D dashboards.
That said, data that is 3D in nature can often benefit from a 3D representation~\cite{munzner14visualization}.

Furthermore, a full 3D environment provides benefits not available in a 2D workspace.
For one thing, using \AR{} as a new platform for visualization provides freedom to move and arrange data at different angles, layouts, and locations than on a 2D screen~\cite{marriott18immersiveanalytics}.
Furthermore, the immersive 3D space can also facilitate a collaborative environment~\cite{langner:2021:marvis, ens21uplift} if the same chart can be seen by multiple users.
Stereographic effects may also be beneficial for certain 3D views.
Finally, switching from a 2D display to an augmented 3D world opens up new possibilities in presenting 3D data embedded in the place where it was collected.

\subsection{Evolution of AR Development Platforms}

The development ecosystems used for the situated analytics systems we have surveyed in this paper have evolved radically over the last 25 years.
Between 1997 and 2015, the platforms were mostly bespoke and custom-made for the specialized \AR{} equipment used (e.g.,~\cite{feiner97touringmachine, hollerer99mars, hollerer99sitdocument}).

However, since 2016, game engines such as Unity and Unreal Engine---that have versatile mechanisms for usage with contemporary Virtual and \AR{}/\MR{} equipment and are increasingly affordable---have become an important enabler for the development of contemporary situated analytics systems (e.g., for IATK~\cite{cordeil19iatk}, DXR~\cite{sicat19dxr}, and ImAxes~\cite{cordeil17imaxes}).
Game engines have strong synergies with mobile platform development ecosystems (ARCore and ARKit), and therefore could be considered a relatively unified development platform.
In addition, depending on the deployment hardware, they also support a variety of registration mechanisms---a feature that often required bespoke, often hybrid, implementations in the past.

However, commercial game engines have the drawback of being proprietary software ecosystems controlled by a single vendor.
While situated analytics research has yet to make an \textit{en masse} shift to open-standards web-based technologies~\cite{Roberts2014}, such as that pioneered by Butcher et al.~\cite{butcher20vria} (or by Badam et al.~\cite{Badam2019} for ubiquitous analytics), we believe that this will eventually be necessary.
Unity supporters will note its performance and convenience; we note that these are mostly engineering concerns.
As a case in point, this transition from desktop and native apps to web-based ones has already mostly happened for regular data visualization~\cite{Bostock2011}.

\subsection{Gaps and Opportunities}

One of the key benefits of conducting a survey such as ours is the opportunity to use the classification to not just identify current research trends, but also to identify gaps where no research has been conducted.
Such gaps may represent opportunities for future research that will significantly enrichen and advance the research area.

Based on the discussion above, we can see several opportunities for future growth in this area:

\begin{itemize}
    \item\textbf{Sensemaking support.}
    One important observation from our survey concerns the level of sensemaking support offered by most SA systems, with few offering capabilities for schematizing and none in our survey enabling reporting.
    It is clear that as \SA{} technology matures, more systems will also include the \textsc{Schematize} and \textsc{Report} levels of sensemaking support.

    \item\textbf{Increase analytical capabilities.}
    We think there is significant potential for creating more advanced analytical functionality in future situated analytics tools.
    Human-centered Artificial Intelligence may help us create ``supertools''~\cite{Shneiderman2022}; what about a situated Tableau?
    
    \item\textbf{Richer interaction.}
    Most current situated analytics systems support only limited interaction.
    We think that future situated analytics systems should better harness the mobile setting and the touch, gestural, and voice-based interaction modalities available there.
    
    \item\textbf{Richer data depictions.}
    Similarly, there is significant potential for drawing on existing research from \AR{}/\MR{} to create compelling visual representations, potentially even embedding them into the real-world so that they are virtually indistinguishable from real objects.
    
    \item\textbf{Interoperability.}
    All of the situated analytics systems surveyed are designed for exclusive and focused use. 
    But what if you want to use more than one tool at the same time?
    Future research in this space should focus on building a unified platform where multiple SA systems can co-exist and interoperate.
    
    \item\textbf{Context matters.}
    Few of the situated analytics systems we surveyed use event-based context as a situating trigger---the majority use the setting, object, or physical location to instantiate a situated visualization.
    To be clear, detecting a specific context is a challenging prospect, but as situated and contextual computing proliferates, we think that this will eventually become a key utility for SA: data views that appear implicitly and on-demand without the user having to request them.

\end{itemize}

\subsection{Limitations}

Our classification of situated analytics has several potential limitations.
For one thing, our four design dimensions are clearly not exhaustive, and nor are their values.
For example, our taxonomy assumes that all SA tools are initiated by a trigger, often by a situation. 
However, there are tools that utilize projection maps or physicalizations that present data near the point of interest~\cite{Willet2017_Embedded}; in other words, these are persistently on and not triggered.
We acknowledge that this is a possible scenario, and that our design space does not distinguish these systems.

We also do not claim that our survey corpus is exhaustive; in particular, there is a wealth of situated visualization in the \AR{} domain, some of which potentially could qualify for inclusion in our taxonomy.
Our overall goal with this work was to survey the field of situated analytics from a data visualization viewpoint.
For this reason, we opted to be representative rather than exhaustive in surveying older work from \AR{} prior to 2015.

%% file: 06_conclusion.tex
\section{Conclusion}
\label{sec:conclusion}

Situated analytics in \AR{} is increasingly within reach for a growing share of households as technology improves and becomes more inexpensive.
However, the design space of analytics tools that utilize \AR{} to support such situated analytical tasks is still largely unexplored. 
In this paper, we have elaborated on a rich design space, devised based on a classification of situated analytics systems that involves their situating triggers, view situatedness, data depiction, and data abstraction.
Using this classification, we have identified four archetypes of situated analytics systems. 
We also discuss several design lessons that we derived from our survey of the literature.